# The physics of hearing: fluid mechanics and the active process of the inner ear

**Tobias Reichenbach\* and A. J. Hudspeth**

Howard Hughes Medical Institute and Laboratory of Sensory Neuroscience,
The Rockefeller University, New York, NY 10065-6399, USA

E-mail: reichenbach@imperial.ac.uk and hudspaj@rockefeller.edu

Running head: The physics of hearing

## Abstract

Most sounds of interest consist of complex, time-dependent admixtures of tones of diverse frequencies and variable amplitudes. To detect and process these signals, the ear employs a highly nonlinear, adaptive, real-time spectral analyzer: the cochlea. Sound excites vibration of the eardrum and the three miniscule bones of the middle ear, the last of which acts as a piston to initiate oscillatory pressure changes within the liquid-filled chambers of the cochlea. The basilar membrane, an elastic band spiraling along the cochlea between two of these chambers, responds to these pressures by conducting a largely independent traveling wave for each frequency component of the input. Because the basilar membrane is graded in mass and stiffness along its length, however, each traveling wave grows in magnitude and decreases in wavelength until it peaks at a specific, frequency-dependent position: low frequencies propagate to the cochlear apex, whereas high frequencies culminate at the base. The oscillations of the basilar membrane deflect hair bundles, the mechanically sensitive organelles of the ear's sensory receptors, the hair cells. As mechanically sensitive ion channels open and close, each hair cell responds with an electrical signal that is chemically transmitted to an afferent nerve fiber and thence into the brain. In addition to transducing mechanical inputs, hair cells amplify them by two means. Channel gating endows a hair bundle with negative stiffness, an instability that interacts with the motor protein myosin-1c to produce a mechanical amplifier and oscillator. Acting through the piezoelectric membrane protein prestin, electrical responses also cause outer hair cells to elongate and shorten, thus pumping energy into the basilar membrane's movements. The two forms of motility constitute an active process that amplifies mechanical inputs, sharpens frequency discrimination, and confers a compressive nonlinearity on responsiveness. These features arise because the active process operates near a Hopf bifurcation, the generic properties of which explain several key features of hearing. Moreover, when the gain of the active process rises sufficiently in ultraquiet circumstances, the system traverses the bifurcation and even a normal ear actually emits sounds. The remarkable properties of hearing thus stem from the propagation of traveling waves on a nonlinear and excitable medium.

\* Present address: Department of Bioengineering, Imperial College London, London SW7 2AZ, United Kingdom

# Contents

## Glossary of terms

- *Apex of the cochlea*—the low-frequency region of the cochlea, at the top of its snail-like spiral.
- *Base of the cochlea*—the high-frequency region of the cochlea, at the bottom of its snail-like spiral.
- *Basilar membrane*—an elastic strip inside the cochlea that supports the organ of Corti and oscillates in response to sound.
- *Cochlea*—the inner ear; the organ responsible for sensitivity to sound.
- *Collagen*—the principal protein component of connective tissue, which occurs in elastic fibers that bind tissues together.
- *Deiters' cell*—the supporting cell between an outer hair cell and the basilar membrane.
- *Endolymph*—the $K^+$-rich, $Ca^{2+}$-poor solution in the scala media.
- *Epithelium*—a flat sheet of cells that are interconnected with one another and separate different liquid-filled compartments of the body.
- *Hair bundle*—the mechanosensitive organelle of a hair cell, comprising a cluster of several stereocilia and a single kinocilium.
- *Hair cell*—the mechanically sensitive cell of the cochlea; similar cells also detect acceleration in the organs of the vestibular system.
- *Helicotrema*—the liquid-filled connection between the scala vestibuli and scala tympani at the cochlear apex.
- *Hensen's cell*—a specific type of supporting cell within the organ of Corti.
- *Inner hair cell*—the type of cochlear hair cell that forwards electrical signals to the brain.
- *In vitro*—literally "in glass"; characteristic of experimental conditions outside an animal.
- *In vivo*—literally "in life"; characteristic of conditions in a largely intact animal.
- *Kinocilium*—a constituent of a hair bundle, resembling the cilia that provide motility to sperm and in the airways of the respiratory system.
- *Membrane potential*—the difference in voltage across the fatty membrane surrounding a cell.
- *Myosin*—a motor protein related to that responsible for the contraction of muscles.
- *Nerve fibre*—the extension of a nerve cell that connects to other neurons; synonymous with axon.
- *Organ of Corti*—a strip of cells on the basilar membrane that includes the hair cells.
- *Otoacoustic emission*—sound produced within and broadcast by the cochlea.
- *Outer hair cell*—the type of cochlear hair cell that provides mechanical amplification.
- *Oval window*— the elastic, membrane-covered opening in the cochlea's scala vestibuli that connects to the tiny bones of the middle ear.
- *Perilymph*—the $Na^+$-rich solution in the scala vestibuli and scala tympani.
- *Prestin*—the motor protein occurring in the membranes of outer hair cells and responsible for somatic motility.
- *Reissner's membrane*—an elastic strip inside the cochlea that separates the scala vestibuli from the scala media.
- *Round window*—the elastic, membrane-covered opening in the cochlea's scala tympani.
- *Scala*—one of three liquid-filled compartments inside the cochlea; the scala vestibuli, scala media, and scala tympani are delimited by Reissner's membrane and the basilar membrane.
- *Somatic motility*—the piezoelectric process by which an outer hair cell changes length in response to altered membrane potential; synonymous with electromotility.

- *Stereocilium*—One of the cylindrical extensions from the top or apical surface of a hair cell and the site of mechanoelectrical transduction; a component of the hair bundle.
- *Soma*—a cell body, especially of a hair cell or nerve cell.
- *Supporting cell*—an epithelial cell that adjoins a hair cell and provides metabolic assistance.
- *Synapse*—a cellular structure through which a hair cell or neuron can transmit an electrical signal to another neuron, usually through the intervention of a chemical signal.
- *Tectorial membrane*—an acellular gel in the cochlea that attaches to the hair bundles of outer hair cells and deflects them during oscillations of the basilar membrane.

# 1. Introduction

The performance of the human ear would be as remarkable for a carefully engineered device as it is for a product of evolution. The frequency response of a normal human ear extends to 20 kHz, and that of an ear in a whale or bat can exceed 100 kHz. Measured at the eardrum, an ear is sensitive to mechanical stimuli of picometer dimensions. The dynamic range of human hearing encompasses 120 dB of sound-pressure level (SPL), a millionfold range in input amplitude and a trillionfold range of stimulus power. Explaining how the ear meets these technical specifications is a major challenge for biophysics.

In this review we describe the biophysical principles through which the ear achieves its remarkable performance. A main focus is the active process of the cochlea, which provides tuned mechanical amplification of weak signals. Our presentation involves theoretical descriptions of the main biophysical aspects as well as illustrations and descriptions of key experiments.

We begin our discussion at the level of the whole inner ear, which employs specialized hydrodynamics to separate a complex sound mechanically into its distinct frequency components (Section 2). The active process is introduced through a historical overview of its discovery (Section 3). We then delve into the cellular and molecular components that mediate mechanotransduction, that is, the transformation of the mechanical sound signal into electrical action potentials in neurons and thus the language of the brain (Section 4). Mechanotransduction is achieved in specialized hair cells that also exhibit mechanical activities that underlie the active process. From there we return to the active hydrodynamics of the inner ear, starting from a description of the active micromechanics of the organ of Corti that houses the hair cells (Section 5) and continuing to active wave propagation (Section 6). Important insight into the cochlea's active fluid dynamics comes from otoacoustic emission, sound that is produced inside the inner ear through the active process and emitted into the ear canal (Section 7). We conclude with a brief overview of future research goals.

## 1.1. Structure of the ear

The external ear acts as a funnel to collect airborne sound vibration that the ear canal transmits to the eardrum (figure 1(a)). The middle ear consists of the three smallest bones in the body—the malleus, incus, and stapes—that together convey the eardrum's vibration to the oval window, an elastic opening in the bony casing of the cochlea or inner ear.

Vibration of the oval window elicits waves within the cochlea. Whereas sound propagates in air as longitudinal pressure waves, the relevant signals inside the cochlea travel as surface waves on two elastic structures, the basilar membrane and Reissner's membrane, that partition the inner ear into three parallel, liquid-filled scalae (figure 1(b),(c)). Because the liquid may be regarded as incompressible, the volume displaced by the moving oval window must be compensated by the motion of liquid elsewhere; this is accomplished at the elastic round window. Detailed descriptions of the cochlea's hydrodynamics and its capacity to spatially separate different frequencies are the subject of Section 2.

How is sound-induced vibration inside the cochlea transduced into electrical signals in nerve fibers? Specialized sensory receptors called hair cells are situated along the basilar membrane in the organ of Corti (figure 1(c)). Each of these cells has at its apex a hair bundle consisting of rigid, parallel cylindrical protrusions termed stereocilia. An appropriately directed displacement of this bundle opens mechanosensitive ion channels, allowing cations such as $K^+$ and $Ca^{2+}$ to flow into the cell and depolarize it. Because a hair cell is connected to the basilar membrane in such a way that vibration of the membrane displaces the hair bundle, sound elicits an electrical signal inside the cell. When the change in the membrane potential is large enough, it triggers the release of chemical neurotransmitter at the cellular base and elicits action potentials in the associated auditory-nerve fibers.

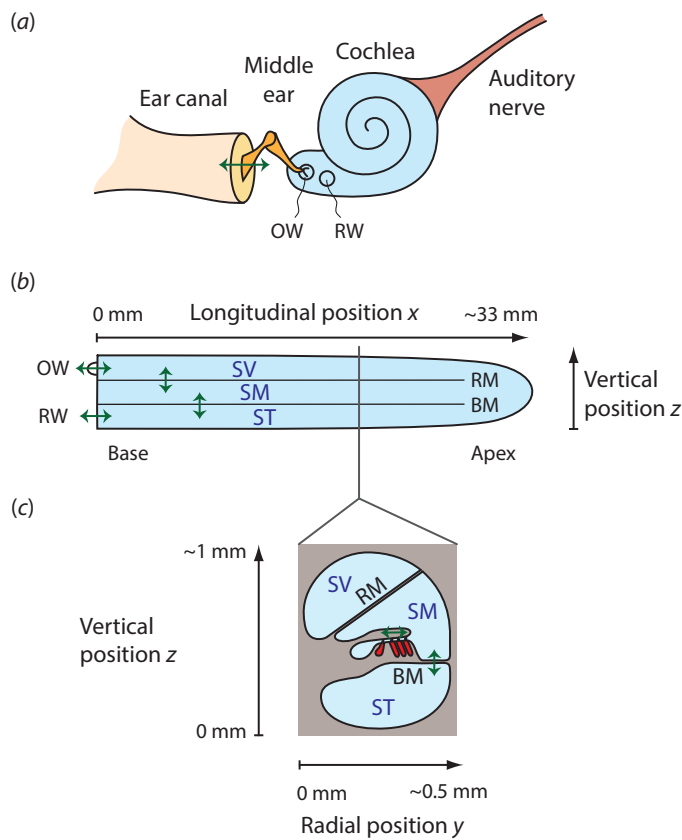

**Figure 1.** Structure of the ear. (*a*) Sound propagating through air as a compressive wave is funneled onto the eardrum by the external ear and the ear canal. The middle ear conveys the eardrum's resulting vibration (green arrow) to the inner ear, or cochlea, where it oscillates the elastic oval window (OW). A second elastic membrane, the round window (RW), can also oscillate to compensate for the resulting fluid displacement inside the cochlea. (*b*) A longitudinal and (*c*) a transverse section of the uncoiled cochlea show its interior structure. Two elastic membranes, the basilar membrane (BM) and Reissner's membrane (RM), delineate three liquid-filled chambers: the scala vestibuli (SV), scala media (SM), and scala tympani (ST). The mechanosensitive hair cells (red) are embedded in the organ of Corti on the basilar membrane.

The influx of $K^+$ and $Ca^{2+}$ through the hair bundle's mechanotransduction channels is enhanced by an additional cochlear specialization. The hair bundles are bathed in endolymph, a $K^+$-rich solution that fills the scala media. In contrast, the scala vestibuli and scala tympani contain perilymph whose ionic composition is dominated by $Na^+$ instead of $K^+$. Because endolymph has an enhanced potential of 80-120 mV as compared to perilymph—the so-called endocochlear potential—and the somata of hair cells are negatively polarized at about -50 mV, a potential difference of 150 mV across the stereociliary membranes drives cations through the mechanotransduction channels.

The physiological function of Reissner's membrane now becomes apparent. Like the basilar membrane, it is covered by a layer of epithelial cells that provides the electrical insulation required to sustain the endocochlear potential. In contrast to the basilar membrane, however, its displacement does not serve a physiological function. We discuss in chapter 7 evidence that vibrations of Reissner's membrane play a role in otoacoustic emission, that is, sound produced by the cochlea.

Remarkably, hair cells not only transduce mechanical signals into electrical ones, but also amplify weak mechanical stimuli through the production of mechanical forces. Section 3 gives an overview of the discovery of this active process and Section 4 describes the motile machinery of hair cells in more detail. How exactly the activity of hair cells influences the mechanics of the cochlea remains incompletely understood. A description of fundamental considerations regarding this question and of recent research are the scope of Sections 5-7.

## 2. Fluid dynamics of the passive cochlea

Oscillatory mechanical vibration can elicit different types of waves in liquid media. Compression and expansion of a fluid, for instance, results in longitudinal sound waves. As another example, surface waves can arise at the deformable interface between two media. Such waves occur on the basilar membrane inside the cochlea and transmit mechanical vibration to the hair cells. This section provides a detailed discussion of those waves. We start with a review of sound waves; the mathematical formalism developed thereby is then extended to describe surface waves on the basilar membrane.

### 2.1. Sound propagation in a liquid medium

The Navier-Stokes equation for the velocity $\boldsymbol{u}$, pressure $p$, and density $\rho$ of an inviscid fluid reads

$$\rho \partial_t \boldsymbol{u} + \rho (\boldsymbol{u} \boldsymbol{\nabla}) \boldsymbol{u} = -\boldsymbol{\nabla} p \ . \tag{2.1}$$

For oscillatory motion at an angular frequency $\omega$, the quadratic velocity term $(\boldsymbol{u} \boldsymbol{\nabla}) \boldsymbol{u}$ owing to convective acceleration can be ignored if the displacement amplitude $a$ is much smaller than the wavelength $\lambda$. Indeed, the amplitude of the fluid velocity $u$ is $a\omega$, such that the term $\partial_t \boldsymbol{u}$ in equation (2.1) is of the order $a\omega^2$, whereas the term $(\boldsymbol{u} \boldsymbol{\nabla}) \boldsymbol{u}$ is of the order $a^2 \omega^2 / \lambda$. When $a << \lambda$ it follows that the quadratic term is small, $(\boldsymbol{u} \boldsymbol{\nabla}) \boldsymbol{u} << \partial_t \boldsymbol{u}$.

The vibrational amplitudes of sound are tiny. As an example, the displacement associated with a very loud sound of 100 dB SPL at a frequency of 1 kHz is several micrometers for sound in air or around 2 nm for sound in water. The corresponding wavelengths are several orders of magnitude larger, about 300 mm in air and more than 1 m in water. The surface waves on the basilar membrane described below have wavelengths of the order of 1 mm, whereas the displacement is below 1 μm even for loud sounds. We can therefore safely ignore the quadratic term $(\boldsymbol{u} \boldsymbol{\nabla}) \boldsymbol{u}$ in the Navier-Stokes equation.

Because we consider only small displacements, the density changes are also small and occur around a mean density $\rho_0$. To leading order we can approximate $\rho \partial_t \boldsymbol{u} \approx \rho_0 \partial_t \boldsymbol{u}$. The Navier-Stokes equation (2.1) thus becomes an equation of momentum that relates an acceleration—that is, a temporal change in the velocity—to a force, in this instance owing to a pressure gradient:

$$\rho_0 \partial_t \boldsymbol{u} = -\boldsymbol{\nabla} p \ . \tag{2.2}$$

The equation of continuity informs us that a temporal change in density must be related to a net flux:

$$\partial_t \rho = -\rho_0 \nabla \boldsymbol{u} \ . \tag{2.3}$$

On the other hand, a density change follows from a local pressure change owing to the liquid's compressibility $\kappa$:

$$\partial_t \rho = \rho_0 \kappa \partial_t p \ . \tag{2.4}$$

We thus obtain

$$\nabla \boldsymbol{u} = -\kappa \partial_t p \ , \tag{2.5}$$

which, combined with equation (2.2) for the fluid's momentum, yield a wave equation for the pressure:

$$\Delta p = \rho_0 \kappa \partial_t^2 p \ . \tag{2.6}$$

Velocity and density follow through equations (2.2) and (2.4) and obey analogous wave equations. Equation (2.6) describes the propagation of sound through local compression of the liquid medium, producing a wave of wavelength

$$\lambda = 2\pi / (\omega \sqrt{\rho_0 \kappa}) \tag{2.7}$$

that progresses at the speed

$$c = 1 / \sqrt{\rho_0 \kappa} \ . \tag{2.8}$$

## 2.2. Surface waves on the basilar membrane

Waves of a different type are key in the cochlea (de Boer, 1980, 1984, 1991, Lighthill 1981, Pickles 1996, Ulfendahl 1997, Robles and Ruggero 2001). Sound signals propagate as surface waves on the elastic basilar membrane with a much smaller wavelength and hence speed than a sound wave moving through a fluid. To understand the physics of a surface wave, consider a two-dimensional section along the uncoiled cochlea in which the longitudinal position $x$ measures the distance from the base toward the apex. The vertical position $z$ denotes the distance from the resting position of the basilar membrane (figure 2(*a*)). A positive value $z$ therefore denotes a position inside the upper chamber, which combines scala media and scala vestibuli. The unstimulated basilar membrane lies at $z=0$ and negative values of $z$ occur in the lower chamber, the scala tympani. We assume that both chambers have an equal height $h$, such that the upper and lower walls are located at respectively the positions $z=h$ and $z=-h$.

In combining the scala vestibuli and scala media into a single compartment we ignore Reissner's membrane. At least within the basal and middle turns of the cochlea we can justify this simplification, for the impedance of Reissner's membrane there is well below that of the basilar membrane (Bekesy 1960). In the apical turn, however, the impedances of the two membranes become comparable with largely unknown consequences for cochlear hydrodynamics. In subsection 7.4 we discuss a cochlear model that includes both membranes.

Let $p^{(1)}$ and $p^{(2)}$ be the pressures in respectively the upper and lower chambers and denote by $\boldsymbol{u}^{(1)}$ and $\boldsymbol{u}^{(2)}$ the corresponding velocities of fluid elements in the vertical dimension. The wave that we are seeking is influenced critically by the impedance of the basilar membrane. Because the impedance depends on the frequency of stimulation—through contributions from inertia, viscosity, and stiffness—we consider stimulation at a single angular frequency $\omega$. The pressures and velocities then read $p^{(1,2)} = \tilde{p}^{(1,2)} e^{i\omega t} + c.c.$ and $\boldsymbol{u}^{(1,2)} = \tilde{\boldsymbol{u}}^{(1,2)} e^{i\omega t} + c.c.$ Here and in subsequent appearances, "$c.c.$" represents the complex conjugate and the tilde denotes a Fourier coefficient.

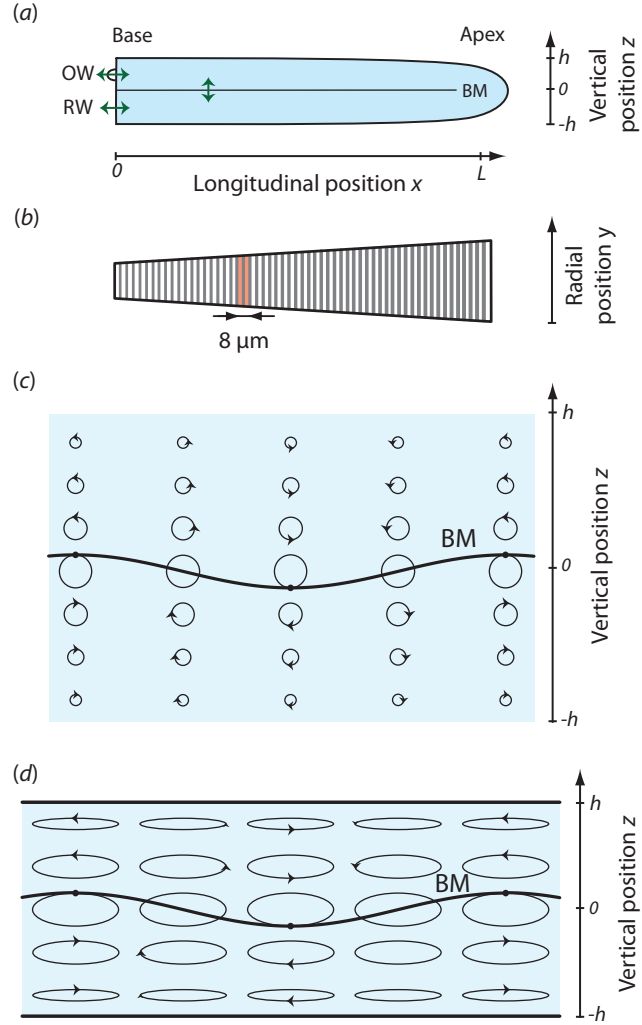

**Figure 2.** Wave propagation in the cochlea. (*a*) Two-dimensional model of the uncoiled cochlea. The scala vestibuli and scala media are represented as a single chamber, separated from the scala tympani by the basilar membrane (BM). (*b*) Surface view on the basilar membrane. Collagen fibers (grey) run from the neural to the abneural edge. The longitudinal coupling between them is small, and a membrane segment (red shading) of the width of one hair cell, about 8 $\mu$m, can accordingly be considered as mechanically uncoupled from neighboring segments. (*c*) In the case of a small wavelength, $\lambda << h$, the trajectories of liquid particles are circular and their amplitudes decay exponentially away from the membrane. Each arrow's head denotes the position of a particle at a given time. (*d*) In the instance of a large wavelength, $\lambda >> h$, motion occurs predominantly in the longitudinal direction.

The pressures obey the wave equation (2.6) derived above, which for the Fourier coefficients takes the form

$$\Delta \tilde{p}^{(1,2)} = -\omega^2 \rho_0 \kappa \tilde{p}^{(1,2)} \ . \tag{2.9}$$

Boundary conditions arise at the cochlear walls, where the vertical velocity must vanish: $\tilde{u}_z^{(1)}\big|_{z=h} = \tilde{u}_z^{(2)}\big|_{z=-h} = 0$ . Equation (2.2) for the fluid momentum provides a linear relation between the

vertical velocity and the $z$-derivative of the pressure: $\partial_z \tilde{p}^{(1,2)} = -i\omega\rho_0 \tilde{u}_z^{(1,2)}$, such that the vertical variation in the pressure vanishes at the cochlear walls:

$$\partial_z \tilde{p}^{(1)}\Big|_{z=h} = \partial_z \tilde{p}^{(2)}\Big|_{z=-h} = 0 \ . \tag{2.10}$$

Another boundary condition concerns the vertical velocity $\tilde{V} = \tilde{u}_z^{(1,2)}\Big|_{z=0}$ of the basilar membrane, which is related to the pressure difference across the membrane through its acoustic impedance $Z$:

$$\tilde{V} = \frac{1}{Z}\left(\tilde{p}^{(2)} - \tilde{p}^{(1)}\right)\Big|_{z=0} \ . \tag{2.11}$$

Such a linear relation between pressure difference and membrane velocity may be assumed because the amplitudes of the sound vibrations are tiny, as argued above. In a living cochlea, however, the mechanical activity of hair cells amplifies the basilar membrane's displacement and yields a nonlinear relation between membrane velocity and pressure difference; we discuss this in subsection 6.3.

The basilar membrane's vertical velocity also follows from the pressures through equation (2.2) for the fluid momentum, which yields $-i\omega\rho_0 \tilde{V} = \partial_z \tilde{p}^{(1)}\Big|_{z=0} = \partial_z \tilde{p}^{(2)}\Big|_{z=0}$ and hence

$$\partial_z \tilde{p}^{(1)}\Big|_{z=0} = \partial_z \tilde{p}^{(2)}\Big|_{z=0} = -\frac{i\rho_0\omega}{Z}\left(\tilde{p}^{(2)} - \tilde{p}^{(1)}\right)\Big|_{z=0} \ . \tag{2.12}$$

The hydrodynamics of the two-dimensional cochlear model is therefore described by equation (2.9) with boundary conditions **Error! Reference source not found.** and (2.12). Although these equations are linear in the pressures, an analytical solution is complicated by the longitudinal variation in the impedance, $Z = Z(x)$. In the next subsection we discuss how the model can nevertheless be solved through an approximation. Here we proceed with an analysis of the simpler case in which the basilar-membrane impedance is spatially constant; the solution will then inform the subsequent approximation.

Consider stimulation at a single angular frequency $\omega$, and make the following ansatz for the pressures $\tilde{p}^{(1)}$ and $\tilde{p}^{(2)}$:

$$\begin{aligned}
\tilde{p}^{(1)} &= \hat{p}^{(1)} \cosh[k(z-h)]e^{-ik_x x} + c.c., \\
\tilde{p}^{(2)} &= \hat{p}^{(2)} \cosh[k(z+h)]e^{-ik_x x} + c.c.
\end{aligned} \tag{2.13}$$

The wave vector $k_x$ implies that the wave travels in the longitudinal direction at a phase velocity $c = \omega / k_x$. This wave speed must be distinguished from the vertical velocity $V$ at which any element of the basilar membrane vibrates. The coefficient $k$ defines a length scale $1/k$ over which the amplitude decays in the vertical direction.

The boundary conditions (2.10) at the cochlear walls are satisfied with the above ansatz, and equation (2.9) requires that

$$k_x^2 = k^2 + \rho_0 \kappa \omega^2 \ . \tag{2.14}$$

The boundary condition (2.12) at the membrane can be written in the matrix form

$$(A - \zeta I)\hat{p} = 0 \tag{2.15}$$

in which the vector $\hat{p} = \left(\hat{p}^{(1)}, \hat{p}^{(2)}\right)^T$ contains both pressure amplitudes, $A = \begin{pmatrix} 1 & -1 \\ -1 & 1 \end{pmatrix}$ is a 2x2 matrix, $I$ is the 2x2 identity matrix, and $\zeta = iZk\tanh[kh]/(\rho_0\omega)$.

The matrix equation (2.15) is fulfilled when $\zeta$ is an eigenvalue of the matrix $A$ and when $\hat{p}$ is a corresponding eigenvector. The matrix $A$ has two eigenvalues and two corresponding eigenvectors,

$$\zeta_{(1)} = 0, \quad \boldsymbol{e}_{(1)} = \begin{pmatrix} 1 \\ 1 \end{pmatrix};$$

$$\zeta_{(2)} = 2, \quad \boldsymbol{e}_{(2)} = \begin{pmatrix} 1 \\ -1 \end{pmatrix}.$$

(2.16)

There accordingly exist two solutions and hence two modes of wave propagation in the cochlea.

In the first solution, which follows from $\zeta_{(1)}$ and $\boldsymbol{e}_{(1)}$, the pressures are the same on both sides of the membrane. The membrane's displacement thus vanishes, as does the vertical variation in pressure and velocity ($k$=0). The wave emerges from longitudinal variation only and represents the conventional sound wave derived in the previous subsection (equation (2.6)). For this wave the two-chamber architecture of the cochlea is inconsequential, for the wave elicits no basilar-membrane displacement. Although this wave mode has no known importance for the physiological functioning of the inner ear, it might be involved in the backward propagation of otoacoustic emissions (Section 7).

The second solution is physiologically relevant. The pressure changes on both sides of the membrane are equal in magnitude but of opposite sign. The membrane displacement does not vanish, and the quantity $k$ is determined by the dispersion relation

$$k \tanh[kh] = -2i\rho_0 \omega / Z .$$

(2.17)

The wave vector $k_x$ follows from equation (2.14). For realistic values of the cochlear impedance $Z$, however, the quantity $k^2$ greatly exceeds $\rho_0 \kappa \omega^2$ and hence $k_x \approx k$. The wavelength follows as $\lambda = 2\pi / k$ and is considerably smaller than that of the compressional wave, equation (2.7).

Let us pause to see how the membrane's impedance $Z$ shapes the wave. The basilar membrane consists of parallel collagen fibers that run radially between the neural and the abneural sides of the temporal bone (Figure 2(b)). The longitudinal coupling between the fibers is weak: its space constant has been estimated as only 20 $\mu m$ (Emadi *et al* 2004) or as between 10 $\mu m$ and 50 $\mu m$, depending on the longitudinal position in the cochlea (Naidu and Mountain 2001). The membrane may therefore be treated as a longitudinal array of thin, uncoupled segments. Each segment possesses a mass $m$, friction coefficient $\xi$, and stiffness $K$ and responds to a pressure difference across it according to Newton's equation of motion. Those can be written as

$$Z\tilde{V} = \tilde{p} ,$$

(2.18)

in which $Z$ denotes the acoustic impedance of the membrane segment that we assume to have an area $A$ (figure 2(b)):

$$Z = \left( i\omega m + \xi - iK / \omega \right) / A .$$

(2.19)

Under what conditions does this impedance yield a propagating wave? The real part of the wave vector $k$ encodes the wavelength and its imaginary part describes the amplitude change owing to damping. For a traveling wave the wave vector must accordingly have a non-vanishing real part, which necessitates a positive real part of the left-hand side of the dispersion relation (2.17). The right-hand side of this equation has a positive real part only when the imaginary part of the impedance (2.19) is negative, that is, when the impedance is dominated by stiffness. In the opposite case, when mass dominates stiffness, the wave vector is purely imaginary; an evanescent wave then arises. Both conditions occur in the inner ear. Indeed, in subsection 2.5 below we show that the cochlea ingeniously achieves frequency selectivity by employing a resonance at the transition from stiffness- to mass-dominated impedance.

The dispersion relation (2.17) remains constant upon changing the sign of the wave vector, so a solution with a certain wave vector $k$ implies that $-k$ satisfies the dispersion relation as well. It follows that a given wave can propagate both forward and backward along the membrane.

Regarding the wavelength, two limiting cases are important (Lighthill 1981). First, when the wavelength is much less than the height of the channels, that is when $kh \gg 1$, we can approximate $\cosh[k(z-h)] \approx e^{k(h-z)}/2$ for $z > 0$ and $\cosh[k(z+h)] \approx e^{k(z+h)}/2$ for $z < 0$ in the ansatz for the pressures, equation (2.13). The fluid velocity and pressure then decay exponentially with the vertical distance from the membrane. Water particles undergo circular motion, which for a forward-traveling wave is clockwise below and anticlockwise above a basilar membrane oriented as in figure 2(c). The motion near the cochlear walls is essentially zero, so those boundaries have no influence on the wave. The dispersion relation (2.17) simplifies to

$$|k| = -2i\rho_0 \omega / Z .$$ (2.20)

This situation corresponds to the deep-water approximation for surface waves.

In the second limiting case, the wavelength greatly exceeds the height of the chambers: $kh \ll 1$. The dispersion relation is then given approximately by

$$k^2 = -2i\rho_0 \omega / (Zh) .$$ (2.21)

Water particles move on elliptical trajectories, for which motion occurs predominantly in the longitudinal direction (figure 2(d)). The longitudinal component of velocity barely depends on the vertical position. As shown in the following subsection, the vertical variation can therefore be neglected and the wave treated by a one-dimensional model for the longitudinal variation alone.

## 2.3. One-dimensional model

The wavelength of the basilar-membrane wave basal to its resonant position usually exceeds a few millimeters whereas the height of the channels is 0.5 mm or less. Much can therefore be learned about the behavior of the traveling wave from the long-wavelength approximation, $kh \ll 1$. Because the variation in the vertical direction is then small, the hydrodynamics can be described through a one-dimensional wave equation for the longitudinal variation alone.

Let us start from equation (2.5) within the upper chamber:

$$\nabla \tilde{\boldsymbol{u}}^{(1)} = -i\omega\kappa \, \tilde{p}^{(1)} .$$ (2.22)

We integrate this equation over a small volume $v$ that spans the height $h$ of the chamber and extends longitudinally by a small amount $dx$ (figure 3). Employing Gauss's divergence theorem we obtain

$$\oint_s \tilde{\boldsymbol{u}}^{(1)} ds = -i\omega\kappa \int_v \tilde{p}^{(1)} \, dv$$ (2.23)

in which $s$ is the volume's surface. Because the vertical variation of $\tilde{p}^{(1)}$ is small compared to the longitudinal change, equation (2.23) becomes

$$h\left[u_x^{(1)}(x+dx) - u_x^{(1)}(x)\right] - u_z^{(1)}(x, z=0)dx = -i\omega h\kappa \tilde{p}^{(1)}(x)dx .$$ (2.24)

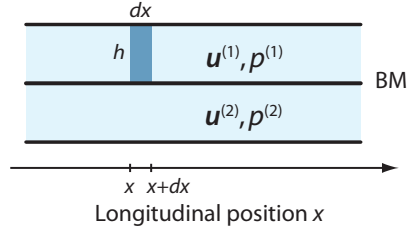

**Figure 3.** One-dimensional model of traveling-wave propagation. Integration over a thin volume (dark blue) within the two-dimensional model that spans the height $h$ of a chamber and extends longitudinally by a small distance $dx$ allows one to derive a one-dimensional wave equation for the pressure difference across the basilar membrane.

The terms in brackets on the left-hand side can be combined to yield a derivative $\partial_x \tilde{u}_x^{(1)}$. We rewrite this derivative with help of the longitudinal volume flow $j^{(1)} = h u_x^{(1)}$. The subsequent term in equation (2.24) **Error! Reference source not found.**contains the vertical velocity near the membrane that follows from the pressure difference through the membrane impedance $Z$, equation (2.11). We thus obtain

$$\partial_x \tilde{j}^{(1)} = \left( \tilde{p}^{(2)} - \tilde{p}^{(1)} \right) / Z - i\omega h \kappa \tilde{p}^{(1)} \,. \tag{2.25}$$

This equation has an intuitive interpretation: a longitudinal change in the fluid's volume flow arises either from a membrane displacement, and thus a change in the height of the chamber, or from a compression of the fluid.

The longitudinal component of equation (2.2) yields another equation that relates volume flow and pressure,

$$i\omega \rho_0 \tilde{j}^{(1)} = -h \partial_x \tilde{p}^{(1)} \,. \tag{2.26}$$

We can use this equation to eliminate the volume flow $\tilde{j}^{(1)}$ in the equation above to arrive at a relation for the pressure alone:

$$\partial_x^2 p^{(1)} = -\omega^2 \rho_0 \kappa \tilde{p}^{(1)} - \frac{i\omega \rho_0}{Zh} \left( \tilde{p}^{(2)} - \tilde{p}^{(1)} \right) \,. \tag{2.27}$$

This is the wave equation (2.6) but with an additional term that results from the membrane's displacement. An analogous computation for the lower chamber yields

$$\partial_x^2 p^{(2)} = -\omega^2 \rho_0 \kappa \tilde{p}^{(2)} + \frac{i\omega \rho_0}{Zh} \left( \tilde{p}^{(2)} - \tilde{p}^{(1)} \right) \,. \tag{2.28}$$

As in the two-dimensional description considered above, these coupled partial differential equations have two distinct solutions. First, the average pressure $\Pi = \left( p^{(1)} + p^{(2)} \right) / 2$ propagates as a fast sound wave through compression of the liquid:

$$\partial_x^2 \tilde{\Pi} = -\omega^2 \rho_0 \kappa \tilde{\Pi} \,. \tag{2.29}$$

With a wavelength at least severalfold the length of the cochlea, this wave traverses the organ in about 20 μs and does not displace the basilar membrane.

Second, the pressure difference between the upper and the lower chambers, $p = p^{(2)} - p^{(1)}$, propagates along the basilar membrane as a slow wave:

$$\partial_x^2 \tilde{p} = \frac{2i\omega \rho_0}{Zh} \tilde{p} \,. \tag{2.30}$$

We have discarded the compressibility, which makes a minor contribution as compared to the basilar-membrane impedance. Equation (2.30) represents the one-dimensional wave equation that we sought; it describes the longitudinal propagation of the traveling wave of basilar-membrane deflection. We verify easily that it obeys the dispersion relation (2.21) for a basilar-membrane wave of great wavelength.

## 2.4. Wentzel-Kramers-Brillouin approximation and energy flow

The basilar-membrane impedance $Z$ in the actual cochlea is not constant but changes systematically with longitudinal position $x$. What are the implications for the hydrodynamics and the basilar-membrane wave?

The Wentzel-Kramers-Brillouin (WKB) approximation provides useful analytical insight into the behavior of a wave propagating in an inhomogeneous medium. The approximation was originally developed for quantum mechanics, in which the Schrödinger equation describes the dynamics of particles travelling in a potential landscape. When the potential varies spatially, the Schrödinger equation can become difficult to solve and closed analytical solutions are typically infeasible. The WKB approximation provides a simple and general way to obtain an approximate analytical solution.

The approximation can be successfully applied to wave propagation in the cochlea (Zweig *et al* 1976, Steele and Taber 1979, Steele and Miller 1980, Reichenbach and Hudspeth 2010a). We illustrate the method in the most relevant case of a long wavelength, in which wave propagation can be described through the one-dimensional equation (2.30).

Let us start from the ansatz

$$\tilde{p} = \hat{p}(x) \exp\left[-i\int_0^x dx' k(x')\right]$$ (2.31)

with the wave's pressure amplitude $\hat{p}(x)$ and the local wave vector $k(x)$. The local wavelength follows as $\lambda(x) = 1/k(x)$. In the WKB approximation we assume that the spatial scale over which the impedance $Z(x)$ changes is large compared to the wavelength. Many cycles of oscillation therefore occur within a segment of the membrane whose impedance is approximately uniform. Only on a larger length scale does a significant impedance change occur and cause variations in amplitude. The wave vector is accordingly much larger than the relative change of pressure amplitude: $k(x) >> \left[\partial_x \hat{p}(x)\right]/p(x)$. The left-hand side of the one-dimensional wave equation (2.30) is therefore dominated by the term proportional to $k^2(x)$:

$$\partial_x^2 \tilde{p}(x) = -k^2(x)\hat{p}(x)\exp\left[-i\int_0^x dx' k(x')\right] + o(k^2).$$ (2.32)

To leading order, the one-dimensional wave equation (2.30) thus yields

$$k^2(x) = -2i\omega\rho_0 / [Z(x)h].$$ (2.33)

This dispersion relation agrees with that obtained for the case of a spatially homogenous membrane impedance, equation (2.21). The local wave vector and wavelength then follow from the local impedance as if the latter did not vary. In particular, the wavelength is proportional to the square root of the impedance, $\lambda \sim \sqrt{Z}$. Of course, this is a consequence of our assumption that the spatial scale at which the membrane impedance changes greatly exceeds the local wavelength. The speed $c(x) = \omega/k(x)$ at which the wave advances longitudinally also follows from the local membrane properties alone:

$$c(x) = \sqrt{\frac{i\omega Z(x)h}{2\rho_0}}.$$ (2.34)

Spatial impedance changes modify the wave's amplitude. Consider equation (2.30) with the ansatz (2.31) to first order in $k$:

$$2ik\partial_x \hat{p}(x) + i[\partial_x k(x)]\hat{p}(x) = 0,$$ (2.35)

which yields $\hat{p}(x) \sim 1 / \sqrt{k(x)}$ and hence

$$\hat{p} = \hat{p}_0 \sqrt{\lambda(x)} \, , \tag{2.36}$$

with an integration constant $\hat{p}_0$. The amplitude thus varies as the square root of the local wavelength. If this wavelength declines owing to impedance changes, then so do the amplitude and speed of propagation. In contrast, when an impedance change increases the local wavelength, the wave's amplitude and speed also rise. The virtue of the WKB approximation is that it quantifies these relations and confirms their general validity.

How does the basilar membrane's vertical velocity depend on the impedance? This velocity $\tilde{V}$ follows from a pressure difference across the basilar membrane through division by the impedance $Z$ (equation (2.11)). We obtain

$$\tilde{V} = \hat{V}(x) \exp\left[-i \int_0^x dx' k(x')\right] \tag{2.37}$$

with an amplitude $\hat{V}(x) = \hat{p} / Z$. Because the amplitude $\hat{p}$ varies as $\hat{p} \sim \sqrt[4]{Z}$, we find that the velocity's amplitude $\hat{V}$ depends on the impedance according to

$$\hat{V} \sim 1 / \sqrt[4]{Z^3} \, . \tag{2.38}$$

The dependence on the local wave vector is thus $\hat{V} \sim 1 / \sqrt{\lambda^3}$. The basilar membrane's vertical velocity increases when the wavelength declines and decreases when the wavelength grows.

The variation in velocity can be understood in a physically more intuitive way by considering the wave's energy flow in the absence of an active process. The energy associated with a segment of the moving basilar membrane is proportional to the impedance $Z$ and the squared vertical velocity $\hat{V}^2$, hence to $\hat{V}^2 Z$. The energy flow is obtained by multiplying the energy by the traveling wave's speed $c = \omega / k$, which through the dispersion relation (2.33) is proportional to the square root of the impedance, $c \sim \sqrt{Z}$ (equation (2.34)). The energy flow thus varies in proportion to $\hat{V}^2 \sqrt{Z^3}$. This value is constant if the amplitude $\hat{V}$ varies as $\hat{V} \sim 1 / \sqrt[4]{Z^3}$, which is also the result of the WKB approximation (equation (2.38)).

## 2.5. Resonance and critical-layer absorption

The cochlea achieves its astonishingly sharp frequency selectivity by combining a traveling wave on the basilar membrane with a local resonance that arises from the interplay of mass and stiffness to set the impedance $Z$ (equation (2.19)). The imaginary part of the impedance vanishes at a resonant frequency

$$\omega_0 = \sqrt{K / m} \, . \tag{2.39}$$

If the cochlear fluids were absent and a thin segment of the membrane were excited directly, it would vibrate with the greatest amplitude at its resonant frequency $\omega_0$. This observation led Hermann Helmholtz to propose in the 19th Century that each segment of the basilar membrane has a distinct resonant frequency, which he termed the proper frequency, with high resonant frequencies near the cochlear base and successively lower ones toward the apex. In his pioneering book *On the Sensations of Tone* Helmholtz concluded:

Consequently any exciting tone would set that part of the membrane into sympathetic vibration, for which the proper tone of one of its radial fibers that are stretched and loaded with the various appendages already described, corresponds most nearly with the exciting tone; and thence the vibrations will extend with rapidly diminishing strength on to the adjacent parts of the membrane... Hence every simple tone of determinate pitch will be felt only by certain nerve fibers, and simple tones of different pitch will excite different fibers.

The material properties of the basilar membrane indeed vary systematically from base to apex. The radial fibers that constitute the basilar membrane are thick and short near the base, conferring a high stiffness, and become progressively longer, thinner, and more compliant towards the apex. The stiffness accordingly decays exponentially with the longitudinal distance from the base (Emadi *et al* 2004). The mass associated with a segment of the membrane is more difficult to measure but may be estimated from the cross-sectional area of the organ of Corti, which increases toward the apex. If we assume an exponential variation with distance, we obtain resonant frequencies that decline exponentially from base to apex (figure 4(*a*)). This accords with the characteristic frequencies of auditory-nerve fibers, that is, the frequencies at which the fibers are activated most easily, also known as "natural frequencies" or "best frequencies." Nerve fibers that originate near the cochlear base exhibit high characteristic frequencies whereas those originating farther apically have characteristic frequencies that decline exponentially with the distance from the cochlear base. According to Helmholtz's resonance theory the characteristic frequency of an auditory-nerve fiber reflects the resonant frequency of the basilar-membrane segment from which it originates.

Quantification of the resonant frequency owing to the stiffness and mass of the basilar membrane supports this idea. In the basal turn of the gerbil's cochlea, for example, a segment of the basilar membrane 8 $\mu$m in length imposes a stiffness of about 1 N·m$^{-1}$ (Emadi *et al* 2004). The cross-sectional area of the organ of Corti at that position is 8,000 $\mu$m$^2$, which implies a mass near 60 ng (Richter *et al* 2000). Because the basilar membrane moves as a rigid beam, however, not all of that mass oscillates at the maximal basilar-membrane displacement. If we assume that the moving mass represents half the total mass of the organ of Corti, we obtain a resonant frequency of about 30 kHz that agrees well with the characteristic frequency of auditory-nerve fibers from the basal turn. In the middle turn, about 7 mm from the base, the basilar-membrane stiffness decreases to about 30 mN·m$^{-1}$ (Emadi *et al* 2004). The cross-sectional area of 18,000 $\mu$m$^2$, however, exceeds that at the base (Richter *et al* 2000). This yields an effective mass of about 70 ng and a resonant frequency of 3 kHz, again in good agreement with the characteristic frequencies of auditory-nerve fibers.

Cochlear hydrodynamics in combination with local basilar-membrane resonance produces frequency selectivity that is even sharper than that obtained through the simple resonance envisioned by Helmholtz. If we rewrite the impedance $Z$ in terms of the resonant frequency $\omega_0$ and ignore friction we obtain

$$Z = -\frac{iK}{\omega}\left(1 - \frac{\omega^2}{\omega_0^2}\right).\tag{2.40}$$

The frequency of stimulation $\omega$ equals the resonant frequency at a certain spatial location $x_0$, $\omega = \omega_0(x_0)$. Basal to that position $x_0$ the resonant frequency of the basilar membrane exceeds the frequency of stimulation. The impedance $Z$ there accordingly has a negative imaginary part and sustains a propagating wave. The impedance diminishes, however, as a wave approaches the resonant position $x_0$. According to the WKB approximation, the local wavelength as well as the speed (2.34) then fall in proportion to $\sqrt{1 - \omega^2/\omega_0^2}$, vanishing at the resonant position (figure 4(*b*)). The amplitude of the vertical basilar-membrane velocity diverges as $1/\sqrt[4]{\left(1 - \omega^2/\omega_0^2\right)^3}$. On the apical side of the resonant position the impedance has a positive real part that implies an evanescent wave. Because its amplitude decays exponentially, the wave cannot progress much beyond the resonant position.

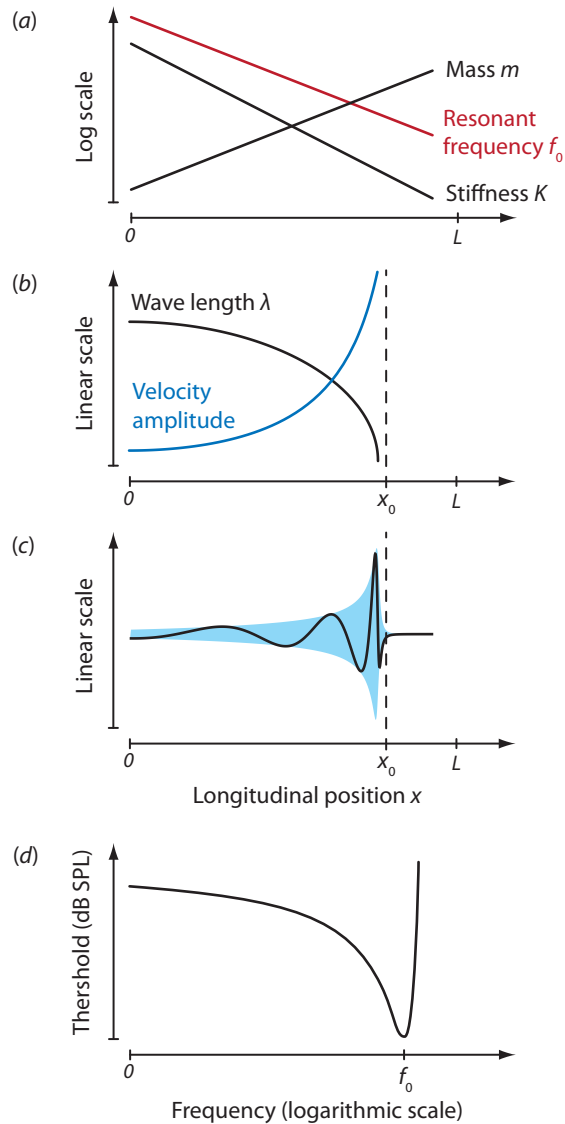

**Figure 4.** Critical-layer absorption. (*a*) At each longitudinal position, the mass and stiffness of the basilar membrane define a resonant frequency $f_0$. Because the basilar-membrane stiffness decreases exponentially and the mass increases exponentially from base to apex, the resonant frequency follows an exponential map from high frequencies near the base to low frequencies near the apex. (*b*) According to the WKB approximation, the basilar-membrane wave elicited by a single frequency slows upon approaching the resonant position $x_0$, whereas the amplitude of the displacement diverges. (*c*) Numerical solution of a one-dimensional model confirms the predictions of the WKB approximation. The wave's instantaneous displacement is shown as a black line and its envelope is represented by blue shading. (*d*) A schematic threshold tuning curve depicts the frequency tuning of an auditory-nerve fiber from the high-frequency region of the cochlea.

Viscosity reduces the peak near the resonance and renders the velocity there finite. An example of a numerical solution of a one-dimensional cochlear model based on equation (2.30) and with realistic values for mass, damping, and stiffness is shown in figure 4(*c*). Because of viscous damping, the wave's peak occurs slightly basal to the resonant position. The smaller the damping, the larger is the peak displacement and the closer it approaches the resonant position. In the following, we refer to the position at which the wave elicited by a stimulus at a single frequency and of small amplitude peaks as that frequency's characteristic place. Alternative terms in the literature are "natural place" and "best place."

The fluid mechanist James Lighthill (1981) referred to the combination of a traveling wave with a resonance as critical-layer absorption. Because the wave slows upon approaching its resonant position and because its amplitude increases accordingly, most of its energy is dissipated in a critical layer around the characteristic position.

A living cochlea exhibits intriguing behavior at the characteristic position. For stimulation at a low sound-pressure level, the peak displacement is much larger than that in a passive cochlea: near the resonant position the impedance nearly vanishes. This occurs because the ear's active process counteracts the effect of viscous damping to enhance the response. The membrane's response becomes highly nonlinear near the resonance, probably because the system operates close to a critical point, a Hopf bifurcation (Choe et al 1998, Eguíluz *et al* 2000, Camalet *et al* 2000, Hudspeth 2010). Sections 4-7 are devoted to an understanding of these processes.

The sharp decay of the wave's amplitude apical to its peak provides precise frequency information. Consider the response of an auditory-nerve fiber that emerges from a certain position along the cochlea. One can measure experimentally the fiber's tuning curve, that is, the set of least sound-pressure levels that must be applied at various frequencies to elicit a criterion response in the fiber (figure 4(*d*)). The tuning curve displays a minimum at that frequency for which the traveling wave peaks at the fiber's position: at that frequency the least signal suffices to induce action potentials in the nerve fiber. Lower frequencies require stronger stimuli to evoke the criterion response, for the associated traveling waves peak farther apically and the amplitude at the fiber's location is no longer maximal. Higher frequencies, however, produce no response in the fiber, for the waves never reach the fiber's position. Hence the sharp high-frequency cutoff of the tuning curve reflects critical-layer absorption and provides precise frequency information.

Recordings have been performed from auditory-nerve fibers originating at many locations along the cochlea (Kiang and Moxon 1974, Kiang 1984, Temchin *et al* 2008a, 2008b). Interestingly, low-frequency fibers, with characteristic frequencies below 1-2 kHz, have broader shapes and lack the high-frequency cut-off. This accords with measurements of basilar-membrane mechanics near the apex, which have not found a rapid decay of the wave apical to its peak (Cooper and Rhode 1995, Khanna and Hao 1999, 2000, Zinn *et al* 2000). Apical to the middle turn the cross-sectional area of the organ of Corti, and hence the basilar-membrane mass, does not increase (Richter *et al* 2000). Moreover, the stiffness there does not decline greatly (Naidu and Mountain 1998). It therefore appears that basilar-membrane resonance and critical-layer absorption occur only for frequencies higher than a few kilohertz. Discrimination at low frequencies must then function through a distinct mechanism that remains elusive. Understanding frequency discrimination at relatively low frequencies is of great importance, for most of the information in human speech occurs at those frequencies. We have made a theoretical suggestion, which is further described in subsection 4.3, for how a low-frequency resonance might be achieved through mechanical activity by hair cells (Reichenbach and Hudspeth 2010b).

## 3. Historical overview of the active process

In 1947 and 1948 the physicist Thomas Gold published three remarkable papers in which he suggested that the cochlea amplifies signals through tuned active feedback (Pumphrey and Gold 1947, 1948, Gold 1948). Characteristically for Gold's highly original work in areas as diverse as neuroscience, astrophysics, space engineering, and geophysics—"with a stream of elegant and sometimes unconventional ideas that his peers acknowledged for their daring, without always accepting them" (Pearce 2004)—his concept of active feedback was initially ignored by the hearing-research community. Only after the discovery of cochlear nonlinearity and otoacoustic emission, described below, were Gold's ideas revived and made a cornerstone of cochlear mechanics.

### 3.1. Theoretical proposal

How did Gold arrive at the conclusion of an active process in the inner ear at a time when few electrical or mechanical recordings from the cochlea were available? Gold pursued the resonance hypothesis advanced by Helmholtz (1954). Assume that, in response to a pressure difference $\tilde{p}$, each segment of the basilar membrane vibrates independently at a velocity $\tilde{V} = \Delta \tilde{p} / Z$ with the impedance $Z$ given by equation (2.19). Specifying the sharpness of the emerging resonance by the quality factor $Q = \sqrt{Km}/\xi$, the impedance

$$Z = \frac{\xi}{A}\left[1 + iQ\left(\frac{\omega}{\omega_0} - \frac{\omega_0}{\omega}\right)\right] \tag{3.1}$$

depends on the stimulus frequency $\omega$ only through its ratio to the resonant frequency $\omega_0$.

The vibration velocity $\tilde{V}$ of an isolated basilar-membrane segment is maximal at the resonant frequency $\omega_0$, for which it reaches $\tilde{V} = A\Delta\tilde{p}/\xi$. Lower damping therefore leads to a larger velocity. Moreover, the velocity decays both for higher and for lower frequencies $\omega$, and the decrease is stronger the larger the quality factor $Q$. The width of the resultant tuning curve is proportional to the inverse of $Q$ (figure 5(*a*)).

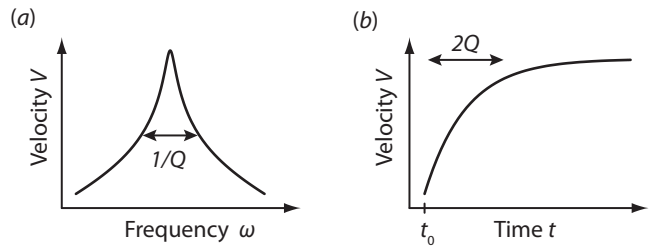

**Figure 5. Quality factor $Q$.** (*a*) The width of an oscillator's tuning curve is inversely proportional to the quality factor $Q$. (*b*) In response to a signal at the resonant frequency that starts at a certain time $t_0$, an oscillator requires a time proportional to $Q$ to reach a response close to the maximum.

The quality factor $Q$ also governs the temporal response of the membrane segment at the beginning or end of a stimulus. If a pressure oscillating at frequency $\omega_0$ starts at time $t_0$, then the membrane velocity $V$ increases from rest to its maximal amplitude $V_{\max}$ according to

$$V = V_{\max}\left[1 - e^{-\omega_0(t-t_0)/2Q}\right]e^{i\omega_0 t}. \tag{3.2}$$

The amplitude approaches its steady-state value in an exponential manner with a time constant of $2Q/\omega_0$ (figure 5(*b*)). After the stimulus ends, the amplitude drops exponentially with the same time constant.

Two distinct regimes arise. When the quality factor $Q$ is less then 1/2, the membrane segment is overdamped: viscous friction is so large that, at the end of a stimulus, the membrane quickly relaxes to its resting position without oscillation. If instead $Q$ exceeds 1/2, damped oscillations occur after a signal ends. Because high values of $Q$ prolong the response, they might degrade the functioning of the cochlea in the time domain.

Gold assumed that the cochlea achieves a compromise between sharp frequency discrimination, which favors a pronounced resonance manifested by a high quality factor $Q$, and a fast response to transients, which requires a low $Q$ value. But where does the optimal compromise lie, in the overdamped or underdamped regime of membrane oscillation?

Gold realized that he could estimate the $Q$ value through psychoacoustic experiments. If each segment of the basilar membrane were to resonate strongly, with a high value of $Q$, its response to a pure tone would accumulate slowly after the onset of a tone (equation (3.2)). Shorter tones would thus result in smaller membrane vibration and be harder to detect. Gold therefore measured the threshold intensities at which human subjects could detect tones of various lengths. Employing equation (3.2) he also calculated how the intensity threshold should depend on the length of the tone for particular values of $Q$ in the auditory system. By comparing his experimental data to the theoretical expectations he obtained reliable estimates for the quality factor $Q$. The values varied from around 30 for frequencies below 1 kHz to 300 for 10 kHz, indicating sharp resonances well within the underdamped regime.

In a subsequent article Gold reasoned that such high values of $Q$ could not arise in a passive cochlea. The damping of a basilar-membrane segment moving in liquid, even if estimated conservatively, should lead to dramatically lower values for the quality factor in the overdamped regime. How, then, could the cochlea achieve its high degree of resonance?

Gold drew an ingenious analogy to the regenerative electronic circuits that had been invented a few decades earlier (Morse 1925, Horowitz and Hill 1989). These circuits start from a simple resonator comprising an inductance $L$, a resistance $R$, and a capacitance $C$ in series with an oscillating input voltage $V_{\text{in}}$ (figure 6(*a*)). If the input voltage $U_{\text{in}}$ oscillates at an angular frequency $\omega$, $U_{\text{in}} = \tilde{U}_{\text{in}} e^{i\omega t} + c.c.$, then so does the circuit current $I = \tilde{I} e^{i\omega t} + c.c.$, and both quantities are related through

$$\frac{\tilde{U}_{\text{in}}}{\tilde{I}} = i\omega L + R - \frac{i}{\omega C}.$$

(3.3)

We recognize a resonance in the current at the angular frequency $\omega_0 = 1/\sqrt{LC}$; the resonance grows sharper as the resistance $R$ declines.

Such a circuit may be used in a radio for detecting an amplitude-modulated signal at a carrier frequency that matches the resonance. Of course, the signal detection of the device improves with a sharper resonant peak. Can one, therefore, lower the circuit's resistance $R$ below the value specified by the material properties of the conducting material?

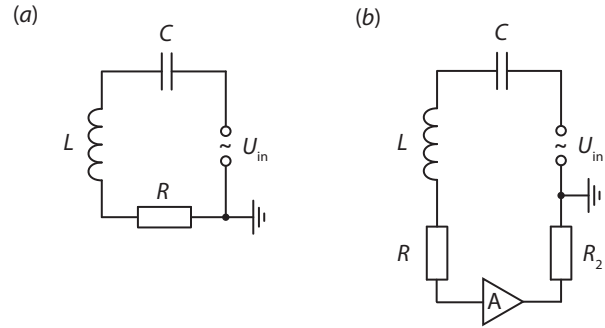

**Figure 6. Regenerative circuit**. (*a*) The inductance $L$, resistance $R$, and capacitance $C$ form a simple resonant circuit that can detect a frequency-tuned input voltage $U_{in}$. (*b*) A regenerative circuit applies feedback, here through an inverting amplifier $A$, to reduce or even cancel damping from the resistance.

Regenerative circuits achieve this goal through active feedback. Consider a resonant circuit in which an inverting amplifier with a subsequent resistance $R_2$ lies in series with the resistance $R$ (figure 6(*b*)). If the amplifier provides a gain $G$, such that an input voltage $U$ yields an output voltage $-GU$, then the circuit's current $I$ follows from

$$\frac{\tilde{U}_{in}}{\tilde{I}} = i\omega L + R - \frac{R_2}{G} - \frac{i}{\omega C} \ . \tag{3.4}$$

The damping is therefore reduced from $R$ to $R$-$R_2$/$G$ and vanishes at a critical value of the gain, $G_c$= $R_2$/$R$. For a gain above this critical value, the net damping is negative and the circuit oscillates spontaneously, that is, in the absence of an input signal!

Gold noted that the resonant circuit may be viewed as an electric analogue of a segment of the inner ear. Both systems detect signals, whether electrical or mechanical, at a specific frequency. In both cases frequency discrimination is achieved by resonance and hindered by damping, either through the circuit's resistance or through the viscosity associated with motion of the basilar membrane. If active feedback in a regenerative circuit can lower and even cancel the resistive damping, he reasoned, might not the ear have evolved an active mechanism to counter the viscous drag?

### 3.2. Otoacoustic emission, nonlinearity, and the Hopf bifurcation

Based on his analogy to regenerative circuits, Gold (1948) drew a far-reaching conclusion from his idea of an active inner ear: "If the feedback ever exceeded the losses, then a resonant element would become self-oscillatory....If this occurred, we should hear a clear note....It is very tempting to suggest that the common phenomenon of 'ringing of the ear' is frequently of this origin, and not always a central nervous disturbance....If the ringing is due to actual mechanical oscillation in the ear, then we should expect a certain fraction of the acoustic energy to be radiated out. A sensitive instrument may be able to pick up these oscillations and so prove their mechanical origin. This would be almost a conclusive proof of this theory, as such a generation of sound on any other basis is exceedingly unlikely."

Gold did not succeed in measuring the predicted emission of sound from the ear. A few decades later, however, David Kemp detected "echoes" from human ears in response to click stimuli (Kemp 1978). By verifying that these signals were absent in control subjects who had hearing losses but no middle-ear deficiencies, Kemp concluded that the acoustic emissions had their origin within the healthy inner ear. Because *oto* is Greek for ear, these signals were subsequently named otoacoustic emissions (OAEs). Kemp (1981) also recorded such emissions in the absence of external stimulation. We know today that these spontaneous otoacoustic emissions can be measured from most human ears. They arise at a set of frequencies that is characteristic for a given ear and have therefore been proposed as a means of biometric identification (Swabey *et al* 2004).

Earlier in the 1970s William Rhode discovered a compressive nonlinearity in the basilar membrane's response (Rhode 1971). Measuring from living squirrel monkeys, he found that the velocity of vibration at the characteristic place grew sublinearly with the intensity of stimulation (figure 7). After the death of an experimental animal, the velocity, especially at low intensities, fell to much lower values. The membrane's velocity in the dead cochlea grew linearly with increasing sound intensity. Even in living animals, the response to stimulation away from the characteristic frequency was much lower in amplitude and varied linearly with the intensity of stimulation.

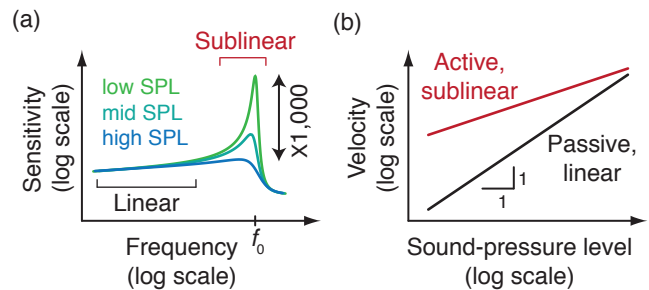

**Figure 7. Schematic diagrams characterizing compressive nonlinearity**. (*a*) The sensitivity of the basilar membrane is defined as the membrane's vertical velocity divided by the applied sound pressure. For frequencies much below the resonant value $f_0$, the sensitivity is independent of the pressure and the membrane's response is linear. Near the resonant frequency, however, a sublinear response emerges: the sensitivity at low levels of stimulation can exceed that at high sound-pressure levels by a thousand times or more. (*b*) The membrane velocity near the resonant frequency depends nonlinearly on the sound intensity for a living cochlea. The response drops and becomes linear in a dead ear.

Rhode's measurement in the living cochlea of basilar-membrane vibration greatly enhanced with respect to that in an impaired cochlea evidenced an active process in the healthy ear. Because of the difficulty of measuring the tiny, sound-induced vibrations in living animals, however, confirmation of Rhode's measurements came only years later (LePage and Johnstone 1980, Rhode 1980, Sellick *et al* 1982, Robles *et al* 1986). These subsequent observations coincided with Kemp's measurements of otoacoustic emissions and with the discovery of active processes in hair cells (Section 4).

How does the nonlinearity arise in the response of the basilar membrane, and what does it signify? When describing the hydrodynamics of a passive cochlea in Section 2, we assumed a linear relation between pressure difference and membrane velocity (equation (2.11)). In many physical systems such a relation holds for small pressures and velocities, whereas nonlinearities become important for larger ones. In the case of the cochlea, however, the acoustic impedance $Z$ that linearly relates pressure and velocity displays a resonant frequency $\omega_0$ at which its imaginary part vanishes. Its real part is specified by the damping. If the ear's active process counteracts the damping so as to exactly cancel it, then the impedance and therefore the linear response vanishes entirely. The basilar membrane's response to sound stimulation becomes nonlinear even for small vibrations.

The mathematical field of nonlinear dynamics that can describe such a situation flowered only in the 1960s and 1970s, when the rise of powerful computers made numerical investigations feasible. It is accordingly unsurprising that Gold did not foresee the implications of his ideas concerning cochlear nonlinearity. Originating in studies of hair-bundle activity, such a mathematical analysis has more recently suggested that each segment of the basilar membrane operates near a Hopf bifurcation (Choe *et al* 1998, Eguíluz *et al* 2000, Camalet *et al* 2000, Duke and Jülicher 2003, Kern and Stoop 2003, Magnasco 2003). In the language of dynamical-systems theory, a bifurcation denotes a qualitative transition in the behavior of a dynamical system in response to a graded change in the value of a control parameter (Wiggins 1990, Strogatz 1994). In the instance of the supercritical Hopf bifurcation, the transition extends from underdamped resonance to spontaneous oscillation.

What precisely is a Hopf bifurcation and how can it arise in cochlear mechanics? Equation (2.11) describes the response of an isolated segment of the basilar membrane to a pressure difference $p$. The linear part with the impedance $Z$ follows from Newton's equation of motion for the temporal development of the membrane displacement $X$:

$$m\partial_t^2 X + \xi \partial_t X + KX = Ap \, . \tag{3.5}$$

Employing the velocity $V = \partial_t X$ we may recast this second-order ordinary differential equation as two first-order differential equations:

$$\partial_t \begin{pmatrix} X \\ V \end{pmatrix} = \begin{pmatrix} 0 & 1 \\ -K/m & -\xi/m \end{pmatrix} \begin{pmatrix} X \\ V \end{pmatrix} + \begin{pmatrix} 0 \\ Ap/m \end{pmatrix} . \tag{3.6}$$

In the case of an underdamped oscillation, $\xi < 2\sqrt{Km}$, the matrix on the right-hand side has two eigenvalues $\lambda$ and $\bar{\lambda}$ that are each other's complex conjugates. By diagonalizing the matrix we can find a variable transformation $(X,V)^T \rightarrow (Y,\bar{Y})^T$ with a complex variable $Y$ such that the equation takes the form

$$\partial_t \begin{pmatrix} Y \\ \bar{Y} \end{pmatrix} = \begin{pmatrix} \zeta & 0 \\ 0 & \bar{\zeta} \end{pmatrix} \begin{pmatrix} Y \\ \bar{Y} \end{pmatrix} + \begin{pmatrix} P \\ \bar{P} \end{pmatrix} . \tag{3.7}$$

This is achieved with the eigenvalue $\zeta = -\xi/(2m) + i\sqrt{4Km - \xi^2}/(2m)$, the complex variable $Y = V/2 + i(KX + \xi V/2)/\sqrt{4Km - \xi^2}$, and forcing through the transformed pressure $P = Ap\left(1 + i\xi/\sqrt{4Km - \xi^2}\right)/(2m)$. The dynamics may then be described by a single complex equation for $Y$:

$$\partial_t Y = (\zeta_r + i\omega_*)Y + P \, . \tag{3.8}$$

Here we have decomposed the complex eigenvalue $\zeta$ into its real and imaginary parts, $\zeta = \zeta_r + i\omega_*$, in which $\zeta_r$ and $\omega_*$ are real numbers. Nonlinearities can appear through an additional function $f_{\text{nonlinear}}(Y,\bar{Y})$ that depends nonlinearly on $Y$ and $\bar{Y}$:

$$\partial_t Y = (\zeta_r + i\omega_*)Y + f_{\text{nonlinear}}(Y,\bar{Y}) + P \, . \tag{3.9}$$

The function $f_{\text{nonlinear}}(Y,\bar{Y})$ may be represented as a Taylor series in which the lowest-order terms are quadratic: $Y^2$, $Y\bar{Y}$, and $\bar{Y}^2$. Cubic terms such as $Y^3$ or $Y^2\bar{Y}$ constitute the next order. One can show that a nonlinear transformation $Y \rightarrow W$ exists such that $Y$ and $W$ agree to leading order and that, to third order, the equation (3.9) is transformed to

$$\partial_t W = (\zeta_r + i\omega_*)W - (\nu + i\sigma)|W|^2 W + P \; ; \tag{3.10}$$

the parameters $\nu$ and $\sigma$ are real. A constructive proof may be found, for example, in Wiggins (1990). Note that the quadratic terms have disappeared, and of the cubic terms only that proportional to $|W|^2 W$ remains. In the absence of forcing, the equation is therefore invariant under the $U(1)$ transformation $W \rightarrow We^{i\varphi}$. If higher-order terms were to be included, only those that obey this symmetry would be relevant, such as $|W|^4 W$ or $|W|^6 W$.

Let us first discuss equation (3.10) in the absence of external stimulation, $P = 0$. Linear stability analysis shows that the only fixed point of the dynamics, that is the only value $W_*$ for which the temporal variation vanishes, is $W_* = 0$. This fixed point is stable for a negative control parameter $\zeta_r < 0$. Without an external signal, the membrane segment rests at that position. For positive values of $\zeta_r$, however, the fixed point becomes unstable. When the control parameter $\nu$ is positive, the nonlinear term in equation (3.10) counteracts the instability of the linear term and leads to a stable limit cycle with an amplitude $\sqrt{\zeta_r / \nu}$ and frequency $\omega_* - \zeta_r \sigma / \nu$. The amplitude decreases as $\zeta_r$ diminishes and vanishes at the critical value $\zeta_r^{(c)} = 0$. This situation represents a supercritical Hopf bifurcation and equation (3.10) without external forcing is known as its normal form.

In the vicinity to the bifurcation point $\zeta_r^{(c)} = 0$ the membrane responds nonlinearly to an applied acoustic stimulus. Let the pressure $P$ oscillate at an angular frequency $\omega_*$ and with a Fourier coefficient $\tilde{P}$. We consider the membrane's Fourier coefficient $\tilde{W}$ at the same frequency. The linear part of equation (3.10) then vanishes and the nonlinear part yields an amplitude $|\tilde{W}| = \sqrt[3]{|\tilde{P} / (\nu + i\sigma)|}$. The membrane's displacement and velocity thus increase as the cubic root of the growing stimulus level, corresponding to a compressive nonlinearity. The growth exponent of 1/3 closely matches that measured from the cochlea (Ulfendahl, 1997; Robles and Ruggero, 2001).

Such a nonlinear oscillator also produces distortion. Consider equation (3.9) for the variable $Y$ and assume only one nonlinear quadratic term $Y^2$:

$$\partial_t Y = (\zeta_r + i\omega_*)Y + \beta Y^2 + P \tag{3.11}$$

with a complex coefficient $\beta$. Denote by $\tilde{V}(\omega)$ and $\tilde{P}(\omega)$ the Fourier transforms of respectively the oscillator's velocity and the applied force. Because the Fourier transform of the product $V^2$ is the convolution $\tilde{V} * \tilde{V}$, the Fourier transform of equation (3.11) is

$$i\omega \tilde{Y}(\omega) = (\zeta_r + i\omega_*)\tilde{Y}(\omega) + \beta \int_{-\infty}^{\infty} d\omega' \tilde{Y}(\omega')\tilde{Y}(\omega - \omega') + \tilde{P}(\omega) \; . \tag{3.12}$$

If the oscillator is stimulated at two distinct frequencies $\omega_1$ and $\omega_2$, its linear response has a component at those two frequencies, $Y = \tilde{Y}(\omega_1)e^{i\omega_1 t} + \tilde{Y}(\omega_2)e^{i\omega_2 t} + c.c.$ The quadratic nonlinearity, however, elicits a force at additional frequencies. According to equation (3.12) a force results, for example, at a frequency $\omega$ for which $\omega' = \omega_1$ and $\omega - \omega' = \omega_2$, and hence at $\omega = \omega_1 + \omega_2$. Cubic nonlinearities produce analogous cubic distortion frequencies such as $2\omega_1 - \omega_2$ and $2\omega_2 - \omega_1$.

Such distortion tones actually occur in the cochlea. Their discovery dates to the Italian baroque violinist and composer Giuseppe Tartini in the 18$^{\text{th}}$ Century. When playing two tones $\omega_1$ and $\omega_2$ simultaneously on his instrument, he perceived linear combinations such as $\omega_2 - \omega_1$ and $2\omega_1 - \omega_2$ (Tartini 1754, 1767). Although training and experience were required for observers to perceive these "phantom" tones, a few composers have subsequently invoked them in their work (Campbell and Greated 2002). The distortion tones that can sustain a melody are not produced by any musical instrument, but result from the inner ear's nonlinearity!

Psychoacoustic measurements demonstrated that distortion is particularly prominent during weak acoustic stimulation (Goldstein 1967), indicating once again that cochlear mechanics is nonlinear at low sound-pressure levels. Indeed, immediately after Kemp's discovery of otoacoustic emissions following click stimuli, distortion products were measured both as acoustic signals in the ear canal and as responses of the auditory nerve (Kemp 1979, Kim 1980, Kim *et al* 1980). To ensure that distortions were not generated by potential nonlinearities in the equipment, the two frequencies in these experiments were supplied to the ear canal through independent oscillators, amplifiers, and speakers. Because the response of the middle ear is linear even for high sound intensities, the measured distortion clearly arose within the cochlea. Direct proof that the inner ear generates nonlinear distortion came through recordings of distortion signals from the basilar membrane (Robles *et al* 1991, 1997) and finally from the hair bundles of individual hair cells (Jaramillo 1993, Barral and Martin 2012).

## 4. Hair cells and hair bundles

A few decades ago the cochlea was viewed as a mechanical and hydrodynamic system for transmitting acoustic signals to hair cells that simply interpreted these inputs as electrical responses. As is apparent from the foregoing discussion, however, hair cells have proven to make a far more important contribution to cochlear function. The mechanical activities of these cells account for the hallmarks of the cochlea's active process: amplification, frequency tuning, compressive nonlinearity, and spontaneous otoacoustic emission. Because each of these features has a correlate in the behavior of individual hair cells, *in vitro* investigations of hair cells from model organisms have provided valuable insights into the cochlea's operation.

### 4.1. Structure of the hair cell

Although they vary in their dimensions and appearance, all hair cells in vertebrate organisms have a common underlying structure. Each is an epithelial cell with a cylindrical or flask-shaped cell body and is joined at the perimeter of its upper or apical surface to a ring of supporting cells. Derived embryologically from the same precursors as hair cells, supporting cells provide mechanical anchorage for the sensory cells and participate in metabolic tasks such as the uptake of excess $K^+$. In many instances—though woefully not in the case of the mammalian cochlea—supporting cells can divide and differentiate to replenish lost hair cells.

Hair cells are not neurons. They originate from a part of the embryo distinct from those that yield neural progenitors. Moreover, hair cells lack axons and dendrites, the characteristic extensions from nerve cells. In several ways, however, hair cells nonetheless resemble neurons. They express numerous proteins that are characteristic of nerve cells but not of epithelia. The surface membranes of hair cells contain a variety of ion channels that shape the electrical responses to mechanical stimulation. Finally, the basal surfaces of hair cells bear specialized chemical synapses. When excited by mechanical stimulation, a hair cell releases packets of the neurotransmitter glutamate at these so-called ribbon synapses. The glutamate activates receptors on the nerve terminals attached to the hair cell's base; the ensuing depolarization excites action potentials. These electrical signals then propagate along the afferent axons into the brain to initiate the appropriate behavioral response.

Although the details lie beyond the scope of this review, it is noteworthy that the cell bodies of hair cells contribute significantly to the processing of sensory information. In many receptor organs the interplay of $Ca^{2+}$ and $K^+$ channels provides an electrical resonance that tunes hair cells to specific frequencies of stimulation (Hudspeth and Lewis 1988a, 1988b, Fettiplace and Fuchs 1999). Furthermore, the release of synaptic transmitter by a hair cell can be frequency-selective, providing an additional means of enhancing responsiveness to specific frequencies (Patel *et al* 2012).

The defining feature of a hair cell is its mechanoreceptive organelle, the hair bundle (figure 8). Extending from the flattened apical surface of a hair cell by as little as 1 μm in some organisms and by as much as 100 μm in others, the bundle comprises from about a dozen to over 300 cylindrical protrusions called stereocilia (Howard and Ashmore 1986, Tilney and Saunders 1983). This name, which means "stiff hairs," captures the mechanical essence of these tiny processes. Each stereocilium consists of a core of parallel actin filaments that are cross-linked into a rigid fascicle by several proteins (DeRosier and Tilney 1980, Shin *et al* 2013). Because the number of microfilaments declines from hundreds to tens at the tapered base of the stereocilium, the process is most compliant there. As a consequence, application of a mechanical force to a stereocilium perpendicular to its long axis causes the structure to pivot at its flexible base while remaining largely straight along its shaft.

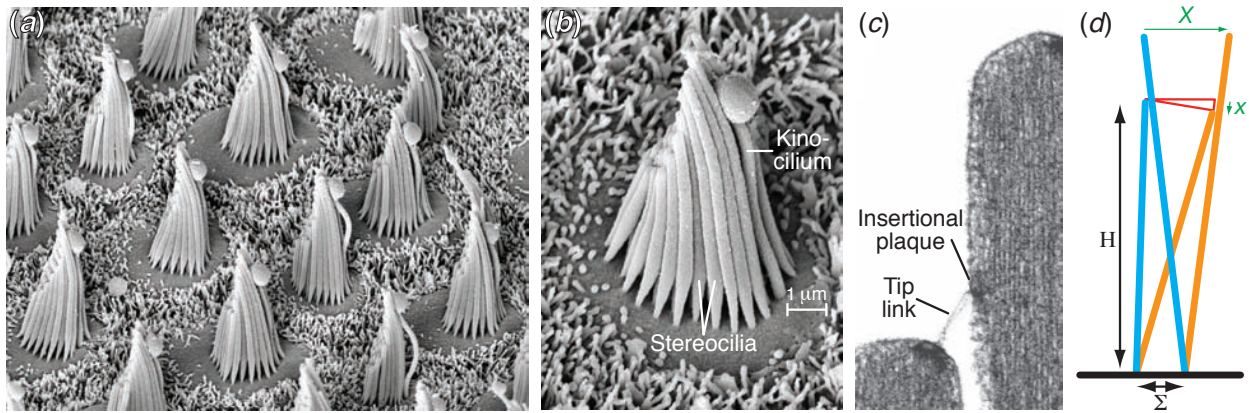

**Figure 8. The hair bundle**. (*a*) A scanning electron micrograph depicts roughly a dozen hair bundles protruding from the apical epithelial surface of the sacculus, a receptor for seismic vibration and airborne sound in the bullfrog's ear. Each bundle stands about 8 μm tall. (*b*) An individual hair bundle comprises approximately 60 actin filament-filled stereocilia and a single microtubule-based kinocilium, the last terminating in a bulbous swelling. Note the bundle's beveled appearance owing to a progression in stereociliary lengths. Microvilli stud the surfaces of the adjacent supporting cells. (*c*) A transmission electron micrograph reveals a tip link extending 120 nm between the upper end of a short stereocilium and the flank of the longest adjacent stereocilium. The insertional plaque at the link's upper insertion contains myosin-1c molecules thought to regulate the tension in the link. (*d*) When the tip of a hair bundle moves by distance $X$, from the blue configuration of the two stereocilia to the orange one, the shearing motion $x$ of a stereociliary tip along the adjacent process is roughly proportional to the ratio of the stereocilia's separation $\Sigma$ to their height H: $\gamma = x/X \approx \Sigma/H$. For the anuran hair bundles illustrated in this figure, $\gamma \approx 0.14$.

The geometrical arrangement of the stereocilia within a hair bundle is key to the bundle's operation as a mechanotransducer. Originating from much smaller microvilli on the cellular surface, the stereocilia remain packed in a hexagonal array throughout their development. In every instance, however, the stereocilia are not of equal length along each of the three hexagonal axes. Instead, the stereocilia display a monotonic decrease in length, often by very regular increments, along one axis. Orthogonal to that axis, all the stereocilia in each rank are of equal length. With its top edge beveled like a hypodermic needle, the hair bundle thus manifests elements of both hexagonal and mirror symmetry. As we shall see, this configuration plays an important role in the organelle's mechanosensitivity.

The cross-sectional configuration of a hair bundle is correlated with the sensory modality to which the associated hair cell responds. The sensory organs of the vestibular labyrinth detect linear and angular accelerations at relatively low frequencies, from less than a hertz to a few hundred hertz. The lateral-line organs of fishes and some aquatic amphibians detect movements of the aqueous environment over a similar frequency range. The hair bundles in each of these instances are roughly circular in cross-section. In contrast, the acoustically responsive hair bundles of birds, and more strikingly those of mammals, are compressed along the hexagonal axis running from the tallest to the shortest stereocilia. In the chicken, the hair bundles sensitive to the highest frequencies, about 5 kHz, are some 35 files of stereocilia in width but only eight stereociliary ranks in depth (Tilney and Saunders 1983). The hair bundles of outer hair cells in the high-frequency region of the mammalian cochlea can be 40 stereocilia across but only three rows deep (Kimura 1966). As discussed below, this arrangement probably facilitates hearing at frequencies of tens of kilohertz by reducing the necessity for mechanical stimuli to propagate through the hair bundle.

At least three types of proteinaceous connections extend between contiguous stereocilia. Just above their tapers, the stereocilia of many species display basal links that run along each of the three hexagonal axes. Shorter horizontal top links join the stereocilia near their tips, again in all directions. Links of both types are thought to stabilize hair-bundle structure by restricting the separation of stereocilia and preventing twisting motions. The most important connections are tip links, fine protein filaments that join each stereocilium to the longest adjacent process (Pickles *et al* 1984). These links run along only a single hexagonal axis that corresponds to the bundle's plane of mirror symmetry. Each tip link comprises four cadherin molecules, members of a large family of proteins responsible for intercellular adhesion. The upper two-thirds of a link is formed of two parallel cadherin-23 molecules; the lower third consists of a parallel dimer of protocadherin-15 molecules (Kazmierczak *et al* 2007). The dissimilar cadherin molecules are joined at their distal tips, the amino termini, by $Ca^{2+}$-dependent intermolecular bonds (Sotomayor *et al* 2012).

Intact tip links are required for mechanoelectrical transduction. If the links are severed, for example by exposure to very low $Ca^{2+}$ concentrations that destabilize cadherins, transduction vanishes immediately (Assad *et al* 1991). Regeneration of the links accompanies the restoration of mechanosensitivity (Zhao *et al* 1996, Indzhykulian et al 2013). Physiological evidence suggests that a pair of transduction channels occurs at the lower insertion of each tip link, on the top surface of the shorter stereocilium (Beurg *et al* 2009). It thus seems highly probable that each strand of a tip link contacts a transduction channel, either directly or through one or more linking proteins, and that hair-bundle deflections are detected when the link pulls open the channel's molecular gate.

*4.2. Mechanical properties of the hair bundle*

Most models for transduction, adaptation, and the other activities of hair bundles invoke only a single degree of freedom, the deflection of the bundle's top along its axis of symmetry. A hair bundle can, however, move equally well in the perpendicular direction or at any intervening orientation. Moreover, each of the stereocilia in a hair bundle is potentially capable of independent motion. To what extent is the simplified representation of bundle movement justified?

Each stereocilium is a relatively rigid rod that pivots at its basal insertion (Crawford and Fettiplace 1985, Howard and Ashmore 1986). Although the parameter generally cited in both experimental and theoretical studies is the lateral displacement of the hair bundle's top, the tip of each stereocilium must actually move through an arc. Here the minute dimensions of hair-bundle movement justify the simplification. Although a bundle can endure displacements of a few micrometers (Shepherd and Corey 1994), ordinary stimuli—especially those of high-frequency auditory receptors—evoke bundle motions on the order of ±30 nm or less. Given that most hair bundles are about 5 μm in height, a threshold stimulus of ±0.3 nm represents an angular excursion of only 0.007°, which corresponds to a deflection of the Eiffel Tower's tip by the width of a *petite madeleine*.

The issue of independent movements by individual stereocilia is more complex. In most hair bundles, stimulus forces are applied to the kinocilium at the bundle's tall edge and must propagate from there across the array of stereocilia. The hair bundles of outer hair cells in the mammalian cochlea are unusual in that the tallest stereocilia in each file are directly attached to the gelatinous tectorial membrane, so stimulation delivers force across the bundle's entire width. The stimuli must nonetheless travel from the displaced stereocilia to those of the subsequent, shorter ranks. In every instance there is likely to be some decrement in the magnitude of stereociliary displacement deeper in the hair bundle (Silber *et al* 2004).

Interferometric measurements, analytic calculations, and finite-element modeling have revealed that the separation of stereocilia depends critically on the frequency of stimulation (Kozlov *et al* 2011). During stimulation at low frequencies, the elasticity of the stereociliary pivots tends to hold every stereocilium in its resting position. As a result, each shorter stereocilium moves somewhat less than its taller neighbor. The horizontal top connectors between adjacent stereocilia, which are thought to counter stereociliary separation, likely evolved to counter this problem. Stereociliary movements at very high frequencies are instead influenced by inertia. In this regime the force required to accelerate the stereocilia is great enough that, when stimuli are applied at a bundle's tall edge, successively shorter ranks lag behind their taller neighbors. The tendency for hair bundles responsive to high frequencies to have only a few ranks of stereocilia might represent an evolutionary adaptation to this challenge.

The hair bundle's most interesting mechanical behavior occurs during stimulation at the intervening frequencies, the interval from about 100 Hz to 10 kHz that corresponds broadly to our range of hearing. Here viscosity plays the leading role (Kozlov *et al* 2011). When any stereocilium is displaced by a stimulus, its shorter neighbor might move in either of two ways. It could separate from the taller stereocilium in response to elastic and inertial forces, a mode that requires sucking liquid into the nanometer-scale space between the adjacent processes. The shorter stereocilium could alternatively remain in contact with the taller at their point of tangency, in which instance both would pivot about their basal insertions and move in concert, shearing along each other without separation. The latter mode in actuality dominates the former, for the drag coefficient associated with the separation mode is three orders of magnitude greater than that for the longitudinal shearing mode. Although the oscillation of a hair bundle in endolymph inevitably dissipates some stimulus energy, a bundle's drag differs little from the Stokes drag on a sphere of comparable size (Denk *et al* 1989, Howard and Hudspeth 1988). Viscosity actually provides an unexpected advantage by holding the hair bundle together during stimulation. The assumption that a bundle moves with a single degree of freedom, then, rests on the evidence that viscosity suppresses stereociliary separation over the range of audible frequencies.

*4.3. Mechanoelectrical transduction*

Like other components of the nervous system, a hair cell employs electrical signals; like other sensory receptors, the cell uses these signals to represent physical stimuli. When a hair cell is stimulated by sound, the potential difference across its surface membrane changes in a graded way dependent upon the amplitude of hair-bundle motion (Hudspeth and Corey 1977). Threshold stimuli elicit responses estimated at 100 µV, whereas the largest responses might reach 60 mV *in vivo* (Johnson *et al* 2011).

The electrical response of any sensory cell, the so-called receptor potential, results from the flow of current through ion channels in the cell's membrane. A hair cell is endowed with mechanically sensitive channels termed mechanoelectrical-transduction channels. The magnitude and variance of the transduction current suggests that each cell has relatively few such channels, on the order of one hundred to a few hundred, or only a few per stereocilium (Hudspeth 1983; Holton and Hudspeth, 1986). Comparison of a hair cell's total transduction current with the current through individual channels suggests that there are likely two channels at the lower insertion of each tip link (Beurg *et al* 2009). Although there is as yet no evidence of a direct connection between a tip link and the channels, the fact that a link has twofold symmetry—it is a dimer of protein dimers—suggests that each of the link's two strands contacts a single channel.

The polarity of a hair cell's response depends upon the orientation of hair-bundle stimulation. Moving the bundle toward its tall edge elicits a depolarization, or decrease in the resting potential, which is approximately -60 mV. Because the intracellular potential becomes more positive—an excitatory response—such a movement is defined as a positive stimulus. Deflecting the hair bundle in the opposite direction, a negative stimulus, elicits a hyperpolarization: the transmembrane potential grows increasingly negative. For stimulation along the axis of mirror symmetry, the relation between displacement and response is sigmoidal. There is no response when the bundle is moved at a right angle to the axis of symmetry (Shotwell *et al* 1981). Stimulation at other angles produces a response that varies as the cosine of the deviation from the axis of symmetry. The bundle thus resolves any stimulus into two orthogonal components, responding in a graded fashion to the component in the plane of symmetry but ignoring the perpendicular component.

The vectorial responsiveness of hair bundles is unimportant in the auditory system, in which all of the hair bundles are oriented such that their axes of greatest sensitivity accord with the direction of mechanical stimulation. Directional sensitivity plays an important role, however, in the vestibular organs responsive to linear acceleration, which include the human sacculus and utriculus. Here the hair cells lie in two sheets oriented perpendicular to one another. Within each of these organs, the hair bundles assume a range of angular orientations spanning a full 360°. As a consequence, acceleration of the head in any direction optimally excites a coterie of hair cells in at least one of the organs and maximally inhibits another ensemble. Hair cells of the intervening orientations respond to a lesser extent or not at all. The responses transmitted from these hair cells to the brain uniquely determine the magnitude and direction of acceleration. The labyrinths of the two ears provide largely redundant information. If their outputs differ, however, for example because a unilateral infection causes spurious neural activity, the brain is incapable of resolving the discrepant information and vertigo ensues.

## 4.4. The gating-spring model and gating compliance

The opening and closing of a transduction channel can be described by the gating-spring model, in which it is supposed that each channel's molecular gate opens in response to tension conveyed through an elastic element, the gating spring (Corey and Hudspeth 1983, Howard *et al* 1988, Hudspeth 1992, Markin and Hudspeth 1995). This spring is in turn tensed by deflection of the hair bundle in the positive direction and relaxed by motion in the opposite sense. A channel is supposed to exist in either of two states, open or closed, and to spend a negligible time during the transitions between those states. Each channel's opening and closing is thought to be stochastic and independent of that of other channels. In this model, the energy $U_O$ associated with an open channel is

$$U_O = E_O + \frac{\kappa_{GS}}{2}\left(\gamma X_{HB} + x_C - x_A - d\right)^2, \tag{4.1}$$

in which $\kappa_{GS}$ is the stiffness of a gating spring, $X_{HB}$ is the displacement of the hair bundle, and $d$ is the distance through which the gate moves upon opening (figure 9). $E_O$ is the energy associated with the open channel itself, independent of the elastic energy in the gating spring represented by the second term on the right. The parameter $\gamma$ is a geometric factor that relates the shear between two adjacent stereocilia—and thus the extension of the gating spring—to displacement of the hair bundle's top (figure 8(d)). $x_C$ is the extension of the gating spring when the hair bundle stands in its undisturbed resting position; $x_A$ represents the extent to which the gating spring has relaxed as a result of adaptation.

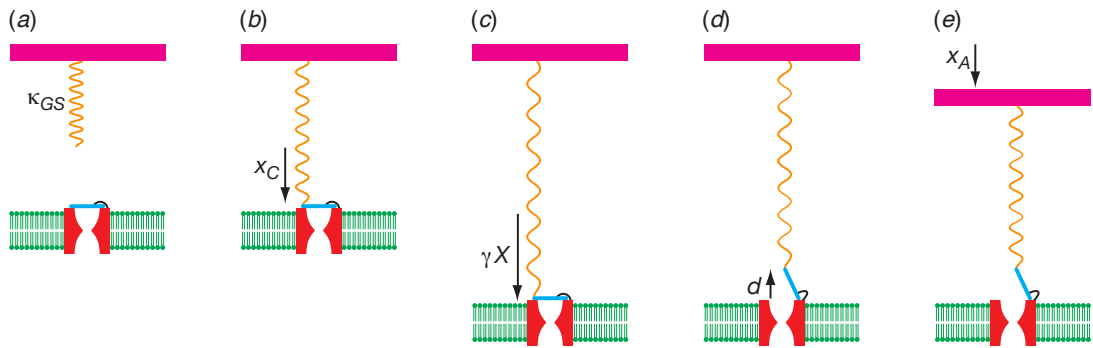

**Figure 9. The gating-spring model of mechanoelectrical transduction**. (*a*) The gating spring (orange) of stiffness $\kappa_{GS}$ represents a tip link and the mechanically compliant structures in series at both of its insertions. The detachment of the spring from the transduction channel is entirely hypothetical. (*b*) When the hair bundle stands in its resting position, the spring extends a distance $x_C$ to contact the molecular gate (blue) of a transduction channel (red) in the stereociliary membrane (green). (*c*) Deflection of the bundle through a distance $X$ elongates the spring by an amount $\gamma X$. (*d*) If the channel opens, the spring relaxes by a distance $d$. (*e*) Finally, movement of the tip link's upper insertion during the adaptation process relaxes the spring by an amount $x_A$.

Because the gating spring elongates by a distance $d$ as the channel shuts, the energy $U_C$ associated with the closed configuration of the channel is

$$U_{\text{C}} = E_{\text{C}} + \frac{\kappa_{\text{GS}}}{2}\left(\gamma X_{\text{HB}} + x_{\text{C}} - x_{\text{A}}\right)^2. \tag{4.2}$$

The energy difference $\Delta U$ upon channel opening therefore proves to be a linear rather than a quadratic function of hair-bundle displacement:

$$\Delta U = U_{\text{O}} - U_{\text{C}} = \Delta E - \kappa_{GS}d\left(\gamma X_{\text{HB}} + x_{\text{C}} - x_{\text{A}} - \frac{d}{2}\right) = -Z_{\text{GS}}\left(X_{\text{HB}} - X_0\right). \tag{4.3}$$

Here $\Delta E = E_{\text{O}} - E_{\text{C}}$, $Z_{\text{GS}} = \gamma\kappa_{\text{GS}}d$, and $X_0$ subsumes a combination of the other parameters. The open probability $P_{\text{O}}$ then follows from the Boltzmann relation,

$$P_{\text{O}} = \frac{1}{1 + e^{-Z_{\text{GS}}(X_{\text{HB}} - X_0)/(k_{\text{B}}T)}}. \tag{4.4}$$

This relation (figure 10(a)) fits the experimental data from the hair cells of numerous receptor organs in a variety of species (Hudspeth and Corey 1977, Howard and Hudspeth 1988). It is apparent that $X_0$ represents the hair-bundle displacement at which the channel's open probability is one-half. The parameter $Z_{\text{GS}}$, which describes the sensitivity of gating to mechanical stimulation, is called the single-channel gating force: it represents the change in the force at a hair bundle's top when a channel opens or closes.

If the open probability of transduction channels depends upon the tension in a gating spring, it follows by mechanical reciprocity that the gating of the channels must affect the tension borne by the spring. Suppose that this force is $f_{\text{O}}$ when the channel is open and $f_{\text{C}}$ when it is closed. The time-averaged force $f_{\text{GS}}$ in each gating spring is then

$$f_{\text{GS}} = P_{\text{O}}f_{\text{O}} + \left(1 - P_{\text{O}}\right)f_{\text{C}} = P_{\text{O}}\kappa_{\text{GS}}\left(\gamma X_{\text{HB}} + x_{\text{C}} - x_{\text{A}} - d\right) + \left(1 - P_{\text{O}}\right)\kappa_{\text{GS}}\left(\gamma X_{\text{HB}} + x_{\text{C}} - x_{\text{A}}\right)$$
$$= \kappa_{\text{GS}}\left(\gamma X_{\text{HB}} + x_{\text{C}} - x_{\text{A}}\right) - P_{\text{O}}\kappa_{\text{GS}}d. \tag{4.5}$$

Because a hair bundle moves as a unit, its $N$ transduction elements lie more-or-less in parallel. Owing to the mechanical advantage afforded by the bundle's geometrical arrangement, the total force produced by the ensemble of gating springs, as measured at the top of the hair bundle, is $N\gamma f_{\text{A}}$. The actin pivots at the bases of the stereocilia, which resist deflection of the hair bundle from its resting position $X_{\text{SP}}$ in the absence of gating springs, make a contribution proportional to their combined stiffness $K_{\text{SP}}$. The total external force $F$ required to deflect a bundle by a distance $X_{\text{HB}}$ is accordingly

$$F = N\gamma\kappa_{\text{GS}}\left(\gamma X_{\text{HB}} + x_{\text{C}} - x_{\text{A}}\right) - NP_{\text{O}}Z_{\text{GS}} + K_{\text{SP}}\left(X_{\text{HB}} - X_{\text{SP}}\right). \tag{4.6}$$

For large excursions in either direction, this relation is linear in the displacement $X_{\text{HB}}$. Near the bundle's resting position, however, the gating of channels interposes a nonlinearity between the two linear regimes (figure 10(b)).

When the bundle stands at rest, the extent of adaptation $x_{\text{A}}$ and the hair-bundle deflection $X_{\text{HB}}$ are by definition zero. The resting open probability $P_{\text{OR}}$ is then set by the opposition of tension in the tip links and flexion of the stereociliary pivots:

$$P_{\text{OR}} = \frac{N\gamma\kappa_{\text{GS}}x_{\text{C}} - K_{\text{SP}}X_{\text{SP}}}{NZ_{\text{GS}}}. \tag{4.7}$$

The hair bundle thus resembles a strung bow in which the tension along the bowstring balances the flexion of the bow itself.

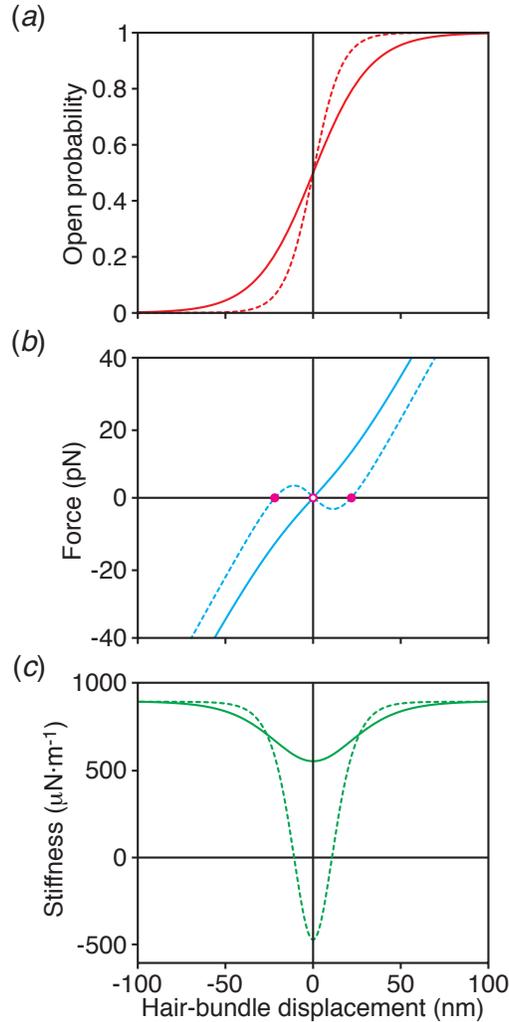

**Figure 10. Schematic representations of responses of a hair bundle to mechanical deflection**. (*a*) The displacement-open probability relations are determined from the responses of a hair bundle to rapid deflections from its resting position. The steepness of the relations depends upon the single-channel gating force; here the values are $Z_{GS}$ = 0.25 pN (continuous curve) and $Z_{GS}$ = 0.50 pN (dashed curve). The open probability is 0.5 in the absence of stimulation. (*b*) The displacement-force relations for the same parameter values reveal that a sufficiently large gating force produces a negative slope stiffness and consequent instability. Each plot depicts the force applied to a bundle by a stimulus fiber after the hydrodynamic drag force has vanished and before adaptation has shifted the relation. The force produced by the hair bundle is equal and opposite to that shown. For the dashed curve, the bundle has two stable fixed points (solid dots) and an unstable fixed point (open dot). (*c*) The hair bundle's stiffness approaches a constant value for large excursions in either direction, but decreases over the range in which channels open and close as a result of gating compliance. For large values of the gating force, the bundle's stiffness actually becomes negative.

For frequencies up to several tens of kilohertz, the hair bundle's inertia has a negligible effect. When a stimulus fiber of stiffness $K_{SF}$ and drag coefficient $\xi_{SF}$ is applied to a bundle with drag coefficient $\xi_{HB}$ and the fiber's base is deflected by an amount $\Delta$, the equation of motion is therefore

$$\left(\xi_{\mathrm{SF}} + \xi_{\mathrm{HB}}\right)\frac{dX_{\mathrm{HB}}}{dt} = K_{\mathrm{SF}}\left(\Delta - X_{\mathrm{HB}}\right) - N\gamma\kappa_{GS}\left(\gamma X_{\mathrm{HB}} + x_{\mathrm{C}} - x_{\mathrm{A}}\right) + NP_{\mathrm{O}}Z_{\mathrm{GS}} - K_{\mathrm{SP}}\left(X_{\mathrm{HB}} - X_{\mathrm{SP}}\right). \tag{4.8}$$

Now consider the slope stiffness of the hair bundle, which is obtained by differentiating the external force $F$, equation (4.6), with respect to displacement:

$$K_{\mathrm{HB}} = \frac{dF}{dX_{\mathrm{HB}}} = N\gamma^2\kappa_{GS} + K_{\mathrm{SP}} - NZ_{GS}\frac{dP_{\mathrm{O}}}{dX_{\mathrm{HB}}} = K_{\infty} - \frac{NZ_{\mathrm{GS}}^2}{k_{\mathrm{B}}T}P_{\mathrm{O}}\left(1 - P_{\mathrm{O}}\right). \tag{4.9}$$

Here $K_{\infty}$ represents the asymptotic stiffness encountered upon large displacements of the bundle in either the positive or the negative direction (figure 10($c$)). The final term, the gating compliance, reflects the effect of channel opening and closing. Gating reduces the stiffness of the bundle near its resting position, with the greatest effect at $P_{\mathrm{O}}=1/2$. Most remarkably, for sufficiently large values of $Z_{\mathrm{GS}}$ the gating compliance can actually confer a negative slope stiffness on the bundle (Denk *et al* 1992). This behavior, which has been observed experimentally (Martin *et al* 2000, Le Goff *et al* 2005), renders the bundle mechanically unstable and plays a critical role in the ear's active process.

Three fundamental aspects of the gating-spring model remain to be confirmed. First, what is the identity of the mechanoelectrical-transduction channel in hair cells? Recent evidence suggests that the transmembrane channel-like proteins TMC1 and TMC2 are constituents, for variants of the transduction channel express distinct single-channel conductances and $Ca^{2+}$ selectivities (Kawashima *et al* 2011, Kim and Fettiplace 2012, Pan *et al* 2013). Because mechanically activated currents of reversed displacement sensitivity persist after knockout of the genes encoding TMC1 and TMC2, however, it remains possible that those proteins are necessary to position some other channel molecules (Kim *et al* 2013).

A second issue is the relationship between the two transduction channels thought to reside at the base of each tip link. Can those channels gate independently of one another, or if not, do they exhibit positive or negative cooperativity that would require modification of the gating-spring model (Markin and Hudspeth 1995)? Electrical recordings from hair cells demonstrate apparent single-channel currents (Ricci *et al* 2003, Pan *et al* 2013), but it remains possible that each opening event represents the activation of two highly synchronized channels and that the lower-conductance states observed occasionally (Crawford et al 1991) actually correspond to the gating of individual channels.

A final uncertainty lies in the identity of the gating spring itself. The gating-spring theory was originally proposed as an abstract model meant to explain the kinetics of transduction-channel gating (Corey and Hudspeth 1983). Although the subsequent discovery of the tip link offered an attractive candidate for the gating spring (Pickles *et al* 1984), evidence from electron microscopy and molecular modeling suggests that the link's cadherin chains are too stiff to serve in that role (Kachar *et al* 2000, Sotomayor *et al* 2012). Moreover, high-resolution measurements of thermally driven hair-bundle movements associate complex viscoelastic behavior with the gating springs (Kozlov *et al* 2011). It is possible that there is significant compliance at the upper insertion of each tip link, where myosin-1c molecules help to anchor the link to the actin cytoskeleton (Howard and Spudich 1996). Perhaps more likely is flexibility at the lower end of the tip link, where electron microscopy often reveals the membrane at the stereociliary tip to be elevated into a "tent," apparently by tension in the tip link (Powers *et al* 2012). In keeping with this hypothesis, certain adaptive properties of hair bundles (see below) are well fit by a model including a compliant element whose stiffness depends on the intracellular $Ca^{2+}$ concentration just inside each transduction channel (Bozović and Hudspeth 2003, Martin *et al* 2003).

*4.5. Slow adaptation by hair bundles*

The narrow operating range of the hair cell's transduction process poses a problem: how does the cell avoid saturation of its responses when confronted with stimuli that deflect the hair bundle far in the positive or negative direction? Many sensory systems adapt to sustained stimulation. The photoreceptors of the visual system, for example, implement gain control through feedback in a complex biochemical cascade. Because mechanoelectrical transduction by hair cells is direct, however, that strategy is inappropriate; hair cells have instead evolved a novel, mechanical form of adaptation.

Adaptation is apparent when one records a hair cell's electrical response during a protracted deflection of the hair bundle in the positive direction. The cell immediately undergoes a depolarization, which relaxes toward the baseline level over a few tens of milliseconds (Eatock *et al* 1987). A common experimental procedure is to voltage-clamp the cell, using negative feedback to hold the membrane potential at a constant level while monitoring the transmembrane current. Adaptation then appears as a time-dependent diminution in the initial influx of cations through the transduction channels. This approach resolves two components of adaptation: the first, fast adaptation, occurs within a millisecond or so; the second, slow adaptation, displays a time constant for approximately exponential relaxation of several tens of milliseconds (Howard and Hudspeth 1987, Wu *et al* 1999, Kennedy *et al* 2003).

Slow adaptation is thought to result from resetting the lengths of tip links to raise or lower the tension that they bear (Howard and Hudspeth 1987; Assad and Corey 1992, Hudspeth and Gillespie 1994, Gillespie and Cyr 2004). In this formulation, the upper insertion of each tip link is attached to a molecular motor, the insertional plaque, that runs along actin tracks in the taller stereocilium of the pair (figure 11). The motor strives continually to ascend, but stalls at some point owing to the tension that it develops in the attached tip link. The same tension maintains the open probability of the associated transduction channels at a nonzero resting value (equation (4.7)). When the hair bundle is deflected in the positive direction, the shearing motion between adjacent stereocilia increases the tension in the tip links. Additional transduction channels open transiently, but the molecular motor then slips down the stereocilium and some channels reclose—the process of adaptation. Because the rate of adaptation in this direction depends upon the extracellular concentration of $Ca^{2+}$ and the transmembrane potential that governs the ion's influx (Assad *et al* 1989, Hacohen *et al* 1989), $Ca^{2+}$ may speed adaptation by entering stereocilia through open transduction channels and facilitating the downward sliding of the motors. A negative stimulus has the inverse effect: deflection of the hair bundle by a stimulus force relaxes the tip link transiently, closing the channels; within a few tens of milliseconds, however, the motor ascends, restores the tension, and reopens some of the channels. Adaptation in either direction has the valuable property of restoring a hair bundle's mechanosensitivity and thus averting a saturated response.

To model the process of slow adaptation, suppose that the upper insertion of each tip link is subject to two opposing forces. The tension in the link, which reflects the hair bundle's deflection, tends to pull the insertion downward. The opposing, upward force is contributed by a molecular motor that consumes the biological energy source adenosine triphosphate (ATP). The motor's activity is sensitive to the local $Ca^{2+}$ concentration, which in turn depends upon the ion's influx through the transduction channels. The extent of adaptation $x_A$, which corresponds to downward motion of the tip-link insertion along the stereocilium, then evolves according to a relation such as (Assad and Corey 1992, Nadrowski *et al* 2004, Fredrickson-Hemsing *et al* 2012)

$$\xi_A \frac{dx_A}{dt} = \kappa_{GS}(\gamma X_{HB} + x_C - x_A - P_O d) - (1 - P_O)f_M. \qquad (4.10)$$

Here $f_M$ is the force produced by an active adaptation motor and $\xi_A$ is a phenomenological drag coefficient that arises from the kinetics of the motor, whose constituent myosin molecules attach and detach from the underlying actin filaments at velocity-dependent rates. The greater the open probability and ensuing $Ca^{2+}$ concentration in the stereociliary cytoplasm, the less force is produced by myosin and the more rapidly tip-link tension effects adaptation.

Several lines of evidence support this model of adaptation. First and foremost, the reclosure of transduction channels during adaptation to a positive stimulus is accompanied by a mechanical relaxation of the hair bundle (figure 11). This behavior is thought to reflect the falling tension in tip links as the associated insertional plaques descend. As anticipated if molecular motors are involved, adaptation is blocked upon the removal of ATP or its replacement by nonhydrolyzable analogs (Gillespie and Hudspeth 1993). A similar effect ensues upon infusion into the cytoplasm of phosphate, a product of ATP's hydrolysis, or related ions (Yamoah and Gillespie 1996).

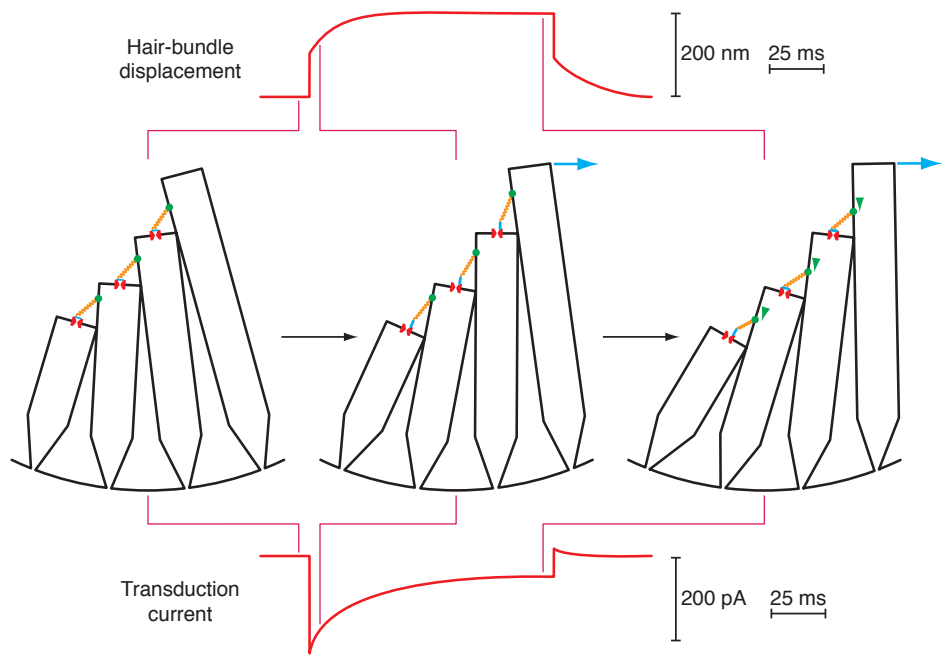

**Figure 11. Schematic diagram of the inferred mechanism of slow adaptation by hair cells**. From left to right, the schematic diagrams at the center portray a hair bundle at rest, shortly after the onset of a force pulse (blue arrow) in the positive direction, and after the completion of slow adaptation. The transduction channels at the bases of the tip links are initially closed, but open during stimulation in response to the shear between adjacent stereocilia. As the insertional plaques at the upper insertions of the tip links slide downward (green arrowheads), however, the tension in the links declines and most of the channels reclose. The electrical response portrayed at the bottom accordingly manifests a transient peak of inward current, which is negative in sign, followed by a roughly exponential decline during adaptation. As shown at the top, the bundle's displacement in response to a constant force displays a progressive relaxation in the positive direction during the adaptation process.

Stereocilia contains several forms of myosin, the only motor protein known to operate along actin filaments. Mutation of myosin-7a in mammalian hair cells perturbs transduction by displacing the range of hair-bundle positions at which channels open far positive to the bundle's resting position (Kros *et al* 2002). This isozyme might participate in adaptation or could represent the extent spring thought to hold each insertional plaque near its resting position (Yamoah and Gillespie 1996). In the hair bundles of frogs, myosin-1c clusters at the expected site of adaptation, the insertional plaques at the upper ends of the tip links (García *et al* 1998, Steyger *et al* 1998). Site-directed mutation of this myosin isoform indicates that it is both necessary and sufficient for normal adaptation. Myosins of class 1 possess several properties that render them particularly suitable for the adaptation apparatus (Batters *et al* 2004). In particular, they cling tenaciously to actin filaments when subjected to stalling forces, a useful feature for the maintenance of tip-link tension (Greenberg *et al* 2012). Consistent with the requirements of the model, these isoforms additionally display sensitivity to $Ca^{2+}$, reducing their attachment lifetime with an increase in the ion's concentration (Adamek *et al* 2008, Lewis *et al* 2012). Finally, myosins-1 exhibit a two-step working stroke in which back-and-forth transitions between bound states might provide rapid force steps (Veigel *et al* 1999, Batters *et al* 2004).

Although experimental results accord well with the model of slow adaptation, no experiment has yet proven that the upper insertions of tip links actually migrate during the process. The expected movements are quite small: owing to the geometrical gain of the hair bundle, slow adaptation to a huge stimulus of 500 nm is expected to involve a displacement of the tip link's insertion along the stereocilium of only 55 nm or so. Detecting such a motion by confocal microscopy might be possible, but challenging; recently developed superresolution techniques should suffice.

Experimental measurements have confirmed the capacity of adaptation to shift the displacement-force relation of a hair bundle (Le Goff *et al* 2005). This phenomenon can explain the bundle's ability to amplify at least low-frequency mechanical stimuli. As a hair bundle oscillates in response to stimulation, adaptation continuously offsets its range of negative stiffness (figure 12). As a consequence, the hair bundle does not act as a linear elastic element; instead, the moving bundle displays a phase lead with respect to the stimulus and does work not only against hydrodynamic dissipation but also against the stimulus fiber itself (Martin and Hudspeth, 1999).

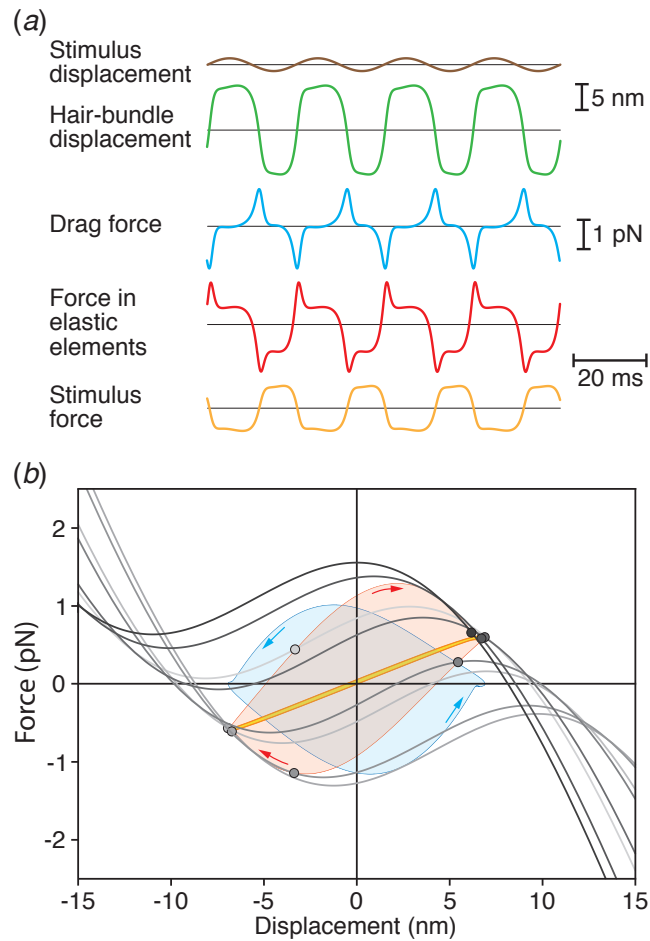

**Figure 12. Modeling the role of adaptation in amplification by the hair bundle.** (*a*) When a piezoelectric stimulator deflects the base of a flexible glass fiber sinusoidally with an amplitude of 1 nm (top trace), the hair bundle attached to the fiber's tip undergoes an amplified and somewhat distorted oscillation with an amplitude of 7 nm at its characteristic frequency of 41 Hz (second trace). The motion of the fiber and bundle through the aqueous medium is retarded by hydrodynamic drag (third trace). The elastic elements of the hair bundle—the gating springs and stereociliary pivots—exert a force (fourth trace) that opposes this drag. Although the stimulus fiber entrains the hair bundle's motion, the fiber actually provides a force (fifth trace) oriented in a direction opposite to the bundle's movement. (*b*) During each cycle of oscillation, the displacement-force relation of the hair bundle migrates back-and-forth as a result of adaptation. Here the relation is shown in eight successive positions separated in phase by 45° and depicted in progressively paler shades of gray, starting from the top. Because these curves display the force exerted by the hair bundle against an external load, they are opposite in sign with respect to the displacement-force relations in figure 10(*b*). Plotting the force owing to hydrodynamic drag against the hair bundle's position yields the blue curve, which encloses an area proportional to the energy dissipation per cycle. The red curve, which represents the force provided by the hair bundle's elastic components, bounds an area representing the energy supplied by the active process during each cycle. The points along this curve show the forces applied at the eight indicated positions of the displacement-force relation. Note that the circulations in the two curves are of opposite directions (arrows), reflecting dissipation by drag

and energy input by adaptation. Because the force delivered through the elastic elements nearly cancels that owing to dissipation, the displacement-force relation for the stimulus fiber shown by the orange curve encompasses only a small area, which represents the work done over a cycle by the stimulus fiber. To achieve the same oscillation in the absence of the active process, the fiber would need to supply a far greater energy equivalent to that lost to hydrodynamic friction.

### 4.6. Fast adaptation by hair bundles

As its name implies, slow adaptation serves principally to immunize mechanoelectrical transduction against static and low-frequency hair-bundle deflections that would otherwise saturate responsiveness. Less clear is the significance of fast adaptation, which was characterized initially under the descriptive term $Ca^{2+}$-dependent channel reclosure. Although this process may also have an adaptive function, an intriguing possibility is that it contributes to the hair cell's active process at frequencies exceeding those accessible to myosin-based motility.

Fast adaptation involves the reclosure of transduction channels and a concomitant diminution in the transduction current on a millisecond or sub-millisecond timescale (Howard and Hudspeth 1997, Kennedy *et al* 2005). In association with this electrical phenomenon, the hair bundle displays a biphasic mechanical response, first moving in the positive direction and then twitching in the opposite direction (Benser *et al* 1996). The effect is graded in intensity, initially growing with an increase in the stimulus amplitude, but non-monotonic: for hair-bundle deflections exceeding about 40 nm, the response declines progressively to zero. The maximal amplitude of the twitch in hair cells from the frog is approximately 20 nm, but the value might be even greater in the larger bundles of the mammalian cochlea. Taking into account the geometrical gain factor $\gamma$ relating tip-link extension to hair-bundle movement, this value implies a displacement at the tip links of only 3 nm. Excursions of this sort are typical of the steps made by molecular motors.

The key uncertainty about fast adaptation is whether it actually requires an explanation. When a frog's hair bundle is abruptly displaced into its region of dynamical instability, it leaps to a new fixed point as rapidly as viscous drag permits. The bundle's subsequent relaxation during slow adaptation then results in the characteristic waveform described above (Tinevez *et al* 2007). If this mechanism proves ubiquitous then fast adaptation may be explained by the combination of gating compliance, which imposes a dynamic instability, and slow adaptation, which poises the bundle on the brink. Repriming the system for each cycle of activity would then proceed at the relatively slow pace of myosin-based movement. This mechanism would pose a conundrum, however, for the process must do at least some work on a cycle-by-cycle basis to amplify inputs. Could myosin-1c or any other myosin isozyme accomplish this at frequencies approaching 100 kHz?

An alternative possibility is that fast adaptation represents a specialized and perhaps unique form of cellular motility. The slow steps in myosin activity include the binding of ATP and especially the docking of myosin's bulky head to an actin filament. If myosin's head were to remain fixed to the filament on a short timescale, however, successive cycles of activity could occur at a much shorter latency. In this model the myosin head would simply rock back-and-forth to accomplish amplificatory work. Exactly this occurs in the flight muscles of insects, which can operate at frequencies exceeding 1 kHz (Pringle 1967). Oscillation would be further accelerated if the energy of amplification were not supplied by ATP, whose stereospecific binding is inevitably prolonged, but stemmed instead from the binding of $Ca^{2+}$. Ions entering through a transduction channel might swiftly bind to a protein that regulates myosin's activity, such as calmodulin, or perhaps to another protein altogether. This binding would evoke a reconfiguration of the relaxation element that would pump energy into the hair bundle's motion (Choe *et al* 1998, Bozović and Hudspeth 2003). The process would ultimately draw its power from ATP, which would maintain the $Ca^{2+}$ gradient through the activity of the $Ca^{2+}$ pumps that densely stud the stereociliary membrane (Yamoah *et al* 1998). Note that no single pump molecule would be required to keep pace with motility on a cycle-by-cycle basis; instead, a large ensemble of pumps would cooperate to maintain the time-averaged $Ca^{2+}$ concentration in the stereociliary cytoplasm at a suitably low level.

There is no evidence regarding the identity of the putative relaxation element responsible for fast adaptation. Proteins such as the spasmin of *Vorticella* possess the hypothesized property of changing conformation rapidly in response to $Ca^{2+}$ binding (Amos *et al* 1975, Misra *et al* 2010), but no candidate substance with this property has been identified at stereociliary tips. Moreover, $Ca^{2+}$ may not be required for fast adaptation in mammalian cochlear hair cells (Peng et al 2013). Owing to the submillisecond time scale of fast adaptation and its restricted displacement range of only a few nanometers, experimental measurement of the molecular rearrangements involved is even more problematical than for slow adaptation.

### 4.7. Spontaneous oscillations of hair bundles

When hair bundles operate in the intact ear, they are almost always loaded by accessory structures, such as tectorial or otolithic membranes, that convey mechanical stimuli to the bundles. After having been separated from their accessory structures by enzyme treatment, hair bundles *in vitro* may oscillate spontaneously (Fettiplace and Crawford 1985, Martin *et al* 2003). The phenomenon is especially prevalent when the isolated receptor organ is maintained in a two-compartment experimental chamber so that the cellular surfaces are bathed in physiologically appropriate media, namely endolymph for the apical surfaces and perilymph for the basal ones (Martin and Hudspeth 1999, Strimbu *et al* 2010). The hair bundles from the extensively studied sacculus of the bullfrog oscillate at frequencies of a few tens of hertz and with amplitudes of several tens of nanometers. The oscillation waveforms vary widely, from small sinusoidal motions to large, biphasic relaxation oscillations.

With a suitable choice of parameter values, a model comprising equation (4.8) and equation (4.10) provides realistic simulations of spontaneous hair-bundle oscillations (Bozović and Hudspeth 2003, Martin *et al* 2003, Nadrowski *et al* 2004, Frederickson-Hemsing *et al* 2012). The model successfully encompasses the effects of experimental alterations in the stiffness load on a bundle, the offset from the bundle's resting position, or the extracellular $Ca^{2+}$ concentration. The phase space of the model system displays regions of monostability and bistability as well as a locus of spontaneous oscillation bounded by Hopf bifurcations (Nadrowski *et al* 2004, Ó Maoiléidigh *et al* 2012). Because parts of the phase space can be parameterized by experimentally accessible variables such as the elastic load and offset displacement imposed on a hair bundle (figure 13), it should be possible to test how well the model captures the behavior of actual hair bundles.

Although the accessory structures normally attached to hair bundles are soft protein gels, their stiffness is of the same order of magnitude as that of hair bundles (Benser *et al* 1993). As a consequence, the spontaneous oscillation of hair bundles is largely suppressed *in vivo* (Strimbu *et al* 2010). This observation suggests that the relatively large, low-frequency oscillations observed for unencumbered hair bundles *in vitro* are an artifact of the recording situation. Working against the elastic load of accessory structures, native hair bundles may instead operate well to the right in the phase diagram, near the head of the fish-shaped locus of spontaneous oscillation (figure 13). Bundles are thought to use an as yet uncharacterized feedback to hold themselves at the brink of criticality (Camalet *et al* 2000). Because the rate and amplitude of spontaneous oscillation are influenced by drugs that affect the cytoplasmic concentration of the second messenger cyclic AMP, this adjustment might involve protein phosphorylation (Martin *et al* 2003).

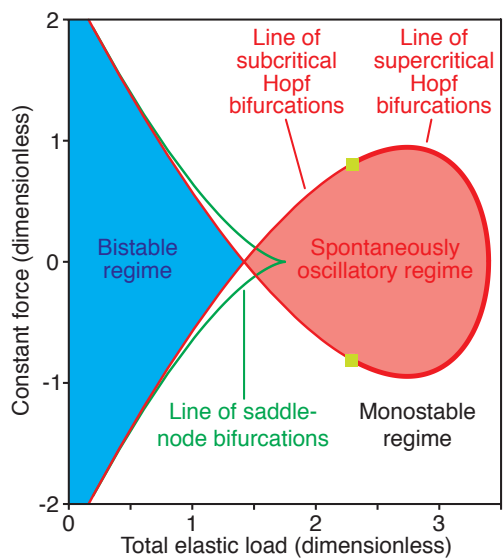

**Figure 13. Phase diagram for the behavior of a hair bundle**. The parameters of the analysis are the elastic load on the hair bundle, including both the bundle's internal stiffness and that of an attached stimulus fiber, and the force with which the bundle is offset from its resting position. Within the region of spontaneous oscillation, movement to the right is associated with progressively smaller oscillations of progressively increasing frequency. Bautin points (yellow squares) mark the transitions between supercritical and subcritical Hopf bifurcations.

Spontaneous hair-bundle oscillations seem a likely source of spontaneous otoacoustic emissions. Nonmammalian tetrapods display from as few as one to more than ten emission peaks (Manley 2006). Although it is not yet possible to demonstrate how many hair cells contribute to each peak, the upper bound set by dividing the number of cells by the number of emission peaks ranges from a handful to a hundred (Manley and Gallo 1997, Gelfand *et al* 2010). Modeling indicates that multiple hair cells can be mutually entrained in their oscillations through either viscous or elastic coupling between adjacent hair bundles (van Hengel et al 1996, Vilfan and Duke 2008, Gelfand *et al* 2010, Braun 2013).

The energy dissipation associated with spontaneous emission is surprisingly small. In the *Anolis* lizard, for example, each hair cell sensitive in the range 1-8 kHz has an emission power of about 140 aW, or some 20-140 zJ per cycle of oscillation (Manley and Gallo 1997). The hair bundle of such a cell possesses about 40 tip links, and therefore that number of motors, each of which is estimated to contain about 50 myosin-1c molecules (Gillespie *et al* 1993). With an efficiency of 50%, the hydrolysis of an ATP molecule by a single myosin molecule liberates about 40 zJ. It is apparent that, even if a given myosin molecule were active only once in every several cycles, the ensemble could readily supply the power necessary to sustain the observed emissions.

*4.8. Somatic motility of outer hair cells*

An epithelium characteristically rests upon a firm foundation of connective tissue. The auditory receptor organ of a reptile or mammal is unusual in that its underpinnings have been greatly attenuated, leaving only a thin cellular layer suspended between two liquid-filled chambers. This configuration may have evolved because separation of the basilar membrane from the bone and cartilage of the skull produced a structure capable of resonating at different frequencies along its length, and thus of superior frequency resolution. At the same time, however, the development of a resonant basilar membrane posed a challenge: how might the active process that had arisen to amplify the activity of individual hair bundles drive a far larger structure, the entire basilar membrane and sensory epithelium? At least in mammals, and perhaps in other taxa as well (Beurg *et al* 2013), evolution apparently solved this problem through the development of a second component of the active process, somatic motility or electromotility.

Somatic motility is a prominent property of the specialized outer hair cells that are unique to mammals (Brownell *et al* 1985, Ashmore 1987, 2008). These long, cylindrical cells have a minimum of intracellular organelles; they also possess few synapses by which to transmit information to nerve fibers. Outer hair cells are instead specialized for mechanical amplification. The surface membrane of each cell is loaded with millions of molecules of the protein prestin, a modified anion transporter that has assumed an amplificatory role (Zheng *et al* 2000).

Prestin is a piezoelectric transducer. Upon changes in the potential across the membrane, the tetrameric protein alters its configuration in the plane of the membrane (Zheng *et al* 2006, Wang *et al* 2010, Hallworth and Nichols 2012). Depolarization causes the molecule's area to decrease, and as a consequence the cell shortens. Hyperpolarization instead increases prestin's membrane area and evokes a cellular elongation. Because the volume of a hair cell is conserved on the short timescale involved in high-frequency transduction, the cellular radius undergoes converse changes in both instances.

The voltage sensitivity of prestin reflects the movement of charged components of the protein in the electrical field across the membrane. With the simplifying assumption that the protein has only two states, compact and extended, the probability $P_E$ of occupying the latter state and hence of cellular elongation is

$$P_E = \frac{1}{1 + e^{-z_P e(V - V_0)/k_B T}}, \tag{4.11}$$

in which $z_P$ is the gating valence associated with movement of prestin's voltage sensor, $e$ the electron charge, $V$ the membrane potential, and $V_0$ the potential at which the probability of extension is one-half. As a result of this charge movement, the total membrane capacitance $C_{OHC}$ of an outer hair cell depends nonlinearly on the membrane potential:

$$C_{OHC} = C_L + \left( \frac{N_P z_P^2 e^2}{k_B T} \right) \frac{e^{-z_P e(V - V_0)/k_B T}}{\left[ 1 + e^{-z_P e(V - V_0)/k_B T} \right]^2} = C_L + \left( \frac{N_P z_P^2 e^2}{k_B T} \right) P_E \left( 1 - P_E \right), \tag{4.12}$$

in which $C_L$ is the linear component owing primarily to membrane lipids and $N_P$ is the number of prestin molecules. Consistent with this formulation, plots of membrane capacitance against potential reveal the expected symmetrical, bell-shaped relation (Santos-Sacchi 1991).

Two aspects of somatic motility are particularly important. First, this motile process is remarkably fast. Outer hair cells can follow sinuosidal voltage changes to frequencies approaching 100 kHz, near the upper limit of mammalian hearing (Scherer and Gummer, 2004, Frank *et al* 1999). By contrast, a myosin-based phenomenon such as slow adaptation can probably operate to only a few kilohertz. The second key feature of somatic motility is the relatively large force that it produces. Whereas active motility by a hair bundle can exert a force of 10 pN or so, the motile soma of an outer hair cell exhibits a piezoelectrical responsiveness of 100 nN·V$^{-1}$, and can thus provide several nanonewtons for receptor potentials in the range of a few tens of millivolts (Iwasa and Adachi 1997). This feature is critical for the mechanical amplification of basilar-membrane motion: at high frequencies, somatic motility is apparently necessary to counter both the inertia of the basilar membrane and organ of Corti and the hydrodynamic drag on and within those moving structures (Ó Maoiléidigh and Jülicher 2010, Meaud and Grosh 2011).

An important topic of contemporary research is the relationship between the two components of the active process. One possibility is that somatic motility has simply supplanted active hair-bundle motility in high-frequency hair cells. When somatic motility is blocked, however, significant manifestations of the active process persist (Chan and Hudspeth 2005a, 2005b, Fisher *et al* 2012). Moreover, somatic motility lacks the strikingly nonlinear features of the active process—features that are demonstrably associated with active hair-bundle motility. It is most likely that the two mechanisms of motility collude in the active process. Somatic motility delivers most of the mechanical power that drives the basilar membrane (Meaud and Grosh 2011, Ó Maoiléidigh and Hudspeth, 2011), whereas active hair-bundle motililty helps to overcome the principal problem associated with piezoelectrical responsiveness, the membrane time constant. Prestin responds to changes in membrane potential, which in turn depend upon the flow of current through transduction channels. As the frequency of stimulation increases, a progressively greater current is required to alter the membrane potential with suitable rapidity. Numerous proposals have been advanced to explain how the problem posed by the membrane's time constant might be overcome. Perhaps the most probable explanation, however, is that the mechanical amplification provided by active hair-bundle motility augments the transduction current at high frequencies, thus extending the range of somatic motility (Ó Maoiléidigh and Hudspeth 2013).

*4.9 Noise and the hearing threshold*

A important corollary of the auditory system's great sensitivity is vulnerability to thermal noise. It is instructive in this regard to compare hearing to vision. The photoisomerization of 11-*cis* retinal, the chromophore that initiates the visual response, requires some 220 zJ or almost $55 \cdot k_B T$ (Yau et al 1979). A single photon of wavelength 505 nm, the peak of sensitivity for rod photoreceptors, delivers an energy of 390 zJ or over $95 \cdot k_B T$ at room temperature. It follows that a photon can reliably evoke a photoisomerization, and more importantly that the mean lifetime of a rhodopsin molecule before thermal isomerization—and a consequent false-positive signal—is around a thousand years! The triggering of a response in a photoreceptor is therefore a reliable indication that a photon has been captured, so a person can report the nearly synchronous arrival of as few as five photons (Hecht et al 1942). In the auditory and vestibular systems, by contrast, classic psychophysical experiments demonstrated that threshold inputs, which must be divided among numerous hair cells, have a total energy content only a few tens of times the level of thermal energy (de Vries 1948, 1949, Sivian and White 1933).

Different kinds of noise occur in the auditory periphery. The brownian motion of air molecules causes pressure fluctuations at the eardrum, for example, and the thermal excitation of water molecules in the cochlea buffets the basilar membrane (Harris 1968). The transduction process by which hair bundles respond to mechanical vibrations and the components of the cochlear active process noisy as well. These processes limit the perception of low-intensity stimuli and can influence the cochlea's nonlinear behavior.

Because most significant sounds are sinusoidal in character and of many cycles' duration, the auditory system can achieve its great sensitivity in part through the resonant averaging of its inputs. Frequency tuning is obviously useful for the identification of sound sources and speech sounds through the analysis of their frequency spectra. A second, less apparent advantage of tuning is the suppression of noise: a sharply tuned resonator responds best at its natural frequency but rejects noise outside its passband. The prevalence of tuning emphasizes the importance of this consideration in auditory systems, which employ at least seven strategies. As already discussed extensively, the resonance of the basilar membrane owing to its mechanical properties makes a major contribution to tuning. In the auditory organs of some amphibians, and probably in the mammalian cochlea, mechanical resonance also occurs in the tectorial membrane (Hillery and Narins 1984, Gummer et al 1996, Cai et al 2004, Ghaffari et al 2007). The damped mechanical resonance of individual hair bundles constitutes another contribution to frequency selectivity (Holton and Hudspeth 1983, Frishkopf and DeRosier 1983, Aranyosi and Freeman 2005). Indeed, in many ears there is a gradient in the mass and stiffness of bundles along the tonotopic axis on which frequency is represented. Another strategy employed by some hair cells, stochastic resonance of hair bundles, actually takes advantage of noise to promote sensitivity to a specific frequency of stimulation (Jaramillo and Wiesenfeld 1998). Active hair-bundle motility is clearly tuned, though as yet we remain unsure of the mechanism by which this is accomplished (Martin et al 2001). After transduction has occurred, the membrane potentials of many hair cells display electrical resonance owing to the interplay between voltage-activated $Ca^{2+}$ channels and $Ca^{2+}$-sensitive $K^+$ channels (Crawford and Fettiplace 1981, Hudspeth and Lewis 1988a, 1988b). In this instance the cells are tuned by gradients in the number of channels, their incorporation of accessory subunits, and their sensitivity to $Ca^{2+}$ and voltage (Art and Fettiplace 1987, Rosenblatt et al 1997, Miranda-Rottmann et al 2012). Finally, in a mechanism unprecedented in the nervous system, even the chemical synapses by which hair cells transmit signals to nerve fibers are frequency-selective (Patel et al 2012). Although it is improbable that any auditory receptor organ employs all seven of these strategies, many are known to use several as a series of cascaded frequency filters.

The root-mean-square motion of a free, passive hair bundle is of the order of 3 nm (Denk et al 1989, Jaramillo and Wiesenfeld 1998), a value in agrement with the expectation from the equipartition relation for a bundle of an observed stiffness around 500 µN·m$^{-1}$ (Hudspeth 1989, Svrcek-Seiler et al 1998). This magnitude of vibration is well in excess of the likely theshold for the effective transduction of stimuli (Martin and Hudspeth 2001). In an intact ear, however, the hair bundles of most receptor organs are coupled together through tectorial or otolithic membranes whose reactive impedance is comparable to the stiffness of a hair bundle (Benser et al 1993, Strimbu et al 2009, 2010, 2012). Because energy can be transferred between cells by this means, the thermal noise of individual hair bundles is attenuated by averaging across a population of bundles.

Application of the fluctuation-dissipation relation reveals that an active hair bundle operates far from thermodynamic equilibrium (Martin et al 2001). At least for low-frequency responses, however, the activity of a bundle is Markovian and accords with a generalized fluctuation-disipation theorem (Dinis et al 2012). When detached from a tectorial or otolithic membrane, such a bundle shows erratic movements that grade smoothly into noisy limit-cycle oscillations (Denk and Webb 1992, Martin et al 2003). A detailed analysis of the sources of bundle noise suggests that the clattering of transduction channels between their open and closed states dominates the contributions of adaptation motors and of the viscous medium around the hair bundle (Nadrowski et al 2004). Moreover, this theoretical work indicates that the presence of noise affects the optimal operating point for a hair bundle in the state space defined by the maximal force of adaptation motors and the Ca²⁺ sensitivity of the adaptation process. At least by the criterion of sensitivity to small stimuli, a bundle should reside near the middle of the locus of spontaneous oscillation rather than along the line of Hopf bifurcations (Nadrowski et al 2004, Han and Neiman 2010). Noise also disrupts the phase coherence of a bundle's response to periodic stimulation (Roongthumskul et al 2013), an effect that is minimized at a similar operating point. Mechanical coupling of hair bundles, as would be expected to occur *in vivo*, should mitigate the effects of noise (Dierkes et al 2008, Barral et al 2010).

In a model of active hair-bundle motility, noise has the effect of renormalizing the linear response function while preserving its functional form (Jülicher et al 2009). To appreciate the effect of noise near threshold, consider a system described by the normal form of the Hopf bifurcation with forcing, equation (3.10), but with an additional noise term:

$$\partial_t W = (\zeta_r + i\omega_*)W - (\nu + i\sigma)\left|W\right|^2 W + \tilde{P}e^{i\omega t} + \xi \;. \tag{4.13}$$

In this Langevin equation the variable $\xi$ represents complex Gaussian white noise: $\left\langle \xi(t)\xi*(t+\tau)\right\rangle = \varepsilon\delta(\tau)$ and $\left\langle \xi(t)\xi(t+\tau)\right\rangle = 0$, in which $\varepsilon$ is the noise strength and $\delta(\tau)$ is the Dirac delta function. Analysis of this equation shows that the response at very low levels of stimulation is entirely governed by the fluctuations: $\left\langle |W|\right\rangle \sim \sqrt{\varepsilon}$. Above a crossover level of stimulus intensity, $\left|\tilde{P}\right| > P_c$, the response then increases monotonically with the level of stimulation. The crossover intensity $P_c$ therefore represents the threshold at which the system begins to detect the external stimulus.

Above the crossover level of stimulation $P_c$ the response of a linear system increases linearly with the stimulation intensity: $\left\langle |W|\right\rangle \sim \left|\tilde{P}\right|$. For a nonlinear system, however, the situation can be more complicated. Assume that the system operates close to the Hopf bifurcation, so that $\zeta_r \approx \zeta_r^{(c)} = 0$, and is stimulated at the resonant frequency $\omega_*$ and at an intensity exceeding the crossover value $P_c$. As in the noiseless case, we then expect the system's response to increase with the cubic root of the stimulation intensity: $|\tilde{W}| \sim \sqrt[3]{|\tilde{P}|}$. Careful investigation reveals, however, that the nonlinear dependence can involve still smaller coefficients; a nonlinear relation $|\tilde{W}| \sim |\tilde{P}|^\alpha$ with $\alpha < 1/3$ is possible (Lindner et al 2009).

Noise also appears at the level of the basilar membrane, which undergoes spontaneous fluctuations of a magnitude near that of the response evoked by a threshold acoustic stimulus (Harris 1968, Nuttall et al 1997). Although this noise behaves in several regards like a bandpass-filtered acoustic response, it is unaffected by obstruction of the middle ear and therefore does not reflect transduction of external noise. Furthermore, the frequency of the measured noise, around 18 kHz, is so great that low-frequency cardiovascular and respiratory sources are unlikely to cause it. Because the strength of the noise is correlated with the sensitivity of a cochlea, it is apparently enhanced by the cochlear active process and thus reflects the activity of hair cells. The ultimate source of the noise, however, remains uncertain (Nuttall et al 1997).

## 5. Micromechanics of the organ of Corti

Resting upon the basilar membrane is the most mechanically complex structure in the body, the organ of Corti (figure 14(*a*)). First investigated by Alfonso Corti (1851), this sensory apparatus is a highly specialized epithelium, a tissue consisting of a layer of cells joined side-to-side in a continuous sheet. Tight junctions form molecular gaskets between the adjacent cells, allowing the epithelium to separate dissimilar liquids without mixing. Adherent junctions between specific cells provide strong mechanical connections that allow the propagation of forces from cell to cell. Nerve fibers enter the organ of Corti along its neural edge, the boundary nearest the center of the cochlea spiral. The opposite, abneural edge contacts a body ridge, the spiral ligament. The organ of Corti comprises at least ten morphologically distinct types of cell, many of which bear eponyms derived from their discoverers. Only four classes of cell are directly germane to an understanding of the transduction process, of which the most important are two varieties of hair cells.

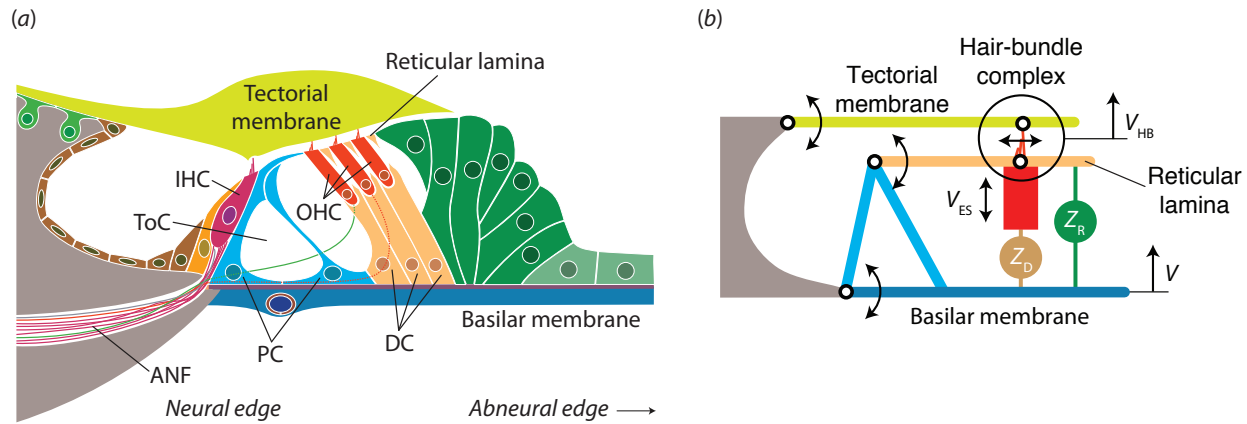

**Figure 14. Micromechanics of the organ of Corti.** (*a*) Two pillar cells (PC) form a stiff triangle on the basilar membrane; the liquid-filled interior space is referred to as tunnel of Corti (ToC). Three outer hair cells (OHC) in series with Deiters' cells (DC) connect the reticular lamina and basilar membrane. Their hair bundles are inserted into the tectorial membrane, whereas that of the inner hair cell (IHC) stands free. Hair cells of both types are innervated by auditory-nerve fibers (ANF). (*b*) In a model for the micromechanics of the organ of Corti, the two main degrees of freedom are the vertical vibration of the basilar membrane and the motion of the complex formed by the reticular lamina, the hair bundles of the outer hair cells, and the tectorial membrane. The components are defined in the text.

## 5.1. Architecture of the organ of Corti

The human organ of Corti is about 33 mm in length and 100-500 μm in width. Because this epithelial strip is broadly similar in structure throughout, one may conceptually divide the organ into transverse segments about 8 μm in length, each of which comprises a full complement of cell types (figure 14(*a*)). In each segment two stiff pillar cells form a solid triangle at the neural edge of the basilar membrane. An inner hair cell abuts the pillar cells on their neural side; on the abneural side stand three outer hair cells, each of which is supported at its base by a Deiters' cell. The apices of the outer hair cells contribute to a rigid, planar structure of actin filaments, the reticular lamina, that extends to the tops of the pillar cells. The tallest stereocilia in the hair bundle of each outer hair cell insert firmly into the overlying tectorial membrane. This acellular, collagenous gel extends in parallel to the reticular lamina from the spiral limbus to the outer hair cells. The tectorial membrane also overlies the hair bundles of the inner hair cells without coupling directly to them.

Inner and outer hair cells differ in several regards. About 95% of the afferent auditory-nerve fibers that transmit information into the brain contact the inner hair cells; only 5% innervate the outer hair cells (Spoendlin 1969). Efferent fibers that send feedback information from the brainstem to the cochlea, however, target predominantly the outer hair cells. The two cell types also exhibit distinct morphologies: inner hair cells have a flask-like shape and are around 30 $\mu$m long, whereas outer hair cells are cylindrical with lengths between 20 $\mu$m at the cochlear base and 80 $\mu$m at the apex.

The differences between the two types of hair cells reflect their distinct functions. Inner hair cells specialize in forwarding information derived from acoustic signals to the brain. Because ablation of outer hair cells significantly elevates the hearing threshold and reduces frequency selectivity, those cells have been recognized as the source of the active mechanical feedback that underlies cochlear amplification (Ryan and Dallos 1975, Kiang *et al* 1976, Dallos and Harris 1978, Harrison and Evans 1979). Consistent with this role, the hair bundles of outer hair cells can exert force and the cell bodies exhibit somatic motility (Kennedy *et al* 2005, 2006, Ashmore 2008). Efferent innervation provides a means for the brain to regulate the amount of amplification (Galambros 1956, Fex 1962, Wiederhold 1970, Mountain 1980, Murugasu and Russell, 1996), presumably in order to suppress noise and to enhance sensitivity to sounds of interest.

### 5.2. Eigenmodes of internal movement

The mechanics within the organ of Corti, commonly referred to as cochlear micromechanics, can be conceptualized through a simplified model with two degrees of freedom (figure 14(*b*); Zwislocki and Kletsky 1979, Allen 1980, Neely and Kim 1986, Mammano and Nobili 1993, Markin and Hudspeth 1995, Nobili and Mammano 1996). The first involves the up-and-down vibration of the basilar membrane at a velocity $V$. The second degree of freedom emerges because the hair bundles of outer hair cells connect the reticular lamina to the tectorial membrane. Owing to geometric constraints, upward or downward deflection of the reticular lamina and tectorial membrane elicits a proportionate shearing of the bundles. We denote by $V_{HB}$ the velocity of the complex formed by hair bundles, reticular lamina, and tectorial membrane. Although both degrees of freedom in principal involve rotatory movements, the angular excursions are so small—no more than a milliradian—that the motions may be regarded as rectilinear.

Two types of forcing each affect a single degree of freedom. First, sound stimulation evokes pressure changes in the three scalae. The pressure in the scala tympani produces a force on the basilar membrane, whereas the pressure in the scala media acts on the organ of Corti. If we assume that the organ is deformable but not compressible, then it transmits the pressure in the scala media to the basilar membrane. Sound stimulation therefore results in a force $F_S$ on the basilar membrane but not directly on the hair-bundle complex. In the following we shall consider stimulation at an angular frequency $\omega$, with $F_S = \tilde{F}_S e^{i\omega t} + c.c.$, and the response at the same frequency. As the second forcing, the hair bundle can produce an active force $F_{HB}$ that is exerted against the tectorial membrane but not directly on the basilar membrane.

If the two degrees of freedom were uncoupled, each would respond only to the force that directly acts on it. In this situation the basilar-membrane velocity $V$ would be related to the sound-induced force $F_S$ through its impedance $Z$, $Z\tilde{V} = \tilde{F}_S$. The velocity $V_{HB}$ of the complex formed by the reticular lamina, hair bundles, and tectorial membrane would follow as $Z_{HB}\tilde{V}_{HB} = \tilde{F}_{HB}$, in which $Z_{HB}$ is the impedance of the hair-bundle complex.

In the organ of Corti the two degrees of freedom are actually coupled. Sound-induced forces can accordingly vibrate the hair bundles, and the forces evoked in the bundles can feed back onto the basilar membrane. Length changes of the outer hair cells can modify the coupling and provide additional force.

The details of these processes govern the micromechanics of the organ of Corti and can produce a variety of intriguing behaviors.

Coupling between the two degrees of freedom partly involves active feedback through the outer hair cells that, in series with Deiters' cells, connect the basilar membrane to the reticular lamina. Let $V_{EM}$ be the velocity at which the outer hair cells change length and $Z_D$ be the impedance of the underlying Deiters' cells. Because the length change occurs against this impedance, it implies an electromotile force $\tilde{F}_{EM} = Z_D \tilde{V}_{EM}$. This force acts with an equal magnitude, but in the opposite direction, on the reticular lamina and basilar membrane. The remainder of the organ of Corti provides additional coupling characterized by an impedance $Z_R$. The velocities of the two degrees of freedom, $V$ and $V_{HB}$, follow from the three forces $F_S$, $F_{EM}$, and $F_{HB}$ through

$$\Gamma \left( \begin{array}{c} \tilde{V}_{HB} \\ \tilde{V} \end{array} \right) = \left( \begin{array}{c} \tilde{F}_{HB} + \tilde{F}_{EM} \\ \tilde{F}_S - \tilde{F}_{EM} \end{array} \right), \tag{5.1}$$

in which the 2x2 matrix $\Gamma$ contains the coupling terms

$$\Gamma = \left( \begin{array}{cc} Z_{HB} + Z_D + Z_R & -Z_D - Z_R \\ -Z_D - Z_R & Z + Z_D + Z_R \end{array} \right). \tag{5.2}$$

The micromechanics of the organ of Corti can be analyzed conveniently by considering the two vibrational eigenmodes associated with the two eigenvectors of the matrix $\Gamma$. For ease of presentation, we consider the simplified situation in which the impedances of the basilar membrane and hair-bundle complex are equal, $Z = Z_{HB}$. The two eigenvalues $\zeta_{(1)}$ and $\zeta_{(2)}$ and the corresponding eigenvectors $e_{(1)}$ and $e_{(2)}$ of the matrix $\Gamma$ are then

$$\zeta_{(1)} = Z, \qquad e_{(1)} = \left( \begin{array}{c} 1 \\ 1 \end{array} \right);$$

$$\zeta_{(2)} = Z + 2Z_D + 2Z_R, \quad e_{(2)} = \left( \begin{array}{c} -1 \\ 1 \end{array} \right). \tag{5.3}$$

The first eigenmode is an in-phase vibration of the hair-bundle complex and the basilar membrane. Because it involves no relative motion between the two degrees of freedom, this motion is not impeded by their coupling, but rather by their intrinsic impedances $Z$. The resonant frequency of this mode therefore accords with that of the isolated basilar membrane or hair-bundle complex.

In the second eigenmode, the two degrees of freedom vibrate in antiphase: their velocities are equal in magnitude but oriented in opposite directions. The resonant frequency of this mode is in general higher than that of the in-phase motion, for the coupling impedances $Z_D$ and $Z_R$ are dominated by stiffness and viscosity but involve negligible inertial contributions.

Because the electromotile force acts on the basilar membrane and the hair-bundle complex in opposite directions, it can excite the antiphase mode of motion but not the in-phase one. This important feature becomes apparent mathematically when we express the motion through the eigenvectors: $\left( \tilde{V}_{HB}, \tilde{V}_{BM} \right)^T = V_{(1)} e_{(1)} + V_{(2)} e_{(2)}$ with the coefficients $\tilde{V}_{(1)} = (\tilde{V}_{BM} + \tilde{V}_{HB})/2$ and $\tilde{V}_{(2)} = (\tilde{V}_{BM} - \tilde{V}_{HB})/2$. Equation (5.1) then becomes

$$\left( \begin{array}{cc} Z & 0 \\ 0 & Z + 2Z_D + 2Z_R \end{array} \right) \left( \begin{array}{c} \tilde{V}_{(1)} \\ \tilde{V}_{(2)} \end{array} \right) = \frac{1}{2} \left( \begin{array}{c} \tilde{F}_S + \tilde{F}_{HB} \\ \tilde{F}_S - \tilde{F}_{HB} - 2\tilde{F}_{EM} \end{array} \right). \tag{5.4}$$

Only the second eigenmode experiences the electromotile force, whereas the forces evoked by sound and by the hair bundles act on both types of vibration. In agreement with this analysis, experiments on an apical segment of the cochlear partition show that the tectorial membrane and basilar membrane move in opposite directions when outer hair cells are stimulated electrically in the absence of a pressure stimulus (Mammano and Ashmore, 1993).

Somatic motility and hair-bundle activity affect the motion of the organ of Corti in distinct ways. Because the sound-evoked force acts only on the basilar membrane, it excites both in-phase and antiphase motions. Somatic motility can then amplify the antiphase component, thus altering the ratio between the two components and hence the type of motion. Although the hair-bundle force acts on both vibrational modes, it does so with opposite effects: either it excites in-phase motion and suppresses antiphase motion, or *vice versa*. Hair-bundle activity, then, alters the way in which the organ of Corti moves. A specific combination of hair-bundle and electromotile forces can produce the same mode of motion as a sound force. In the above example, this situation arises when the electromotile force is equal but opposite the hair-bundle force, $\tilde{F}_{EM} = -\tilde{F}_{HB}$.

Recent experimental techniques and results are reviewed in subsection 5.4 below, but details of the movements in the living organ of Corti remain unclear. The respective roles and importances of hair-bundle activity and somatic motility thus remain uncertain as well. The micromechanics of the organ of Corti is further complicated by the dependences of the electromotile and hair-bundle forces on hair-bundle displacement. The electromotile force $F_{EM}$ is proportional to the electrically evoked length change of the outer hair cell, $\tilde{F}_{EM} = Z_D \tilde{V}_{EM}$. For small signals the alteration in length is proportional to the change in membrane potential, which in turn depends linearly on hair-bundle displacement. We may therefore write $\tilde{V}_{EM} = -\alpha \tilde{V}_{HB}$ with a coefficient $\alpha$. The minus sign arises because deflection of a hair bundle in the positive, excitatory direction depolarizes the cell and shortens it. Because the membrane potential is regulated by ion channels that can provide delay and feedback, however, the coefficient $\alpha$ varies with frequency and is generally complex. The mechanical activity of the hair bundle is likewise evoked by bundle displacement (subsection 4.6). In theoretical models, interactions between the two mechanical activities can produce a variety of complex behaviors (Ó Maoiléidigh and Jülicher 2010, Reichenbach and Hudspeth 2010b, Meaud and Grosh 2011). As a distinct feature set out in the proceeding subsection, somatic motility not only can amplify a type of motion, but also can influence the coupling between the basilar membrane and the hair-bundle complex.

*5.3. Reciprocity breaking and unidirectional amplification*

Consider an amplifier in which a signal acts at the input to elicit a response at the output (figure 15(*a*)). In the case of mechanical coupling, Maxwell's reciprocity theorem states that a force applied at the input produces a displacement at the output that equals the displacement at the input caused by an equal force at the output. In other words, the forward coupling equals the backward coupling. In the situation of the inner ear, the basilar membrane represents the input, for it experiences a sound-evoked force. The output corresponds to the displacement of the hair bundle. The system's mechanical reciprocity is then manifest in equations (5.1) and (5.2): in the matrix $\Gamma$, the two off-diagonal terms that describe the coupling from the basilar membrane to the hair bundle and back agree.

Reciprocity can be adverse to an amplifier's operation. When amplification produces nonlinear behavior and thus distortion (subsection 3.2), backward coupling has the undesirable feature of perturbing the input signal. An ideal amplifier operates unidirectionally: although the forward coupling must remain, the backward coupling from the output to the input should be eliminated (figure 15(*b*)).

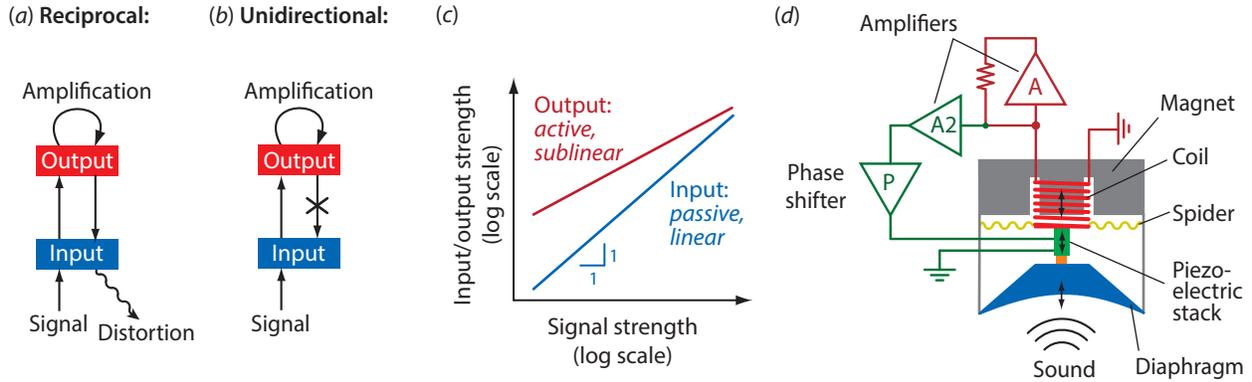

**Figure 15. Unidirectional amplification**. (*a*) A naïve amplifier functions through reciprocal coupling: the output feeds back to the input with the same strength as the input couples to the output. When the output is amplified, it can produce distortion that is then emitted backward to the input. (*b*) Active feedback can reduce the backward coupling to zero, such that coupling occurs unidirectionally from the input to the output. No distortion then arises at the input. (*c*) When amplification is unidirectional, the output response can be enhanced and nonlinear, whereas the input response remains unchanged and linear. (*d*) We have implemented unidirectional amplification in an active microphone described in the text.

Active feedback such as that resulting from somatic motility can violate the reciprocity of a passive system and provide unidirectional coupling (Reichenbach and Hudspeth 2010a, 2011). To demonstrate this effect, we start from the electromotile force $\tilde{F}_{EM} = Z_D \tilde{V}_{EM}$ in which the velocity $V_{EM}$ is proportional to the hair-bundle displacement through a complex coefficient $\alpha$: $\tilde{V}_{EM} = -\alpha \tilde{V}_{HB}$ (subsection 5.2). The velocities of the two degrees of freedom, $V$ and $V_{HB}$, now follow from the sound-evoked force, $F_S$, and the hair-bundle force, $F_{HB}$, through

$$\Lambda \begin{pmatrix} \tilde{V}_{HB} \\ \tilde{V} \end{pmatrix} = \begin{pmatrix} \tilde{F}_{HB} \\ \tilde{F}_S \end{pmatrix}. \tag{5.5}$$

The matrix $\Lambda$ contains the coupling terms as well as the active feedback,

$$\Lambda = \begin{pmatrix} Z_{HB} + (1+\alpha)Z_D + Z_R & -Z_D - Z_R \\ -(1+\alpha)Z_D - Z_R & Z + Z_D + Z_R \end{pmatrix}. \tag{5.6}$$

The off-diagonal term $-Z_D - Z_R$ is unchanged from equation (5.2); it encodes the forward coupling from the sound force that acts from the basilar membrane to the hair bundle. The other coupling term $-(1+\alpha)Z_D - Z_R$ that describes the reverse coupling of hair-bundle forces onto the basilar membrane, however, is altered through the active feedback. In consequence, the forward and backward coupling between the basilar membrane and the hair-bundle complex differ. As an extreme case of this nonreciprocity, the backward coupling can vanish at a certain strength and timing of the feedback, represented by a critical value $\alpha_* = -1 - Z_R / Z_D$ (figure 15(*b*)).

Unidirectional coupling might allow efficient and undistorted amplification in the cochlea. Although a sound-evoked force would be transmitted to the hair bundles, the force evoked there would not feed back to the basilar membrane. Hair-bundle activity would therefore have to counteract only the viscous damping associated with its own motion, but not that involving the basilar membrane. The distortion emerging from nonlinear hair-bundle dynamics would not be emitted from the basilar membrane and would accordingly not contaminate the sound signal. Furthermore, the hair bundles would exhibit a resonance that is not influenced by the material properties of the basilar membrane. As set forth in subsection 5.4 below, this mechanism might be important in the apical, low-frequency region of the cochlea.

Characteristic nonlinear behavior emerges when amplification is unidirectional. Active feedback can cancel the viscous part in the output impedance so that the output's response approaches a Hopf bifurcation and becomes nonlinear at its resonance (figure 15($c$)). Because the active feedback does not couple to the input, the input impedance remains unchanged and the response linear. For low signal strength, the output response exceeds the input signal by a large amount. Measurement of this behavior in the cochlea would provide an unequivocal proof of unidirectional amplification. The hair-bundle motion would be amplified and display a compressive nonlinearity, whereas the basilar membrane would respond linearly to sound intensity, with only a passive vibration. Some measurements from the cochlear apex indeed point in this direction (subsection 5.4).

To demonstrate that unidirectional coupling can be useful in engineering as well, we constructed an active microphone that employs this principle (Reichenbach and Hudspeth 2011). We started from a dynamic-coil microphone in which a diaphragm is attached to a coil that moves in a magnetic field (figure 15($d$), Eagle 1997). Sound vibrates the diaphragm and thereby causes oscillation of the coil that electromagnetically induces a voltage. We then positioned a piezoelectric stack between the coil and diaphragm. The voltage driving this element was controlled by the voltage in the coil but adjusted in magnitude and phase such that it provided unidirectional coupling between the diaphragm and coil. Amplifying the coil's voltage and feeding it back amplified the coil's motion. Because of unidirectional coupling the diaphragm was not affected by the coil's vibration and any distortion produced within the coil was not emitted from the microphone. We thus obtained a nondistorting mechanical sensor that was ultrasensitive near a specific resonant frequency.

The device also serves as an illustration for the putative micromechanics of the organ of Corti at the cochlear apex: because it collects the sound force, the diaphragm corresponds to the basilar membrane. The mechanical signal is converted to an electrical response within the coil, which is accordingly the hair bundle's analogue. The piezoelectric element imitates the electromotile behavior of the outer hair cells.

### 5.4. Measurements from the cochlear base and apex

Laser interferometry readily measures basilar-membrane displacements only a few nanometers in amplitude. A beam of coherent, monochromatic light is split into two beams, one of which is reflected from a sample whose vibrations are to be investigated. The other beam, which serves as a reference, is shifted in frequency by a small amount $\Delta f$ by reflection from an oscillating mirror. When the beams are subsequently recombined, the frequency difference produces a temporal oscillation between constructive and destructive interference and hence beating at the frequency $\Delta f$. The phase of the beating reflects the phase shift between the two beams that emerges from the difference in path length. Vibration of the structure in question thus produces an oscillating phase change whose magnitude encodes the amplitude of vibration. In scanning laser interferometry an experimenter observes the vibration at different positions on the sample; their relative phases can evidence propagating waves.

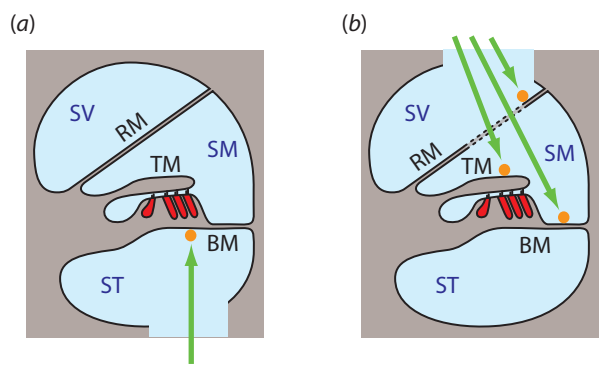

**Figure 16. Laser-interferometric measurements from the cochlea *in vivo*.** (*a*) The base of the cochlea can be accessed from the scala tympani (ST) through either the round window or a hole drilled in the temporal bone. Interferometric vibrometry employs a laser (green) focused on the basilar membrane (BM). The reflectance of the latter may be enhanced by a glass or plastic bead (orange). (*b*) Near the cochlear apex, a hole may be drilled to access the scala vestibuli (SV) and scala media (SM). Interferometric measurements can then be obtained from the basilar membrane (BM) and the tectorial membrane (TM) either by monitoring through the transparent Reissner's membrane (RM) or by opening the latter. The vibrations of Reissner's membrane itself can be measured from attached beads.

Applying laser interferometry to measure the vibration of the organ of Corti requires optical access. Because the cochlea is encased in the temporal bone, the hardest of the body, and because it is wound in a spiral shape, such access is not obtained easily. This technical difficulty is compounded by the fact that the cochlea's active process is compromised by even modest surgical damage. Most of the successful measurements have been obtained from the cochlear base, with only a few from the apex. The intermediate turn or turns remain inaccessible.

The base of the cochlea can be approached either through the transparent round window or by drilling a hole through the cochlear wall. In both cases the basilar membrane is viewed from the scala tympani (figure 16(*a*)). Measurements from a single point on the basilar membrane during stimulation at different frequencies have demonstrated critical-layer absorption (subsection 2.5) as well as a compressive nonlinearity in the basilar membrane's response (subsection 3.2). Scanning along the midline of the basilar membrane has also confirmed the existence and verified the properties of traveling waves (Ren 2002, Ren *et al* 2011, Fisher *et al* 2012). Interestingly, the local wavelength at a given cochlear position changes negligibly between an active and a passive preparation. In Section 6 we show that the local wavelength follows from the imaginary part of the cochlear partition's impedance. The active process accordingly has little influence on the impedance's imaginary part. Scanning across the basilar membrane has shown that each radial segment of the membrane moves approximately in phase (Rhode and Recio 2000, Ren *et al* 2011, Fisher *et al* 2012). This finding is important in that some theoretical models had predicted that the electromotile force would yield a bimodal radial deformation of the basilar membrane (Mammano and Nobili 1993, Nobili and Mammano 1996).

At the cochlear base, the basilar membrane blocks the view of the organ of Corti and precludes measurements of the micromechanics with an ordinary interferometer. Recently, however, researchers have applied optical-coherence tomography to the inner ear (Chen *et al* 2007, Wang and Nuttall 2010). As opposed to the coherent, monochromatic light of a conventional laser interferometer, optical-coherence tomography employs light of multiple wavelengths for which coherence is limited to a very short distance, typically a few micrometers. Interference between the sample beam and the reference beam of an interferometer then occurs only when two path lengths differ by less than this coherence length. In this way one can record from a specific depth within a semitransparent organ. When this technique is applied to the organ of Corti, both the basilar membrane and the reticular lamina reflect enough of the incoming light to permit interferometric measurement. Such recordings from animals *in vivo* have shown that, at the low sound-pressure levels for which the active process provides the greatest amplification, the reticular lamina vibrates twice as much as the basilar membrane and its vibrations lead those of the basilar membrane by 60° (Chen *et al* 2011, Zha *et al* 2012). The vibrations of both structures are much smaller *post mortem* and have the same amplitude and phase. The result from this passive regime accords with the theoretical considerations from subsection 5.2, in which a sound-evoked force near the resonance defined by the impedance $Z$ evokes primarily in-phase motion of the basilar membrane and hair-bundle complex. The measurement from the living organ of Corti disagrees with the antiphase motion anticipated naïvely when somatic motility is the principal force of amplification.

Measurement of displacement alone cannot fully characterize cochlear mechanics or micromechanics, which requires simultaneous determination of the conjugate variable force. The two quantities together define a system's impedance as well as its vibrational energy. To measure force in the cochlea, Elizabeth Olson has developed a miniature pressure sensor that consists of a hollow-core glass fiber with a thin diaphragm across its tip (Olson 1998, 1999). Acoustic pressure deflects the diaphragm, yielding a measurable optical signal. Applied near the cochlear base, this apparatus quantifies the pressure wave as well as its decay as a function of distance from the basilar membrane. Simultaneous interferometric recording of the basilar-membrane velocity yields the impedance of the cochlear partition (Dong and Olson 2009).

Because the apical region of the cochlea has been investigated less, even fundamental aspects of its mechanics remain controversial (Robles and Ruggero 2001). Apical recordings are made through an artificial opening into the scala vestibuli (figure 16(*b*)). In this way, the micromechanics of the organ of Corti is accessible, for the motions of both the basilar and tectorial membranes can be measured in a living animal. Early recordings found that the tectorial membrane vibrates up to 900-fold as much as the basilar membrane (figure 17(*a*), ITER 1989, Khanna and Hao 1999a, 1999b). Although distortion hinted at nonlinear behavior, the displacements at the fundamental frequencies scaled linearly with the applied sound-pressure level. Subsequent studies, however, found that the tectorial and basilar membranes undergo comparable displacements with a partly linear, partly compressive, or partly expansive nonlinearity (Cooper and Rhode 1995, Rhode and Cooper 1996, Zinn *et al* 2000).

Despite these controversies, the apical measurements agree in that they do not find the strongly compressive nonlinearity that characterizes the mechanics in the basal region. The data further evidence a broad tuning without the sharp decay on the resonance's high-frequency side that emerges near the base (figure 17(*a*)). The latter result agrees with observations regarding the tuning of individual auditory-nerve axons. The fibers that emerge from the basal region of the inner ear are tuned sharply, with the high-frequency cutoff that characterizes critical-layer absorption (figure 4(*d*) and figure 17(*b*)). The fibers from the cochlear apex, however, display broader tuning and lack a sharp threshold increase at high frequencies. The phenomenon of critical-layer absorption evidently does not operate in the apical portion of the cochlea, and the mechanism of frequency tuning remains elusive there.

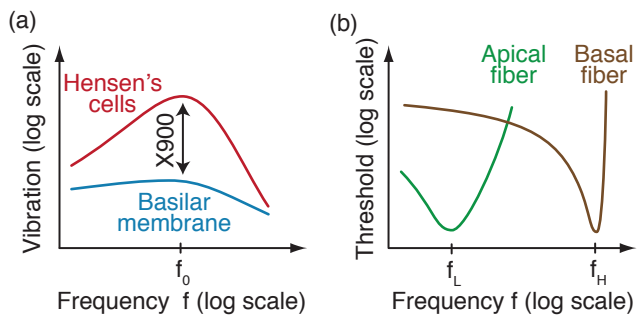

**Figure 17. Schematic diagrams of observations from the cochlear apex.** (*a*) Some interferometric experiments have found that the motion of Hensen's cells in the organ of Corti, which presumably resembles that of the tectorial membrane and the hair bundles, exceeds the motion of the basilar membrane by orders of magnitude near the characteristic frequency $f_0$. The tuning is broad and lacks a sharp high-frequency cutoff. (*b*) The tuning curve of an auditory-nerve fiber from the apex (green, with a low characteristic frequency $f_L < 1$ kHz) differs significantly from that of a basal fiber (brown, with a high characteristic frequency $f_H > 5$ kHz). As implied by the mechanical tuning curve, tuning at low frequencies is much broader.

We have proposed that unidirectional amplification provides frequency selectivity in the low-frequency region. Although the basilar membrane presumably does not resonate at frequencies below about 1 kHz (Naidu and Mountain 1998), unidirectional coupling allows an independent resonance of the hair-bundle complex to amplify its own motion. The vibration of the complex would then greatly exceed that of the basilar membrane, as has been observed in some recordings (figure 17(*a*); ITER 1989, Khanna and Hao 1999a, 1999b).

If the mechanics of the cochlea differs at its basal and apical extremes, what occurs at intermediate positions? Recordings from auditory-nerve fibers in chinchilla reveal a transition between critical-layer absorption near the base and apical mechanics in two steps, one at 4-5 kHz and another near 2 kHz (Temchin *et al* 2008a). Direct measurements of cochlear mechanics from the corresponding middle region of the cochlea, however, seem infeasible at present. *In vitro* preparations of the cochlea in which a cochlear segment is isolated and the endocochlear potential is reestablished through externally applied current might elucidate this issue (Fridberger and Boutet de Monvel 2003, Chan and Hudspeth 2005, Jacob *et al* 2011).

## 6. Active waves on the basilar membrane

Local amplification through hair-cell activity yields only a moderate gain. As an example, individual hair bundles enhance their motion by about tenfold (Martin and Hudspeth 2001). Even though measurements *in vitro* probably underestimate the full capacity of a single cell, the gain *in vivo* is likely of the same order of magnitude. This contrasts with the huge gain observed in the cochlea: increases of membrane vibration by a factor of 1,000 or more have been measured in an intact animal preparation (subsection 3.2). Can local hair-cell activity produce such large cochlear gains?

Mechanical coupling can increase the effectiveness of a cluster of hair cells. The gain of a single hair bundle, as well as that of the cochlea, is largest for low intensities of stimulation. At low signal levels, however, noise becomes important. Below a certain intensity of stimulation, noise obstructs the hair bundle's mechanotransduction and thus limits its amplification. Mechanical coupling of multiple stochastic oscillators restricts the relative motion between them, thereby effectively reducing the noise (Chang *et al* 1997). Theoretical studies have shown that mechanical coupling of hair bundles can indeed lower the noise floor and thereby increase the gain (Dierkes *et al* 2008, 2012). As an example, a nine-by-nine array of hair bundles can theoretically boost the gain of each bundle by more than tenfold. An actual hair cell whose hair bundle had been virtually coupled to model hair bundles displayed an enhanced gain as well (Barral *et al* 2010).

Another type of coupling arises through the cochlear fluids. This hydrodynamic coupling produces on the basilar membrane the surface wave that we have described in subsection 2.2. The propagation of such a wave through a region of local amplification can greatly increase the wave's gain. In the following we describe this gain accumulation together with the physics of active cochlear waves.

### 6.1. Accumulation of gain

A key result of the analysis of surface waves on the basilar membrane of a passive cochlea is equation (2.11), which assumes that the velocity $\tilde{V}$ of the basilar membrane is linearly proportional to the pressure across it, $\tilde{p}^{(2)} - \tilde{p}^{(1)}\big|_{z=0}$, with the frequency-dependent impedance $Z$ of the basilar membrane as the proportionality constant. The micromechanics of the organ of Corti can alter the impedance of the basilar membrane. As set out in subsection 5.2, the organ of Corti may be described through two degrees of freedom, the velocity $V$ of the basilar membrane and the velocity $V_{HB}$ of the hair-bundle complex. Both depend on the sound-induced force $F_S$, the hair-bundle force $F_{HB}$, and the electromotile force $F_{EM}$ through different coupling impedances (equations (5.1) and (5.2)). The force derived from acoustic stimulation follows from the pressure difference across the area $A$ of a basilar-membrane segment: $F_S = A(p^{(2)} - p^{(1)})$. If we assume that the electromotile and hair-bundle forces are proportional to hair-bundle velocity, equation (5.1) yields a linear relation of basilar-membrane velocity to the pressure difference:

$$\tilde{V} = \frac{1}{Z_{act}}\left(\tilde{p}^{(2)} - \tilde{p}^{(1)}\right)\Big|_{z=0} . \tag{6.1}$$

Here $Z_{act}$ is the active impedance that follows from the micromechanics and the hair-cell activity of the organ of Corti.

The active basilar-membrane impedance $Z_{act}$ in general differs from the passive impedance $Z$. How does this influence the propagating wave? As set forth in subsection 2.4, the WKB approximation employs the ansatz

$$\tilde{V} = \hat{V}(x)\exp\left[-i\int_0^x dx'\,k(x')\right] \tag{6.2}$$

for the basilar-membrane velocity. The local wave vector is $k(x) = \pm\sqrt{-2i\omega\rho_0/[Z_{act}(x)h]}$ and the amplitude factor is $\hat{V}(x) \sim 1/\sqrt[4]{Z_{act}^3(x)}$. The velocity's amplitude $|\tilde{V}|$ thus follows both from the magnitude of the amplitude factor $\hat{V}$ and from the imaginary part of the wave vector $k$:

$$\left|\tilde{V}\right| = \left|\hat{V}(x)\right|\exp\left[\int_0^x dx'\,\mathrm{Im}[k(x')]\right] . \tag{6.3}$$

Except close to the resonant position, the amplitude factor $\hat{V}$ is not significantly altered by active feedback. Indeed, the amplitude factor follows from the impedance that is dominated by stiffness up to the resonant position. Mass and stiffness in turn determine the real part of the wave vector $k$ and hence the local wavelength. Because interferometric recordings from the basilar membrane (subsection 5.4) measure little change in the local wavelength between an active and a passive cochlea, the active and the passive impedance approximately agree in their imaginary parts. The same conclusion follows from the wave's behavior near its resonant position that marks the sharp decay of the wave's amplitude: this location remains unchanged between an active and a passive inner ear (figure 7($a$)). Hair-cell activity does not appear to significantly alter the effective mass and stiffness of the cochlear partition, and therefore does not change the amplitude factor $\hat{V}$.

Active force can, however, increase the traveling wave's amplitude through the wave vector's imaginary part that reflects drag. In the instance of a forward-traveling wave, for which the wave vector's real part is positive, a positive drag coefficient in the local impedance leads to a negative imaginary part in the wave vector and hence to a decrease in amplitude. Reduction by the active process of the drag coefficient—potentially even rendering it negative—counteracts this decay in amplitude.

Changes of the local damping brought about by active forces can accumulate. The phase factor in equation (6.2) contains the integral over the local wave vector, such that a reduction of drag over an extended cochlear region produces a cumulative effect on the amplitude. We have recently demonstrated that the active process, although locally of moderate effect, can through this mechanism produce a much larger overall increase in the wave's amplitude (figure 18($a$); Reichenbach and Hudspeth 2010b). Specifically, we considered a reduction of the local damping through the active process by a moderate factor of ten. Without a propagating wave, a segment of the cochlea near its resonant frequency would exhibit a velocity enhanced by only one order of magnitude (equation (3.1)). By traversing an extended region over which the damping is reduced, however, the wave can grow in amplitude by several orders of magnitude.

We recently performed experiments in which we disabled the cochlear amplifier at specific cochlear locations (Fisher *et al* 2012). We inhibited somatic motility with 4-azidosalicylate, a small carboxylic acid that binds reversibly to prestin and disables its functioning. Upon exposure to ultraviolet light the binding becomes irreversible. We exposed the cochlea to 4-azidosalicylate, irradiated a specific location with ultraviolet light, and then removed the free drug. Somatic motility was thus impaired only within the irradiated cochlear location and functioned normally elsewhere.

When scanning longitudinally along the midline of the basilar membrane and measuring its displacement, we observed that the growth in amplitude was diminished in the region of inhibited somatic motility (figure 18($b$)). Reconstructing the membrane impedance revealed that the local gain was indeed eliminated. However, the global gain—that is, the difference in amplitude between the living and dead cochlea—remained substantial. In confirmation of our theoretical expectations, this residual gain resulted from the accumulation of local gain on both sides of the irradiated region.

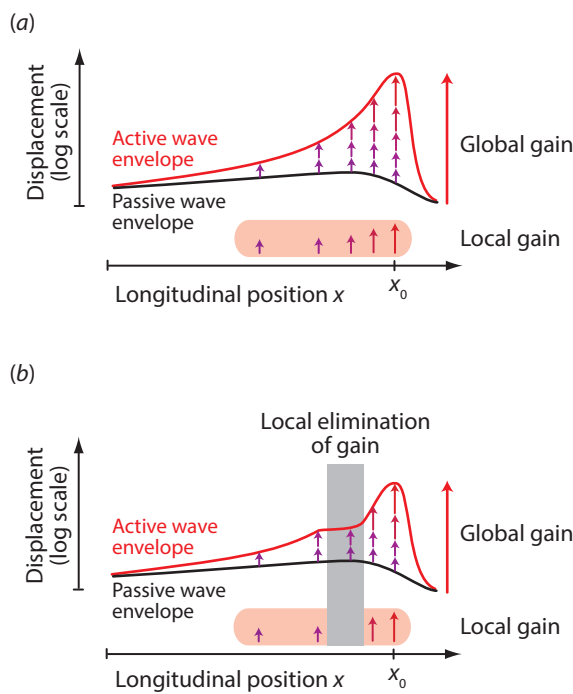

**Figure 18. Schematic diagrams illustrating the accumulation of local gain.** (*a*) A wave that traverses a region of amplification can accumulate local gain to achieve a greatly enhanced global gain. (*b*) When the local gain is eliminated by chemical treatment within a small cochlear region (grey), the global gain is somewhat less than that in a normal cochlea. The global gain is nonetheless significant as a result of the accumulation of local gain over the untreated segments of the basilar membrane.

### *6.2. Allen-Fahey experiment*

The relevant quantity in the above analysis of gain accumulation was the *difference* in drag between an active and a passive cochlear partition. What do we know, however, about the *absolute* value of the membrane's drag coefficient? Can the active process change this from a positive to a negative value? In the latter case, instead of losing energy due to viscous forces, the wave would gain energy when propagating across a region of negative drag.

In 1992 Jont Allen and Paul Fahey conducted a remarkable experiment to measure energy gain in the cochlea. They used two distinct ways to record the distortion frequency $f_d = 2f_1 - f_2$ that was produced by stimulating the cochlea simultaneously at two close frequencies $f_1$ and $f_2$ (subsection 7.1). First, they measured the response of an auditory-nerve fiber that originated from the cochlear position at which the wave produced by the distortion frequency peaked. Second, with a microphone in the ear canal they recorded the airborne vibration at the distortion frequency.

The two recorded signals arose through distinct waves in the cochlea (figure 19). The distortion product was generated in the overlap region of the two primary frequencies $f_1$ and $f_2$ and hence near the peak of the wave elicited by the higher frequency, $f_2$. The lower sideband $2f_1 - f_2$ was smaller than both primary frequencies, such that its characteristic place lay apical to those of $f_1$ and $f_2$. The signal recorded from the nerve fiber at that position thus emerged as the result of a wave at the distortion frequency that traveled apically from its generation site. The microphone in the ear canal, however, detected a signal that must have emerged through a wave that propagated basally. When the two primary frequencies were widely separated, only the forward-traveling wave would experience gain (figure 19). For close primary frequencies, in contrast, the forward-traveling wave would be subject to only a small amount of gain, and the backward-traveling wave might encounter gain as well.

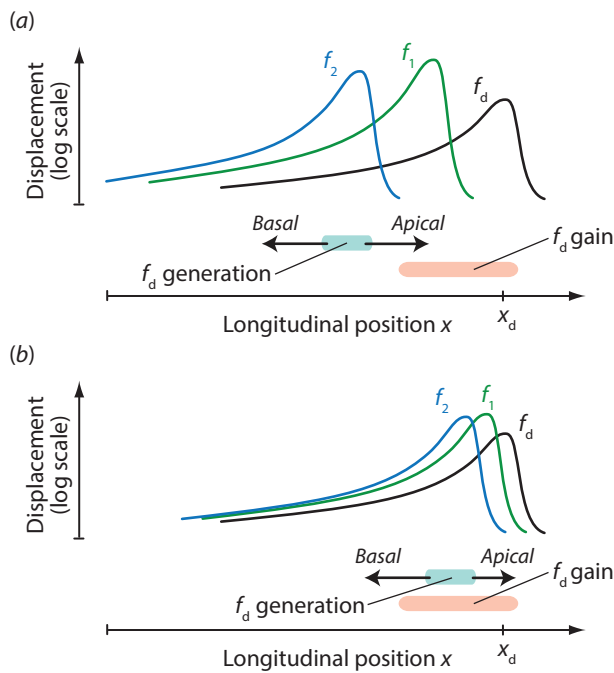

**Figure 19. Schematic representation of an experiment to analyze absolute cochlear amplification.** (*a*) Two primary frequencies $f_1$ and $f_2$ generate the distortion product $f_d = 2f_1 - f_2$ in their region of overlap on the basilar membrane. The wave elicited by the distortion product experiences gain near its characteristic place $x_d$; recording a neuron's activity from that position yields a measure of the wave's peak amplitude. When the primary frequencies are sufficiently distant, the generation site of the distortion frequency lies basal to, and outside, its gain region. (*b*) Nearby primary frequencies produce a distortion product within the region of gain.

Comparing the two circumstances—distant and close primary frequencies—accordingly yields insight into the energy gain: if the cochlea were to raise the wave energy, the ratio of the amplitude of the forward-traveling to that of the backward-traveling wave would increase with a growing distance between the primary frequencies. Allen and Fahey adjusted the primary frequencies and their levels such that the distortion frequency as well as its signal strength at the peak position, as manifested in the firing rate of the corresponding auditory-nerve fiber, was identical in both situations. Surprisingly, they found that the distortion level measured in the ear canal did not depend on the ratio of the primary frequencies. This result suggests that the gain region provides neither net negative nor net positive drag. Because we expect viscous forces at the interface of the fluid and the membrane to dampen wave propagation in a passive cochlea, we conclude that the gain region must provide amplification that almost exactly cancels the effect of this drag.

The experiment by Allen and Fahey, which has been confirmed by various subsequent studies (de Boer *et al* 2005, Shera and Guinan 2007), indicates that the active process tunes the cochlear partition to a critical point, namely to the transition between positive and negative damping. The discussion in subsection 3.2 reveals that this transition represents a bifurcation. Because of its resonance, each segment of the cochlear partition appears to be poised at a supercritical Hopf bifurcation (Duke and Jülicher 2003, Kern and Stoop 2003, Magnasco 2003).

### 6.3. The nonlinear traveling wave

So far we have considered linear wave propagation in the cochlea. The linear approximation seemed justified inasmuch as the amplitudes of the sound-evoked motions are extremely tiny. In subsection 3.2, however, we have described how each segment of the basilar membrane is poised near a Hopf bifurcation, such that it responds nonlinearly to forcing at its characteristic frequency. The Allen-Fahey experiment reviewed above enforced this hypothesis. How, then, does the nonlinear response of the basilar membrane shape the propagating wave?

Consider a nonlinear relation between the velocity $V$ of the basilar membrane and the pressure difference $p^{(2)} - p^{(1)}$ across it:

$$\left( \tilde{p}^{(2)} - \tilde{p}^{(1)} \right)\Big|_{z=0} = Z\tilde{V} + \Theta\tilde{V} * \tilde{V} * \tilde{V} . \tag{6.4}$$

This equation expands the previous linear relationship, equation (2.11), with a cubic nonlinear term weighted by a coefficient $\Theta$. In fact, the right-hand side may be construed as a Taylor expansion of the pressure difference in the velocity's Fourier coefficient $\tilde{V}$. The coefficient of the linear term is the membrane impedance $Z$ (equation (2.19)), whose imaginary part vanishes at the resonant frequency. The real part encodes damping, which may also vanish if the viscous drag is counteracted by the active process. Nonlinear terms then become important, and the leading order is quadratic. As discussed in subsection 3.2, however, a suitable change of variables can eliminate the quadratic term so that the subsequent cubic term governs the response. In the case of the inner ear, the quadratic term is tiny even for the untransformed variables. Indeed, the dominant nonlinearity presumably results from the hair bundle's nonlinear stiffness (equation (4.9)). Because outer hair cells operate near an open probability of 50% (Patuzzi and Rajan 1990, Kirk et al 1997, Bobbin and Salt 2005), which represents a point of inflection in the relation of channel open probability and hair-bundle force to displacement, they do not exhibit a quadratic but only a cubic nonlinearity around their resting position.

The nonlinear basilar-membrane response (equation (6.4)) now replaces the linear equation (2.11). For ease of presentation, we describe the resulting effects within the one-dimensional approximation set out in subsection 3.2. There we had derived equation (2.25) for the fluid's volume flow in the upper cochlear chamber, which can be rewritten using the basilar-membrane velocity $V$ as

$$\partial_x \tilde{j}^{(1)} = \tilde{V} - i\omega h\kappa \tilde{p}^{(1)} . \tag{6.5}$$

A change in volume flow follows from a membrane displacement or from compression of the fluid.

Because the volume flow also follows from the longitudinal pressure change, $i\omega\rho_0\tilde{j} = -h\partial_x\tilde{p}$, we obtain an equation for the pressure and the basilar-membrane velocity alone:

$$\frac{ih}{\omega\rho_0}\partial_x^2\tilde{p}^{(1)} = \tilde{V} - i\omega h\kappa\tilde{p}^{(1)} . \tag{6.6}$$

An analogous relation holds for the lower chamber:

$$\frac{ih}{\omega\rho_0}\partial_x^2\tilde{p}^{(2)} = -\tilde{V} - i\omega h\kappa\tilde{p}^{(2)} . \tag{6.7}$$

As in subsection 2.3, the average pressure $\Pi = (p^{(1)} + p^{(2)})/2$ propagates as a fast sound wave that does not displace the basilar membrane:

$$\partial_x^2 \tilde{\Pi} = -\omega^2 \rho_0 \kappa \tilde{\Pi} \ . \tag{6.8}$$

However, the pressure difference between the upper and the lower chambers, $p = p^{(2)} - p^{(1)}$, does produce a displacement of the basilar membrane:

$$\partial_x^2 \tilde{p} = \frac{2i\omega\rho_0}{h} \tilde{V} - \omega^2 \rho_0 \kappa \tilde{p} \ . \tag{6.9}$$

To solve this equation we must relate the pressure difference $\tilde{p}$ to the basilar membrane's velocity of vibration $\tilde{V}$ through the nonlinear equation (6.4). We obtain a wave equation for the velocity,

$$\partial_x^2 \tilde{V} - \frac{2i\omega\rho_0}{Zh} \tilde{V} = -\frac{\Theta}{Z} \partial_x^2 (\tilde{V} * \tilde{V} * \tilde{V}) \ . \tag{6.10}$$

This relation corresponds to the original, linear equation (2.30), now amended by a cubic nonlinearity.

Numerical solution of the nonlinear wave equation (6.10) shows that the nonlinearity is confined to a relatively small region near the peak of the traveling wave (Duke and Jülicher 2003, Kern and Stoop 2003, Magnasco 2003). The peak displacement in response to signal intensity is strongly nonlinear. The exponent of the nonlinearity generally differs from the value of 1/3 associated with a Hopf bifurcation, for the region of nonlinearity depends on the stimulus intensity. The region is larger for higher levels of stimulation, which slightly increases the exponent of the nonlinearity at the peak (Duke and Jülicher 2003).

## 6.4. Micromechanics of the organ of Corti and wave propagation

The active process that amplifies the traveling wave evidently requires motion within the organ of Corti. The organ is typically modeled in mechanical terms by accounting for the impedances of the basilar membrane, hair bundles, and their coupling (Section 5). However, such a description is valid only when the organ of Corti deforms so as to leave its cross-sectional area unchanged (figure 20). If this condition is not met, alterations in the cross-sectional area must cause longitudinal fluid flows that raise the possibility of hydrodynamical coupling along the organ of Corti.

An electromotile length change of outer hair cells deforms the organ of Corti. Near the apex, this deformation may leave the cross-sectional area constant (figure 20($a$)). The 80 µm-long apical outer hair cells do not stand perpendicular to the basilar membrane, but tilt at an angle of about 45° to insert perpendicularly into the angled reticular lamina. A large liquid-filled space occurs between the outer hair cells and the Hensen's cells, which form an arch reinforced by intracellular polymer cables. This space is the most affected by length changes of outer hair cells, which raises the question whether its cross-sectional area is thereby altered.

We may approximate a cross-section of the liquid-filled space between the outer hair cells and Hensen's cells as a semicircle, with the outer hair cells as its linear edge and the Hensen's cells on its arc. A length change of the outer hair cells leaves the length of the arc constant and hence changes the semicircle into a semi-ellipsoid. Because a circle maximizes the ratio of area to circumference, its area remains constant to first order upon deformation into an ellipsoid of the same circumference. The same applies to the semicircle when transformed into a semi-ellipsoid. The fluid space between the outer hair cells and the Hensen's cells thus retains an essentially constant cross-sectional area.

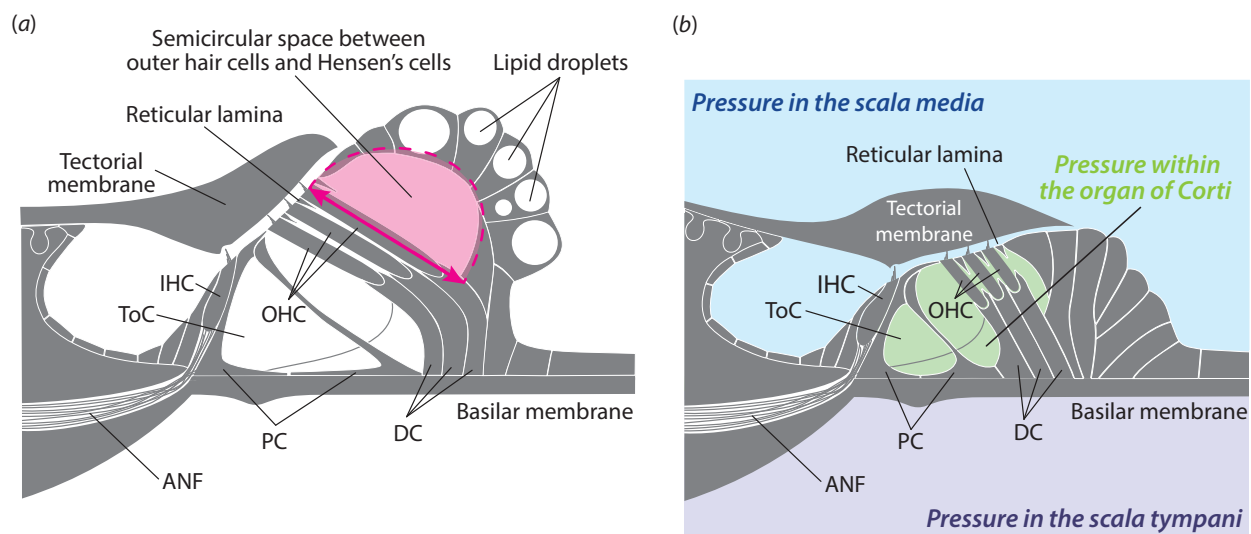

**Figure 20. Liquid-filled spaces within the organ of Corti.** (*a*) Length changes of outer hair cells near the apex predominantly deform the space between them and the arc of Hensen's cells (red shading and red dotted line). Because this space is shaped as a semicircle, its cross-sectional area remains unchanged for small deformations. The Hensen's cells contain large lipid droplets whose function remains unknown. (*b*) In the basal region of the cochlea, length changes of outer hair cells presumably alter the cross-sectional area of the organ of Corti. In addition to the pressures in the scala media and the scala tympani, the pressure within the organ of Corti may then shape wave propagation. Drawings modified from Held (1902); abbreviations as in figure 14.

The organ of Corti in the cochlea's basal region behaves differently (figure 20(*b*)). The reticular lamina there lies parallel to the basilar membrane and the Hensen's cells do not form an arc. Length changes of outer hair cells must alter the distance between the reticular lamina and the basilar membrane and therefore the cross-sectional area of the organ of Corti.

Electrical stimulation of an excised segment from the middle cochlear turn has demonstrated fluid flow in the tunnel of Corti (Karavitaki and Mountain 2007). Microscopic observation revealed longitudinal deflections of the efferent nerve fibers that cross the tunnel of Corti; these movements presumably reflected longitudinal fluid flow in the tunnel. A finite-element simulation showed that longitudinal flow can result when electromotile length changes of the outer hair cells alter the cross-sectional area of the organ of Corti: liquid may be exchanged easily between the tunnel of Corti and the region around the hair cells, for the pillar cells are separated by gaps and do not represent a significant barrier (Zagadou and Mountain 2012).

Because of the fluid environment, internal motion between the reticular lamina and the basilar membrane can give rise to novel types of waves. A model in which the tectorial membrane experiences the pressure in the scala media and the basilar membrane experiences that in the scala tympani exhibits two types of waves, both of which involve internal motion of the organ of Corti (Lamb and Chadwick 2011, 2014). Such distinct wave modes may play a critical role in cochlear amplification, for the cochlear wave appears to transition from one mode to the other upon reaching its peak position. At the cochlear base, the wave elicits vibrations of the basilar membrane and reticular lamina that are in phase and have equal magnitudes (Chen *et al* 2011, Zha *et al* 2012). Near the resonant position, however, the two structures vibrate at different phases and amplitudes. This behavior might reflect a transition between two wave modes as shown in an electrical analogue of the cochlea (Hubbard 1993, Hubbard *et al* 1993). Future investigations of active wave modes may account for the fluid flow within the organ of Corti as well as the longitudinal mechanical coupling within the tectorial membrane that might underlie shear waves (Ghaffari *et al* 2007, Gu *et al* 2008).

## 7. Otoacoustic emissions

Otoacoustic emissions are sounds produced by the inner ear as a result of its active process (subsection 3.2). Because these signals can be measured with a sensitive microphone placed in the ear canal, the cochlea's mechanical activity must in some way propagate back to the middle ear and excite an airborne sound wave. How this reverse transmission occurs within the cochlea, however, remains surprisingly little understood. The issue is further complicated because the backward propagation appears to employ at least two distinct pathways, both of which are currently debated.

### 7.1. Nonlinear combination tones

An important class of otoacoustic emissions includes nonlinear combination tones. When stimulated at multiple nearby frequencies, the ear emits sound at additional frequencies that are linear combinations of the primary tones. As an example, in response to stimulation at two close angular frequencies $\omega_1$ and $\omega_2$ the cochlea produces sound signals at frequencies such as $\omega_1 + \omega_2$ and $2\omega_1 - \omega_2$. These additional tones are termed nonlinear combination tones, for they arise from the inner ear's nonlinearity. Indeed, as set out in subsection 3.2, each transverse segment of the cochlear partition responds nonlinearly to forcing near its resonant frequency. A quadratic nonlinearity $Y^2$ in the membrane displacement $Y$, for example, produces the quadratic distortion frequency $\omega_1 + \omega_2$. To see this, consider the Fourier transformation of a quadratic nonlinearity, equations (3.11) and (3.12). The Fourier component $\tilde{Y}(\omega_1 + \omega_2)$ of the membrane displacement at the angular distortion frequency $\omega_1 + \omega_2$ is then proportional to the product of the Fourier components at the primaries, $\tilde{Y}(\omega_1 + \omega_2) \sim \tilde{Y}(\omega_1)\tilde{Y}(\omega_2)$. It follows that the phase of the emission equals the sum of the primary phases plus a constant:

$$\varphi(\omega_1 + \omega_2) = \varphi(\omega_1) + \varphi(\omega_2) + \text{constant} , \tag{7.1}$$

in which $\varphi(\omega)$ denotes the phase of the Fourier coefficient $\tilde{Y}(\omega)$. The phase combination

$$\hat{\varphi}(\omega_1 + \omega_2) = \varphi(\omega_1 + \omega_2) - \varphi(\omega_1) - \varphi(\omega_2) \tag{7.2}$$

accordingly does not depend on frequency. Although the literature on otoacoustic emission often treats this term as the phase of a distortion product, in the following we refer to it as the *normalized phase*. A similar computation for the case of a cubic nonlinearity reveals that the normalized phase for the cubic emission $2\omega_1 - \omega_2$ reads

$$\hat{\varphi}(2\omega_1 - \omega_2) = \varphi(2\omega_1 - \omega_2) - 2\varphi(\omega_1) + \varphi(\omega_2) . \tag{7.3}$$

Is the normalized phase of an emission as measured in the ear canal also constant? The answer to this question provides clues to how signals propagate into and out of the cochlea. Indeed, equation (7.1) holds only for the phases at the cochlear position where the distortion is generated. The phases of all three signals—the two primaries as well as the distortion product—are different in the ear canal (Figure 21). The signals at the primary frequencies $\omega_1$ and $\omega_2$ progress into the cochlea as waves on the basilar membrane that yield phase delays $\Delta\varphi(\omega_1)$ and $\Delta\varphi(\omega_2)$ at the site where the distortion product is generated. A signal at the distortion frequency $2\omega_1 - \omega_2$ then propagates from that site back to the stapes, which induces an additional phase delay $\Delta\varphi(2\omega_1 - \omega_2)$. The phases $\varphi_{ec}(\omega_1)$, $\varphi_{ec}(\omega_2)$, and $\varphi_{ec}(2\omega_1 + \omega_2)$ in the ear canal then follow as $\varphi_{ec}(\omega_1) = \varphi(\omega_1) - \Delta\varphi(\omega_1)$, $\varphi_{ec}(\omega_2) = \varphi(\omega_2) - \Delta\varphi(\omega_2)$, and $\varphi_{ec}(2\omega_1 - \omega_2) = \varphi(2\omega_1 - \omega_2) + \Delta\varphi(2\omega_1 - \omega_2)$. We obtain the normalized ear-canal phase $\hat{\varphi}_{ec}(2\omega_1 + \omega_2)$ that is determined by the travel times of the waves:

$$\hat{\varphi}_{ec}(2\omega_1 - \omega_2) = \Delta\varphi(2\omega_1 - \omega_2) + 2\Delta\varphi(\omega_1) - \Delta\varphi(\omega_2) + \text{constant} . \tag{7.4}$$

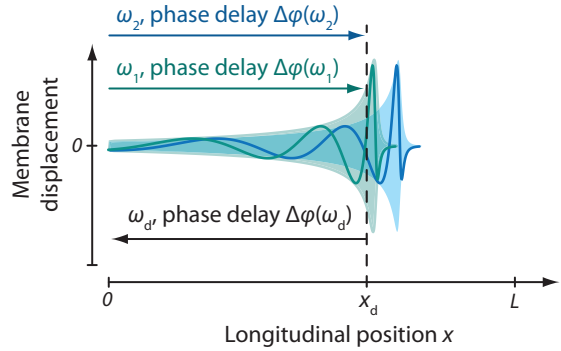

**Figure 21. Schematic representation of the generation and propagation of nonlinear combination tones within the cochlea.** The primary frequencies $\omega_1$ and $\omega_2$ travel into the cochlea as basilar-membrane waves that cover about two cycles from the stapes to the region $x_d$ where the distortion frequency $\omega_d$ is produced. The signal at that frequency then propagates back, experiencing an additional phase delay.

A basilar-membrane wave encompasses about two cycles from the cochlear base to the resonant position (Ulfendahl 1997, Robles and Ruggero 2001). Because a distortion product is generated close to the peaks of the primary waves, the corresponding phase delays $\Delta\varphi(\omega_1)$ and $\Delta\varphi(\omega_2)$ approximate two cycles as well, and in particular are independent of the frequencies. The normalized emission phase in the ear canal thus depends on the frequency only through the delay of the backward propagation:

$$\hat{\varphi}_{ec}(2\omega_1 - \omega_2) = \Delta\varphi(2\omega_1 - \omega_2) + \text{constant} . \tag{7.5}$$

If the distortion product were to propagate backward as a basilar-membrane wave, it would accumulate another two cycles of phase delay. The normalized phase of the distortion signal in the ear canal would then be independent of frequency. How does this expectation compare to measurements?

### 7.2. Two components of otoacoustic emissions

Experimental investigation of the phase of otoacoustic emissions often employs two close frequencies $\omega_1$ and $\omega_2$ and their cubic distortion products $2\omega_1 - \omega_2$ and $2\omega_2 - \omega_1$, for these signals occur at relatively high amplitudes. The frequency dependence of these nonlinear combination tones is typically measured by changing the primary frequencies across a few octaves. The frequency ratio $\omega_2 / \omega_1$ is held constant at values between 1.1 and 1.3 that produce significant overlap of the two traveling waves and therefore strong distortion. Because noise of a physiological origin increases at low frequencies, signals below 500 Hz are difficult to measure and most observations employ frequencies between 1 kHz and 10 kHz (Hall 2000).

Measurements from humans as well as other mammals show that the lower sideband $2\omega_1 - \omega_2$ has a normalized phase that remains approximately constant as a function of frequency. In contrast, the normalized phase of the upper sideband $2\omega_2 - \omega_1$ changes by many cycles—ten or more—when the primary frequencies are swept across a few octaves (figure 22(a)).

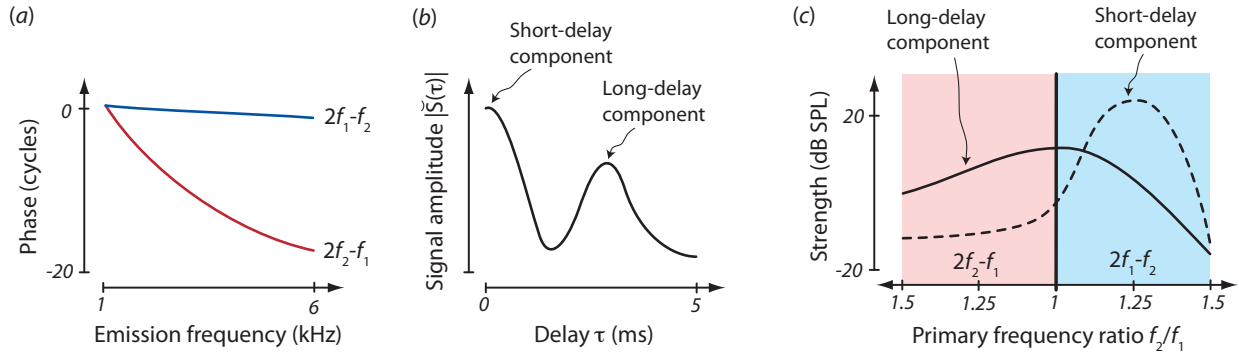

**Figure 22. Schematic illustrations of the two components of distortion-product otoacoustic emission.** (*a*) The normalized phase of the upper-sideband emission at $2f_2 - f_1$ changes by many cycles as the primary frequencies are swept across a few octaves. The normalized phase of the lower sideband $2f_1 - f_2$ remains almost constant. (*b*) A second Fourier transformation $\bar{S}(\tau)$ of the distortion frequency's Fourier coefficient reveals two peaks, one at 0 ms, the short-delay component, and another around 3 ms, the long-delay component. (*c*) The amplitudes of the two components depend on the ratio $f_2 / f_1$ of the primary frequencies and vary between the upper and the lower sideband.

More detailed investigation shows that both the upper and the lower sidebands consist of two components, one whose normalized phase varies with frequency and another whose phase does not (Knight and Kemp, 2000, 2001). Consider a distortion at frequency $\omega_d$ and the corresponding Fourier coefficient $\tilde{S}(\omega_d)$ of the acoustic signal in the ear canal. Denote by $\left|\tilde{S}(\omega_d)\right|$ the amplitude and by $\hat{\phi}(\omega_d)$ the normalized phase of the complex Fourier coefficient. We then consider the normalized signal $\hat{S}(\omega_d) = \left|\tilde{S}(\omega_d)\right| e^{i\hat{\phi}(\omega_d)}$ that has the same amplitude as the original Fourier coefficient but adopts the normalized phase. We can now compute a second Fourier transform, $\bar{S}(\tau)$, with respect to the distortion frequency $\omega_d$:

$$\bar{S}(\tau) = \int d\omega_d \hat{S}(\omega_d) e^{i\tau\omega_d} \ . \tag{7.6}$$

The pseudo-time $\tau$ that appears here is not an actual duration, for the Fourier coefficients $\tilde{S}(\omega_d)$ do not represent all the different frequencies of one recording, but only the signal at the distortion frequency $\omega_d$ acquired when varying the primary frequencies. What, then, does the pseudo-time $\tau$ signify?

The normalized signal $\hat{S}(\omega_d)$ can be obtained from $\bar{S}(\tau)$ through inverse Fourier transformation:

$$\hat{S}(\omega_d) = \frac{1}{2\pi} \int d\tau \bar{S}(\tau) e^{-i\tau\omega_d} \ . \tag{7.7}$$

The contribution at a certain pseudo-time $\tau_0$ has a normalized phase $\hat{\phi}_0(\omega_d) = -\tau_0 \omega_d$ that changes linearly with the distortion frequency. The pseudo-time $\tau_0$ is therefore the group delay, $\tau_0 = -d\hat{\phi}_0(\omega_d) / d\omega_d$, that reveals the rapidity of the phase change.

When the second Fourier transformation (7.6) is performed on data acquired from human ears, only two main components emerge (figure 22($b$)). The first is centered around a delay $\tau = 0$: the normalized phase of this component does not depend on the frequencies of stimulation. For the second component, however, the delay does not vanish, and the corresponding normalized phase varies substantially with the distortion frequency $\omega_d$.

The amplitudes of the two components depend on the ratio $\omega_2 / \omega_1$ of the primary frequencies as well as on the emission frequency (figure 21($c$)). For the upper-sideband emission $2\omega_2 - \omega_1$ the long-delay component exceeds the short-delay component by about 20 dB. The opposite scenario emerges for the lower-sideband signal $2\omega_1 - \omega_2$, for which the short-delay component dominates by about 20 dB. The amplitude of the long-delay component varies little between the upper and lower sidebands. Instead, the difference between the two sidebands arises primarily through the short-delay component that is weak for the upper sideband but much stronger for the lower one.

### 7.3. Green's function and Born approximation

What is the significance of the two components, one with a short and one with a long delay? This question remains debated, and we describe different hypotheses below. For their understanding we must quantify wave generation inside the cochlea through the active process. To this end we employ the methods of Green's functions and perturbation theory that have been developed for quantum field theory. We remain, of course, within classical mechanics; quantum effects do not appear.

For ease of presentation, we consider the one-dimensional approximations to cochlear mechanics set out in subsection 2.3, for the linear case, and in subsection 6.3, for the nonlinear scenario. The nonlinear wave equation (6.10) that we wish to solve contains a cubic nonlinearity on its right-hand side and linear terms on the left-hand side. As a first step in its solution we compute a Green's function, that is, a basilar-membrane velocity $\tilde{V}^{(G,x_0)}$ that satisfies

$$\partial_x^2 \tilde{V}^{(G,x_0)} - \frac{2i\omega\rho_0}{Zh} \tilde{V}^{(G,x_0)} = \delta(x - x_0) \, . \tag{7.8}$$

The cubic nonlinearity of the wave equation (6.10) has been replaced through Dirac's delta function $\delta(x - x_0)$ centered at a cochlear location $x_0$. The Green's function $\tilde{V}^{(G,x_0)}$ accordingly describes the membrane response to periodic forcing at an angular frequency $\omega$ and at a single cochlear location $x_0$.

The Green's function can be computed with the ansatz

$$\tilde{V}^{(G,x_0)} = \int_{-\infty}^{\infty} dk\, g(k) e^{-ik(x-x_0)} \tag{7.9}$$

in which the coefficient $g(k)$ is complex. Because the delta distribution can be represented as

$$\delta(x - x_d) = \frac{1}{2\pi} \int_{-\infty}^{\infty} dk\, e^{-ik(x-x_0)} \, , \tag{7.10}$$

the wave equation (7.8) with forcing at $x_0$ yields the coefficient

$$g(k) = \frac{1}{2\pi L(k)} \, . \tag{7.11}$$

Here we have used the shorthand notation $L(k) = [k^2 + 2i\omega\rho_0 / (Zh)]$. Note that $L(k) = 0$ is the dispersion relation, equation (2.33).

The integral in equation (7.9) can be solved by closing the contour of integration in the complex plane (Figure 23). The residue theorem of complex analysis informs us that only the poles of the integrand inside a closed path contribute to such a path integral. In our situation, poles occur at those values $k_0$ for which $L(k)$ vanishes, and hence at the wave vectors $\pm k_0$ that satisfy the dispersion relation (2.33): $k_0 = \sqrt{-2i\omega\rho_0/(Zh)}$ . If the basilar-membrane impedance $Z$ involves friction, the solutions $\pm k_0$ possess small imaginary parts and lie in the second and fourth quadrants of the complex plane. For a cochlear location basal to the applied force, $x < x_0$ , the amplitude of $e^{-ik(x-x_0)}$ tends to zero in the upper half-plane except when the imaginary part of $k$ vanishes. We can therefore close the integration contour in the upper half-plane. We obtain a contribution from $-k_0$ that describes a retrograde wave traveling from the location of the force back to the cochlear base:

$$\tilde{V}^{(G,x_0)} = -\frac{i}{2k_0}e^{ik_0(x-x_0)} \text{ for } x < x_0 . \tag{7.12}$$

In the opposite case, for a location apical to the forcing, $x > x_0$ , the contour can be closed in the lower half-plane. It then yields a contribution from $k_0$ and thus a wave that travels forward to the apex:

$$\tilde{V}^{(G,x_0)} = -\frac{i}{2k_0}e^{-ik_0(x-x_0)} \text{ for } x > x_0 . \tag{7.13}$$

In the context of quantum theory the corresponding Green's function is known as the Feynman propagator (Peskin and Schroeder 1995).

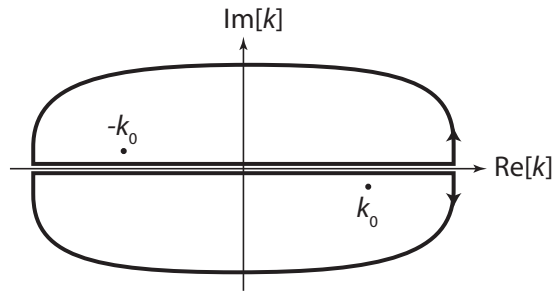

**Figure 23. Integration in the complex plane.** The poles $\pm k_0$ of $L(k)$ , which represent the possible wave vectors, lie in the second and fourth quadrants of the complex plane. Depending on the sign of $x - x_0$ , the contour of integration can be closed in the upper or lower half-plane, yielding a contribution from the pole in respectively the second or the fourth quadrant.

An inhomogeneous wave equation

$$\partial_x^2 \tilde{V}(x) - \frac{2i\omega\rho_0}{Zh}\tilde{V}(x) = \Psi(x) \tag{7.14}$$

with an inhomogeneity $\Psi(x)$ on the right-hand side can now be solved by integrating over the Green's function multiplied by $\Psi(x_0)$ :

$$\tilde{V}(x) = \int dx_0 \tilde{V}^{(G,x_0)}(x)\Psi(x_0) . \tag{7.15}$$

The nonlinear wave equation (6.10) has an inhomogeneity $\Psi(x) = -\frac{\Theta}{Z}\partial_x^2(\tilde{V} * \tilde{V} * \tilde{V})(x)$ and its solution accordingly reads

$$\tilde{V}(x) = -\int dx_0 \frac{\Theta}{Z}\tilde{V}^{(G,x_0)}(x)\partial_{x_0}^2(\tilde{V} * \tilde{V} * \tilde{V})(x_0) . \tag{7.16}$$

Because the inhomogeneity now depends on the velocity, however, the velocity appears not only on the equation's left-hand side but also in the integrand on the right-hand side. The equation accordingly cannot be solved in a simple manner. If we assume, however, that the nonlinearity yields only a small correction $\tilde{V}_1$ to the velocity $\tilde{V}_0$ that emerges in a passive, linear cochlea, we may approximate the nonlinear term through the linear solution $\tilde{V}_0$:

$$\tilde{V}_1(x) = -\int dx_0 \frac{\Theta}{Z} \tilde{V}^{(G,x_0)}(x) \partial_{x_0}^2 (\tilde{V}_0 * \tilde{V}_0 * \tilde{V}_0)(x_0). \tag{7.17}$$

The full solution is then $\tilde{V} = \tilde{V}_0 + \tilde{V}_1$. This approximation was introduced by Max Born in the context of quantum mechanics (Sakurai 1994). There the solution $\tilde{V}_0$ to the linear, homogeneous wave equation represents a free particle that does not interact with others; interactions introduce nonlinear terms and yield scattering of a wave. If the interaction strength is small, such as for the electromagnetic interaction, one can employ the above perturbative approach to compute the first correction $\tilde{V}_1$ to the free solution. Iterating this procedure yields higher-order corrections until the perturbation series converges to the full solution. For our purpose, however, the first perturbation term (7.17) already offers insight into the generation and propagation of otoacoustic emissions.

The Green's function (7.12) and (7.13) together with the Born approximation (7.17) describe the generation and propagation of distortion products. An analysis of these equations confirms that, as set out in subsection 7.1, a distortion such as $2\omega_1 - \omega_2$ is generated in the overlap region of the traveling waves elicited by the two primary frequencies $\omega_1$ and $\omega_2$. A basilar-membrane wave at the distortion frequency then travels back to the stapes and produces a signal in the ear canal. Because of the scaling invariance of the traveling wave, the phase of this emission is approximately independent of frequency. This pathway may accordingly produce the short-delay component of an otoacoustic emission.

Although theory suggests that the short-delay component of distortion-product otoacoustic emissions emerges through reverse basilar-membrane waves, measurements have not detected such signals (Ren 2004, Hea et al 2007, He et al 2008). It remains unclear how else this component of otoacoustic emissions might emerge from the cochlea. Additional types of waves, for example those involving deformation of the cochlear bone, might be involved (Tchumatchenko and Reichenbach 2013). Such waves could also underlie bone conduction, the perception of sound through vibration of the skull that does not require a functional middle ear (Tonndorf 1976).

### 7.4. Waves on Reissner's membrane

How does the long-delay component of a distortion-product otoacoustic emission arise? One hypothesis is that a forward-propagating distortion wave is reflected from inhomogeneities along the basilar membrane (Zweig and Shera 1995, Shera and Guinan 1999, Kalluri and Shera 2001). If only a single inhomogeneity were involved, the phase of the emission would depend on the phase of the wave at the cochlear position of that inhomogeneity. Because the wave's phase at such a fixed cochlear location varies with frequency, the phase of the resulting emission should depend on the frequency as well.

Many inhomogeneities are required, however, for reflection to operate across a range of distortion frequencies. When a distortion wave is reflected by multiple scatterers, it is no longer clear that its phase changes systematically with frequency. Indeed, if a large number of inhomogeneities near the peak of the wave were to cause reflections, the phase of the resulting wave should be roughly independent of the frequency, for the irregularities in the location of the scatterers would approximately average out. Modeling suggests that, under certain circumstances, systematic phase changes may nevertheless result (Zweig and Shera 1995, Kalluri and Shera 2001). Experimental investigations of the matter are difficult because the nature of the hypothesized reflectors remains unclear.

We have recently proposed an alternative means by which the phase-changing component of otoacoustic emissions could arise (Reichenbach et al 2012). This mechanism involves waves on Reissner's membrane, which runs parallel to the basilar membrane from the cochlear base to the apex. Reissner's membrane is often ignored when describing cochlear hydrodynamics, for its compliance, at least near the base, greatly exceeds that of the basilar membrane. It has therefore been assumed that Reissner's membrane follows the fluid motion elicited by the basilar-membrane wave.

Reissner's membrane can sustain its own kind of surface waves. To show this, we have analyzed a two-dimensional cochlear model that describes the inner ear's vertical as well as its longitudinal extent (Figure 24). The pressures within the three chambers delineated by the two membranes constitute three degrees of freedom, and three types of waves can accordingly arise.

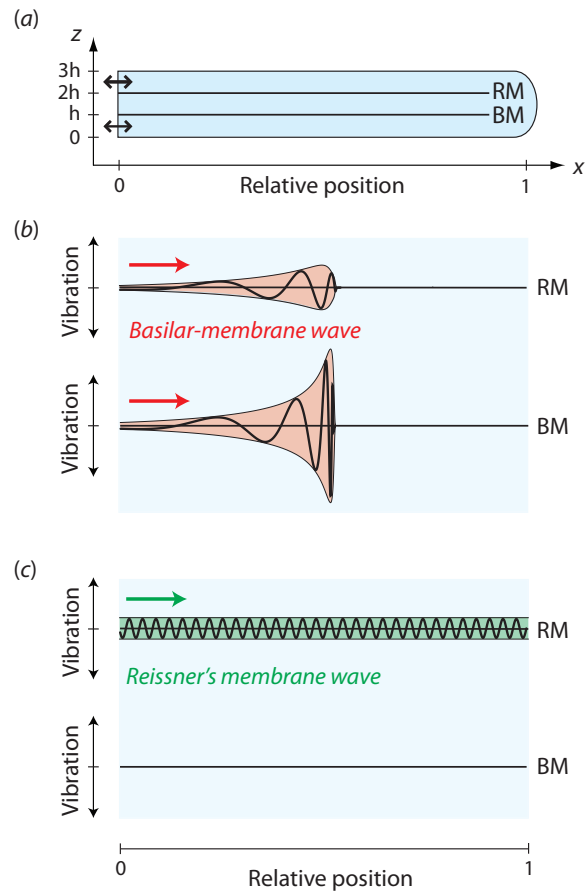

**Figure 24. Schematic diagrams of waves on the parallel basilar and Reissner's membranes.** (*a*) In a two-dimensional model of the inner ear, three chambers of height *h* are delineated by Reissner's membrane (RM) and the basilar membrane (BM). (*b*) The basilar-membrane wave elicits comparable motion of both membranes except in the peak region where the wavelength becomes shorter and fluid coupling between the membranes weakens. (*c*) A short-wavelength wave can propagate exclusively on Reissner's membrane.

First, a fast sound wave can propagate along the cochlea through fluid compression and expansion. For this wave the pressures in all three chambers coincide at each longitudinal position, such that neither membrane is displaced. The wave is therefore not of physiological importance.

Second, for frequencies above a few kilohertz, the well-known traveling wave on the basilar membrane emerges. With the exception of its peak region, its wavelength exceeds the height of the chambers. The fluid velocity accordingly does not vary much with vertical distance (Figure 2), so Reissner's membrane and the basilar membrane move by comparable amounts. Because Reissner's membrane is much floppier, however, the propagation of the wave is dominated by the material properties of the basilar membrane.

Third, a novel surface wave can travel on Reissner's membrane. Because of the membrane's compliance, the length of this wave is much less than the channel's height, at least for frequencies above a few kilohertz. A wave on Reissner's membrane therefore penetrates to only a small degree into the fluid surrounding the membrane and does not elicit motion of the basilar membrane. Because this wave cannot excite hair cells, it appears to be of little importance for the physiological functioning of the inner ear. It can, however, transport a distortion product from its generation site back to the cochlear base. Indeed, the fluid dynamics near the basilar membrane at the distortion frequency does elicit a disturbance that propagates as a wave on Reissner's membrane, both to the cochlear apex and to the base. Laser interferometry in preparations of the chinchilla's cochlea *in vivo* demonstrate that basilar-membrane distortion yields Reissner's membrane waves (Reichenbach et al 2012). Their propagation has been quantified using the approach outlined above, namely Green's functions and the Born approximation, and the dispersion relation has been confirmed by laser-interferometric measurements in several rodent species.

The waves on Reissner's membrane do not exhibit the scaling symmetry of waves on the basilar membrane. Because a wave on Reissner's membrane has a comparatively short wavelength, a distortion product that propagates back to the stapes by this means accumulates a phase delay of many cycles, depending on the frequency. Our numerical computations show that this phase delay matches that of the long-delay component of distortion-product otoacoustic emissions.

For frequencies below a few kilohertz, a traveling wave peaks near the cochlear apex. The stiffness of the basilar membrane there falls to a value near that of Reissner's membrane, which implies that a basilar-membrane wave is shaped by the material properties of both structures. Moreover, the wavelength of the erstwhile Reissner's membrane wave exceeds the height of the chambers, so that disturbance couples to motion of the basilar membrane. The emerging waves can therefore no longer be separated into distinct basilar-membrane and Reissner's membrane waves. Instead, the two wave modes involve the mechanical properties and motion of both membranes. The characteristics of these waves and how they influence apical cochlear mechanics remain open questions.

## 8. Conclusions

The compact, complex, and delicate structure of the cochlea and its location in the hard bone of the skull present formidable challenges to the investigation of auditory transduction. During the past two decades, however, these problems have yielded to progressive technical advances, especially the systematic investigation of hair cells *in vitro* and the improvement of interferometric measurements *in vivo*. Contemporary data are of a quality that permits rigorous comparison of experimental data with theoretical models, which in turn have increased in sophistication and accuracy.

From these efforts has emerged our current picture of the active cochlea. Through biophysical specializations at the molecular, cellular, and organ level, this receptor organ shapes its inputs even as it transduces them into electrical signals. The inner ear as a whole employs sophisticated hydrodynamic effects to spatially separate the different frequency components of a sound. The mechanosensitive hair cells can amplify the vibrational amplitude in a frequency band and boost it by several orders of magnitude, mechanical activities that arise from ion channels and molecular motors within the cells. By operating on the verge of dynamical instability, the cochlea's active process implements the ear's profound amplification, its remarkable frequency selectivity, and its incredible dynamic range.

Although many aspects of cochlear mechanics and hair-cell operation are now well understood, several important issues remain debated. The first is the molecular workings of the mechanotransduction apparatus. Identifying the molecular structure of the mechanotransduction channel, the nature of the gating spring, and its relationship to the tip link will be key steps. Next, several issues of optimality would be worth exploring. How do the number and lengths of the stereocilia in a hair bundle affect its mechanical sensitivity and frequency selectivity? Where does a hair bundle reside *in vivo* with respect to the Hopf bifurcation? Does this operating point correspond to an optimum in the signal-to-noise ratio, the quality factor of tuning, or the sensitivity to threshold stimuli? Third, the implementation of the active process within the organ of Corti, as potentially shaped by a contribution from both hair-bundle forces and electromotile length changes, remains to be clarified. We have outlined some of the potentially important steps toward resolving this issue, including nonlinear mathematical analysis and novel experimental techniques such as optical-coherence tomography. These approaches may reveal distinct versions of the active process in the basal and apical regions of the cochlea. Fourth, the origin and transmission modes of otoacoustic emission remain debated. These issues are important, for otoacoustic emissions are a key signature of the cochlear active process and their behavior can confirm the healthy functioning of the inner ear. Progress will likely originate in refined laser-interferometric measurements coupled with computational modeling of cochlear fluid dynamics.

The ear's astonishing performance can inspire technology, and engineering principles can in turn instruct us about the ear's operation. The initial proposition by Gold that the ear employs active feedback was motivated by the usage of positive feedback in contemporary radio receivers. The concept of a Hopf bifurcation in cochlear mechanics has conversely led to the construction of a bionic ear (van der Vyer 2006). We have shown how a proposed mechanism of unidirectional amplification for apical cochlear mechanics can be applied to construct an ultra-sensitive, non-distorting microphone. Because natural selection and engineering must solve similar problems, we expect that they will continue to mutually inform each other and hence our understanding of the physics of hearing.

## Acknowledgments

The authors thank Dr. D. Ó Maoiléidigh for figure 13 and thank him, Dr. A. Jacobo, Dr. T. Tchumatchenko, Ms. C. Reichenbach, and the three anonymous reviewers for comments on the manuscript. T. R. was supported by a Career Award at the Scientific Interface from the Burroughs Wellcome Fund; A. J. H. is an Investigator of Howard Hughes Medical Institute.